\documentclass[aps,prl,reprint,nobibnotes,nofootinbib]{revtex4-1}
\usepackage[margin=1in]{geometry}
\usepackage{amsmath,amssymb,amsfonts,amsthm,mathtools}
\usepackage{graphicx}
\usepackage{tikz}
\usetikzlibrary{
    arrows.meta,
    positioning,
    calc,
    decorations.pathmorphing
}
\usepackage{pgfplots}
\pgfplotsset{compat=1.18}
\usepackage{bbm}
\usepackage{mathrsfs}
\usepackage{hyperref}
\usepackage{soul}
\usepackage{slashed}
\usepackage{caption}
\hypersetup{colorlinks=true,linkcolor=blue,citecolor=blue,urlcolor=blue}
\usepackage{pgfplots}
\pgfplotsset{compat=1.18}
\usetikzlibrary{arrows.meta,positioning,calc}

\usepackage{tikz}
\usepackage{tikz-feynman}
\usetikzlibrary{arrows.meta,calc,decorations.pathmorphing,decorations.markings}

\begin{document}

\title{A First Bound on the Moffat Energy and Thorium--229 Clock as a Probe of the Nonlocal Time-Energy Structure}

\author{E.~J.~Thompson}
\thanks{Corresponding author (Ethan Thompson)}
\email{thom3471@mylaurier.ca}
\affiliation{Wilfrid Laurier University, Waterloo, Canada, N2L 3C5}

\author{Arvin Kouroshnia}
\affiliation{Department of Physics and Astronomy, University of Waterloo, Waterloo, Ontario N2L 3G1, Canada}

\author{J. W. Moffat}
\affiliation{Perimeter Institute for Theoretical Physics, Waterloo, Ontario N2L 2Y5, Canada}
\affiliation{Department of Physics and Astronomy, University of Waterloo, Waterloo, Ontario N2L 3G1, Canada}

\author{Caspian Chyrak}
\affiliation{Department of Physics and Astronomy, University of Waterloo, Waterloo, Ontario N2L 3G1, Canada}

\author{Gabriel Gervais}
\affiliation{Department of Physics and Astronomy, University of Waterloo, Waterloo, Ontario N2L 3G1, Canada}

\author{H.~A.~Carteret}
\affiliation{Waterloo, Ontario, Canada}

\date{\today}

\begin{abstract}
    The new thorium--229 nuclear clocks give us a new laboratory system in which to test if a modification of the Time-Energy uncertainty principle can be observed. In this paper we derive the nonlocal Time-Energy uncertainty principle and then apply the published $^{229}$Th nuclear clock data as a first probe of the nonlocality scale \(E_M\). We note that since these experiments were not designed to test nonlocal quantum field theory the constraints are interpreted as motivating bounds and not definitive exclusions. By using the direct clock-energy channel, we find conservative lower bounds on \(E_M\) at the tens of MeV scale, with optimistic present-data estimates reaching the hundred MeV scale. Including the known nuclear sensitivity enhancement of the $^{229}$Th transition gives us stronger model dependent bounds in the GeV range, while nuclear-scale reference-energy scenarios can reach the TeV range. The conclusion we draw from this is that nuclear clocks already provide an experimental route from nonlocal time--energy uncertainty to measurable laboratory bounds on non-Planckian nonlocality.
\end{abstract}

\maketitle

\section{Introductory Considerations}

In June of 2026 nuclear clocks were first built and successfully demonstrated by independent research teams in Europe and China~\cite{ToscaniDeCol2026,Huang2026}. Nuclear clocks measure time by tracking the Rabi oscillations between states of the nucleus of an atom, rather than the behavior of its outer electrons such as those used in atomic clocks~\cite{VonDerWense2018,Thirolf2019,VonDerWenseSeiferle2020,Peik2021,BeeksReview2021,Lu2026Review}. Thorium itself was identified more than a century before the discovery of its unusual low-energy nuclear isomer, Berzelius was the first to propose the existence of a new elemental ``earth,'' which he called Thorina, in 1818~\cite{Berzelius1818Thorina}. This original identification was subsequently withdrawn when the material was found to be a compound of yttrium rather than a new element~\cite{Berzelius1825Thorina}. Then in 1828 the Norwegian mineralogist Morten Thrane Esmark discovered an unusual black mineral on the island of L{\o}v{\o}ya near Brevik, Norway. The sample was passed through his father, Jens Esmark, to J\"ons Jacob Berzelius, who chemically analyzed it, established that it contained a previously unknown element, and published the discovery in 1829~\cite{Berzelius1829Thorium}. Berzelius named the new element thorium, after Thor, and named its source mineral thorite. Nuclear clocks use $^{229}$Th specifically, because it has the lowest nuclear excitation energy of any known isotope of any element, as it has isomeric states, one of which $^{229m}$Th$^+$ has a half-life of 29 minutes\footnote{An isomeric state is a metastable excited state of an atomic nucleus, which can persist for timescales ranging from nanoseconds to years before releasing excess energy as gamma rays or internal conversion electrons~\cite{WalkerPodolyak2020,Garg2023}.}~\cite{KrogerReich1976,ReichHelmer1990,Burke1990,HelmerReich1994,Barci2003,Ruchowska2006,Beck2007}. In terms of size, for $^{229}$Th the electron shell is $\sim4\times10^{-10}$ meters. For $^{229}$Th the nucleus is made of 90 protons and 139 neutrons, has a diameter of approximately $1.5 \times 10^{-14}$ meters~\cite{AngeliMarinova2013}, the electron shell of thorium-229 is $2.7 \times 10^{4}$ times larger. Modern optical clocks, together with optical-frequency-comb metrology and SI-traceable frequency standards, provide the precision framework in which nuclear clocks are developed and compared, with applications extending to relativistic geodesy~\cite{Ludlow2015,Diddams2020,Mehlstaubler2018,BIPMSI2019}.

The unusual low-energy structure of $^{229}$Th was first identified indirectly by Kroger and Reich in 1976 through gamma-ray spectroscopy following the alpha decay of $^{233}$U~\cite{KrogerReich1976}. Then subsequent spectroscopic analyses established that the first excited state lies extraordinarily close to the nuclear ground state~\cite{ReichHelmer1990,Burke1990,HelmerReich1994}. Early theoretical work examined the decay channels, optical excitation, and possible metrological applications of this anomalously low-energy nuclear transition~\cite{StrizhovTkalya1991,Tkalya1996}. The first detailed proposal for using the $^{229}$Th transition as an optical nuclear clock was then developed by Peik and Tamm in 2003~\cite{PeikTamm2003}. The isomeric transition was directly detected experimentally in 2016 by von der Wense and collaborators~\cite{VonDerWense2016}.

The development of nonlocal field theory extends from the early formulations of Yukawa, Pais--Uhlenbeck, and Efimov through later work on causality, unitarity, and nonlocal gauge theories~\cite{Yukawa1950I,Yukawa1950II,PaisUhlenbeck1950,Efimov1967,AlebastrovEfimov1973,Efimov1974,AlebastrovEfimov1974,Efimov1977,Krasnikov1987}. We have spent the past year building on and reformulating a class of nonlocal quantum field theory based on entire-functions of the covariant Laplace–Beltrami, or d'Alembertian operator. Plane waves diagonalize the d'Alembertian operator over flat backgrounds, the regulator reduces cleanly to a multiplicative form factor (\(F(\Box) = \exp(\Box / E_M^2)\)) in Minkowski momentum space. Upon Wick rotation to Euclidean space this setup provides exponential ultraviolet (UV) damping in loop integrals. Crucially, it suppresses high-energy infinities without introducing ghost states, new complex poles, or unphysical branch cuts~\cite{Moffat1990,Evens1991,KleppeWoodard1992,Moffat2011Gauge,Moffat2011Gravity,Moffat2019,GreenMoffat2021,Biswas2012,Tomboulis2015,ModestoPivaRachwal2016,ModestoRachwal2017,ChinTomboulis2018,BuoninfanteKoshelevMazumdar2018,KoshelevTokareva2021,Buoninfante2022,Krasnikov2024,MoffatThompson2026Regulators}.

The recent model of NLQFT is similar to the 1990/1991 approach but has some differences. We have focused mainly on solving the mathematical problems of gauge invariance and quantum gravity. Unless it is done carefully, introducing nonlocal smearing into a theory breaks gauge invariance. We resolved this by implementing regulators as an entire function of the covariant Laplace–Beltrami (or d'Alembertian) operator. The difference in approach centers on how gauge covariance and background freedom are treated and mathematically proven. Moffat’s early framework, with Evens, Kleppe, and Woodard, regularized theories by using an intricate perturbative technique to impose gauge invariance~\cite{Moffat1990,Evens1991,KleppeWoodard1992}. To maintain gauge invariance when ``smearing'' a point particle, they had to introduce nonlocality step-by-step through path-ordered exponentials and Mandelstam lines attached to individual interaction vertices. It required complex, diagram-dependent modifications that were incredibly difficult to calculate beyond one loop. The Moffat--Thompson approach moved the nonlocality directly into the baseline kinetic operators using a rigorous background-field formalism where the regulator is implemented cleanly as an entire function of the covariant Laplace–Beltrami operator. This means the nonlocality is built into the fabric of the fields themselves rather than being stitched onto individual interaction vertices as an ad-hoc fix~\cite{MoffatThompson2026Regulators,Thompson2026Covariance}. Related work in this program has examined particle masses, renormalization-group structure, entanglement, localization, phase space, covariance, microcausality, macrocausality, and fermion doubling in entire-function nonlocal theories~\cite{Moffat2021Masses,GreenMoffat2021,LandryMoffat2024,Thompson2026Localization,Thompson2026PhaseSpace,Thompson2026Covariance,Thompson2026Microcausality,Thompson2026SMatrix,MoffatKouroshnia2026Doubling}.

In standard local QFT, a particle detector can theoretically measure a particle at an exact, zero-width point in spacetime. We mathematically model a detector interacting with an ultraviolet-complete NLQFT regulated by an entire function of the d'Alembertian and prove that under an induced equal-time detector response kernel, the measured width of a particle is subject to a strict variance-addition law. Essentially, the intrinsic nonlocality of the universe acts like an inescapable, fundamental structural ``fuzziness'' added directly to any experimental measurement. By combining this variance addition law with the standard Heisenberg uncertainty inequality $\Delta x \Delta p \ge \hbar/2$, we established a modified generalized uncertainty relation~\cite{Heisenberg1927,Kennard1927,Robertson1929,Schrodinger1930}. This mathematical bound behaves precisely like a squeezed state across scales. So as you inject more energy to try and resolve a smaller point (UV), the framework hits a fundamental wall, the field equations contract, establishing a minimal localization length of the order \(L_{M}\) (governed by the nonlocality scale \(E_M\)). Meaning that you can no longer squeeze the position uncertainty any further, meaning spacetime points completely cease to be physical observables at the Planck scale.

This is why squeezed states serve as the ideal test for nonlocal quantum field theory because they translate abstract infinite-derivative mathematics into a physical, measurable minimum noise threshold~\cite{ScullyZubairy1997,BraunsteinVanLoock2005,Weedbrook2012}. By attempting to compress quantum uncertainty, we can look for an inescapable structural ``fuzziness'' in the vacuum that signals the presence of a fundamental nonlocality scale. Calculating the phase evolution of these states allows us to verify that information does not leak faster than light, proving that the model successfully eliminates point singularities without violating macroscopic causality~\cite{AlebastrovEfimov1974,Thompson2026Microcausality,Thompson2026SMatrix}.

So in this paper we will apply the same logic as the previous work with position and momentum uncertainty in NLQFT and apply it to energy and time. The local time--energy uncertainty principle is more subtle than the position--momentum uncertainty principle due to the fact that time is not ordinarily introduced as a self-adjoint operator conjugate to a semibounded Hamiltonian~\cite{Pauli1933,MandelstamTamm1945,AharonovBohm1961,Busch1990I,Busch1990II,Busch2008,DeffnerCampbell2017}. For clocks the relevant quantities are operational and unambiguous and these are exactly the quantities that define the time--energy phase-space response of a clock. If nonlocal quantum field theory is correct and imposes a fundamental temporal or energetic response width, then a sufficiently precise clock should not behave exactly as a purely local theory predicts.

The solid-state clock places \(^{229}\mathrm{Th}\) nuclei in a calcium fluoride crystal and that crystal provides a transparent host for the vacuum-ultraviolet radiation and suppresses unwanted electronic decay channels. But the host crystal is not a passive background; thorium enters the crystal primarily as a tetravalent ion replacing a divalent calcium site, so charge compensation, local defects, strain, electric-field gradients, temperature shifts, and inhomogeneous broadening all affect the measured nuclear line. These chemical and material effects are essential for the clock's operation, but they are also the main obstacle to interpreting any residual linewidth or frequency shift as fundamental physics~\cite{Rellergert2010,Kazakov2012,Dessovic2014,Nickerson2020,Nickerson2021,Pineda2025,Higgins2025,Schaden2025,Terhune2025,Morgan2025,Zhao2026Materials,Zhang2024Films}. For this reason, the present work treats existing solid-state thorium--229 data as a first phenomenological constraint and not as a dedicated discovery experiment.

The reason for using thorium--229 is not only its precision, but because it may also act as an amplifier for high-energy physics, while the observed transition energy is only a few eV, but it arises from a near cancellation among much larger nuclear and electromagnetic contributions so as a result small modifications to nuclear structure, Coulomb energy, or fundamental couplings can be magnified in the final transition frequency. This is the same reason thorium--229 has been proposed as a sensitive probe of variations of fundamental constants and ultralight dark matter~\cite{PeikTamm2003,Flambaum2006,HayesFriar2007,FlambaumWiringa2009,Fadeev2020,Beeks2025,Caputo2025,Delaunay2025,DereviankoPospelov2014,Arvanitaki2015,VanTilburg2015,Hees2016,Safronova2018,Arakawa2026}. In the present work the same amplification mechanism may suggest that a low-energy nuclear clock may be sensitive to a nonlocality scale that is much higher than the clock photon energy itself.

In this paper we use the already published thorium--229 clock data to obtain a preliminary lower bound on \(E_M\), but again since the experiments were not designed to search for nonlocal time--energy broadening, we interpret the result conservatively. We use the published transition frequency, frequency uncertainty, clock reproducibility, and stability information as limits on any unexplained fractional nonlocal contribution~\cite{Zhang2024Frequency,Ooi2025Reproducibility,ToscaniDeCol2026,Huang2026}. Under the leading quadratic correction model the direct clock-energy channel gives bounds at the MeV scale. In our analysis, the direct frequency-uncertainty channel gives \(E_M>18.7\,\mathrm{MeV}\), the conservative reproducibility channel gives \(E_M>22.3\,\mathrm{MeV}\), and the optimistic day-stability channel gives \(E_M>101\,\mathrm{MeV}\). These numbers are not presented as final experimental exclusions, but as the first laboratory-scale estimates of the nonlocality scale from nuclear-clock data. We then consider the nuclear-enhanced interpretation where if the thorium--229 sensitivity enhancement is included phenomenologically, then the same data give stronger model-dependent bounds. The conservative enhanced channel gives \(E_M>1.72\,\mathrm{GeV}\), while the optimistic enhanced channel gives \(E_M>7.78\,\mathrm{GeV}\). And finally, if we assume that the relevant reference energy is not the optical transition energy but a nuclear-scale internal energy, illustrative model-dependent bounds reach the TeV scale, with \(E_M>2.67\,\mathrm{TeV}\) for a \(1\,\mathrm{MeV}\) reference scale and \(E_M>26.7\,\mathrm{TeV}\) for a \(10\,\mathrm{MeV}\) reference scale. These latter bounds are deliberately separated from the direct clock-energy result, because they require a more detailed model of how the nonlocal regulator enters nuclear binding and Coulomb structure.

\section{THE NONLOCAL TIME--ENERGY PRINCIPLE}

We let $A(t)$ be a complex temporal amplitude associated with a clock interrogation, then the variable $t$ is the laboratory time read by the clock apparatus, and $A(t)$ is the temporal envelope, phase, and coherence of the nuclear transition. We normalize the corresponding time-domain probability density by writing:
\begin{equation}
    r(E)=|\widetilde A(E)|^2,
\qquad
\int_{-\infty}^{\infty}r(E)\,dE=1 .
\label{eq:energy_density}
\end{equation}

The mean event time is given by:
\begin{equation}
\overline T
=
\int_{-\infty}^{\infty}t\,q(t)\,dt ,
\label{eq:mean_time}
\end{equation}
and the local time variance is given:
\begin{equation}
(\Delta T)^2
=
\int_{-\infty}^{\infty}
(t-\overline T)^2 q(t)\,dt .
\label{eq:local_time_variance}
\end{equation}
Similarly, the mean energy is given:
\begin{equation}
\overline E
=
\int_{-\infty}^{\infty}E\,r(E)\,dE ,
\label{eq:mean_energy}
\end{equation}
and the local energy variance is given by:
\begin{equation}
(\Delta E)^2
=
\int_{-\infty}^{\infty}
(E-\overline E)^2 r(E)\,dE .
\label{eq:local_energy_variance}
\end{equation}

With the Fourier convention in the ordinary time--energy uncertainty relation takes the form~\cite{MandelstamTamm1945,AharonovBohm1961,Busch1990I,Busch1990II,Busch2008}:
\begin{equation}
\Delta T\,\Delta E \geq \frac{\hbar}{2},
\label{eq:ordinary_time_energy}
\end{equation}
this relation says that a more sharply localized clock event requires a more broad energy bandwidth, while a sharply defined energy requires a long temporal coherence time. In terms of clock spectroscopy this is the usual relation between interrogation time and linewidth.

In nonlocal quantum field theory all observables are at least somewhat delocalized. So we let $F(\Box/E_M^2)$ be an entire-function nonlocal regulator, where $\Box$ is the spacetime d'Alembertian and $E_M$ is the nonlocality energy scale, basically the energy where nonlocalities must be taken into account. The regulator induces a nonlocal response kernel, where in the temporal sector we write this kernel as $\eta_F(t)$. The observed event-time distribution is then the convolution:
\begin{equation}
q_F(t)
=
(\eta_F*q)(t)
=
\int_{-\infty}^{\infty}
\eta_F(t-s)q(s)\,ds
\label{eq:time_convolution}
\end{equation}
where $q(t)$ is the local event-time distribution, while $q_F(t)$ is the distribution actually measured by a nonlocal clock observable.

We assume that the temporal response kernel is normalized by:
\begin{equation}
\int_{-\infty}^{\infty}\eta_F(t)\,dt=1,
\label{eq:eta_normalization}
\end{equation}
centered:
\begin{equation}
\int_{-\infty}^{\infty}t\,\eta_F(t)\,dt=0,
\label{eq:eta_centered}
\end{equation}
and has a finite variance:
\begin{equation}
\tau_F^2
=
\int_{-\infty}^{\infty}t^2\eta_F(t)\,dt < \infty ,
\label{eq:tau_def}
\end{equation}
$\tau_F$ is the nonlocal temporal response width; it is the time-domain analogue of the nonlocal spatial resolution width. Dimensionally, it is controlled by the nonlocality scale:
\begin{equation}
\tau_F
=
\alpha_t\frac{\hbar}{E_M},
\label{eq:tau_scale}
\end{equation}
where $\alpha_t$ is a dimensionless constant determined by the detailed shape of the regulator and the clock response.

The mean of the measured distribution is given by:
\begin{align}
\overline T_F
&=
\int_{-\infty}^{\infty}t\,q_F(t)\,dt
\nonumber\\
&=
\int_{-\infty}^{\infty}dt\,t
\int_{-\infty}^{\infty}ds\,\eta_F(t-s)q(s).
\label{eq:mean_measured_start}
\end{align}
We then set $u=t-s$, so that $t=u+s$. Then we find:
\begin{align}
\overline T_F
&=
\int_{-\infty}^{\infty}ds\,q(s)
\int_{-\infty}^{\infty}du\,(u+s)\eta_F(u)
\nonumber\\
&=
\int_{-\infty}^{\infty}ds\,q(s)
\left[
\int_{-\infty}^{\infty}u\eta_F(u)\,du
+
s\int_{-\infty}^{\infty}\eta_F(u)\,du
\right].
\label{eq:mean_measured_expand}
\end{align}
Then using Eqs.~\eqref{eq:eta_normalization} and \eqref{eq:eta_centered}, this becomes:
\begin{equation}
\overline T_F
=
\int_{-\infty}^{\infty}s\,q(s)\,ds
=
\overline T ,
\label{eq:mean_preserved}
\end{equation}
so a centered nonlocal response kernel does not shift the mean event time.

The measured temporal variance is given by:
\begin{equation}
(\Delta T_F)^2
=
\int_{-\infty}^{\infty}
(t-\overline T_F)^2 q_F(t)\,dt .
\label{eq:measured_time_variance_def}
\end{equation}
Then using $\overline T_F=\overline T$, Eq.~\eqref{eq:time_convolution}, and again setting $u=t-s$, we find:
\begin{align}
(\Delta T_F)^2
&=
\int_{-\infty}^{\infty}ds\,q(s)
\int_{-\infty}^{\infty}du\,
(u+s-\overline T)^2\eta_F(u)
\nonumber\\
&=
\int_{-\infty}^{\infty}ds\,q(s)
\int_{-\infty}^{\infty}du\,
\left[
u^2
+
2u(s-\overline T)
\right.\nonumber \\
&+ \left. (s-\overline T)^2
\right]\eta_F(u).
\label{eq:variance_expand}
\end{align}
The cross term vanishes because the kernel is centered, so:
\begin{align}
(\Delta T_F)^2
&=
\int_{-\infty}^{\infty}u^2\eta_F(u)\,du
+
\int_{-\infty}^{\infty}(s-\overline T)^2q(s)\,ds
\nonumber\\
&=
\tau_F^2+(\Delta T)^2 .
\label{eq:time_variance_addition}
\end{align}
So the measured clock-time width obeys the exact addition law:
\begin{equation}
(\Delta T_F)^2
=
(\Delta T)^2+\tau_F^2 .
\label{eq:boxed_time_variance}
\end{equation}

Combining Eq.~\eqref{eq:boxed_time_variance} with the ordinary time--energy relation \eqref{eq:ordinary_time_energy} gives us:
\begin{equation}
\Delta T_F
\geq
\sqrt{
\frac{\hbar^2}{4(\Delta E)^2}
+
\tau_F^2
}.
\label{eq:nonlocal_time_energy_first}
\end{equation}
Using Eq.~\eqref{eq:tau_scale}, this becomes:
\begin{equation}
\Delta T_F
\geq
\sqrt{
\frac{\hbar^2}{4(\Delta E)^2}
+
\alpha_t^2\frac{\hbar^2}{E_M^2}
}\;\;,
\label{eq:nonlocal_time_energy_bound}
\end{equation}
this is the nonlocal time resolution bound. In ordinary local quantum field theory by increasing energy bandwidth, we can make the event-time width arbitrarily small, but in the nonlocal theory the measured width approaches the finite limit:
\begin{equation}
\lim_{\Delta E\rightarrow\infty}\Delta T_F
=
\tau_F
=
\alpha_t\frac{\hbar}{E_M},
\label{eq:temporal_floor}
\end{equation}
so the nonlocality scale $E_M$ is equivalently a minimum intrinsic clock-time scale.

The energy side can be treated in the same way as we just treated time, we let $\chi_F(E)$ be the nonlocal response kernel in energy space, then the observed energy distribution is given by:
\begin{equation}
r_F(E)
=
(\chi_F*r)(E)
=
\int_{-\infty}^{\infty}
\chi_F(E-E')r(E')\,dE' .
\label{eq:energy_convolution}
\end{equation}
We again assume:
\begin{equation}
\int_{-\infty}^{\infty}\chi_F(E)\,dE=1,
\qquad
\int_{-\infty}^{\infty}E\,\chi_F(E)\,dE=0,
\label{eq:chi_assumptions}
\end{equation}
and:
\begin{equation}
\epsilon_F^2
=
\int_{-\infty}^{\infty}E^2\chi_F(E)\,dE < \infty .
\label{eq:epsilon_def}
\end{equation}
By repeating the same convolution argument as before, this gives us:
\begin{equation}
(\Delta E_F)^2
=
(\Delta E)^2+\epsilon_F^2 ,
\label{eq:energy_variance_addition}
\end{equation}
where $\epsilon_F$ is the nonlocal energy response width, now since clock experiments measure frequency rather than energy directly, we use:
\begin{equation}
E=h\nu=\hbar\omega ,
\label{eq:energy_frequency}
\end{equation}
where $\nu$ is ordinary frequency and $\omega=2\pi\nu$ is angular frequency. Therefore we have:
\begin{equation}
\epsilon_F
=
h\sigma_{\nu,F}
=
\hbar\sigma_{\omega,F},
\label{eq:epsilon_frequency_width}
\end{equation}
where $\sigma_{\nu,F}$ is the nonlocal contribution to the clock linewidth in frequency units.

For the clock line, the observed frequency variance can be written as:
\begin{equation}
\sigma_{\nu,\mathrm{obs}}^2
=
\sigma_{\nu,\mathrm{known}}^2
+
\sigma_{\nu,F}^2 ,
\label{eq:clock_variance_split}
\end{equation}
where $\sigma_{\nu,\mathrm{known}}$ includes the known local contributions, the residual term $\sigma_{\nu,F}$ is the possible nonlocal contribution, in an experiment if no residual linewidth is observed, then the experiment sets an upper bound on $\sigma_{\nu,F}$, and thus on $\epsilon_F$.

The full time--energy phase-space structure is described by the Wigner function~\cite{Wigner1932,Hillery1984}, this given by:
\begin{equation}
W(t,E)
=
\frac{1}{2\pi\hbar}
\int_{-\infty}^{\infty}
d\tau\,
e^{iE\tau/\hbar}
A\left(t-\frac{\tau}{2}\right)
A^*\left(t+\frac{\tau}{2}\right).
\label{eq:time_energy_wigner}
\end{equation}
The covariance matrix of this distribution is:
\begin{equation}
\Gamma^{(TE)}
=
\begin{pmatrix}
(\Delta T)^2 & \mathrm{Cov}(T,E)\\
\mathrm{Cov}(T,E) & (\Delta E)^2
\end{pmatrix},
\label{eq:local_covariance}
\end{equation}
where:
\begin{equation}
\mathrm{Cov}(T,E)
=
\langle TE\rangle-\langle T\rangle\langle E\rangle .
\label{eq:cov_def}
\end{equation}
For a local minimum-uncertainty Gaussian clock mode, the covariance matrix satisfies~\cite{BraunsteinVanLoock2005,Weedbrook2012}:
\begin{equation}
\det \Gamma^{(TE)}
\geq
\frac{\hbar^2}{4}.
\label{eq:wigner_uncertainty}
\end{equation}

The nonlocal response adds a nonlocal covariance matrix, this is represented by:
\begin{equation}
\Sigma_F^{(TE)}
=
\begin{pmatrix}
\tau_F^2 & c_{TE}\\
c_{TE} & \epsilon_F^2
\end{pmatrix},
\label{eq:nonlocal_covariance}
\end{equation}
where $c_{TE}$ parameterizes any nonlocal correlation between the temporal and energetic response channels. The measured covariance matrix is then given by:
\begin{equation}
\Gamma_F^{(TE)}
=
\Gamma^{(TE)}+\Sigma_F^{(TE)} .
\label{eq:measured_covariance}
\end{equation}
If the nonlocal response is diagonal, meaning $c_{TE}=0$, then we have:
\begin{equation}
\Gamma_F^{(TE)}
=
\begin{pmatrix}
(\Delta T)^2+\tau_F^2 & \mathrm{Cov}(T,E)\\
\mathrm{Cov}(T,E) & (\Delta E)^2+\epsilon_F^2
\end{pmatrix}.
\label{eq:diagonal_covariance}
\end{equation}
So the experimentally observed phase-space area becomes:
\begin{equation}
\det\Gamma_F^{(TE)}
=
\left[(\Delta T)^2+\tau_F^2\right]
\left[(\Delta E)^2+\epsilon_F^2\right]
-
\mathrm{Cov}(T,E)^2 .
\label{eq:nonlocal_phase_space_area}
\end{equation}

Equation~\eqref{eq:measured_covariance} is the most apparent experimental statement of the nonlocal time--energy principle, it tells us that the nonlocal scale $E_M$ does not have to appear only as a shift in the central clock frequency, but it can also appear as an irreducible temporal width, an irreducible energy width, an additional dephasing channel, or a residual contribution to the time--energy Wigner covariance of the nuclear transition.

For the first phenomenological bound that we will discuss in this paper, we use a simpler low-energy expansion. The leading clock-frequency correction is written as:
\begin{equation}
\frac{|\delta\nu_{\mathrm{NL}}|}{\nu_{\mathrm{Th}}}
=
|\beta_{\mathrm{NL}}|
\left(
\frac{E_{\mathrm{ref}}}{E_M}
\right)^n ,
\label{eq:phenomenological_frequency_shift}
\end{equation}
where $\nu_{\mathrm{Th}}$ is the thorium--229 nuclear clock frequency, $E_{\mathrm{ref}}$ is the reference energy scale to which the clock is sensitive, $n$ is the leading power in the low-energy regulator expansion, and $\beta_{\mathrm{NL}}$ is a dimensionless coefficient. For an even entire-function correction the leading generic choice is given by~\cite{MoffatThompson2026Regulators,Krasnikov2024}:
\begin{equation}
n=2 .
\label{eq:n_equals_two}
\end{equation}
If the experiment shows no unexplained fractional residual larger than:
\begin{equation}
\delta_{\mathrm{res}}
=
\left|
\frac{\delta\nu}{\nu_{\mathrm{Th}}}
\right|_{\mathrm{max}},
\label{eq:delta_res}
\end{equation}
then Eq.~\eqref{eq:phenomenological_frequency_shift} implies that:
\begin{equation}
|\beta_{\mathrm{NL}}|
\left(
\frac{E_{\mathrm{ref}}}{E_M}
\right)^n
<
\delta_{\mathrm{res}} .
\label{eq:bound_inequality}
\end{equation}
Solving for $E_M$, we find that:
\begin{equation}
E_M
>
E_{\mathrm{ref}}
\left(
\frac{|\beta_{\mathrm{NL}}|}{\delta_{\mathrm{res}}}
\right)^{1/n}.
\label{eq:direct_EM_bound}
\end{equation}

If the thorium--229 nuclear transition amplifies the response by a factor $K$, how this may happen will be discussed later on, the phenomenological shift is instead given by:
\begin{equation}
\frac{|\delta\nu_{\mathrm{NL}}|}{\nu_{\mathrm{Th}}}
=
K|\beta_{\mathrm{NL}}|
\left(
\frac{E_{\mathrm{ref}}}{E_M}
\right)^n .
\label{eq:enhanced_frequency_shift}
\end{equation}
The corresponding enhanced bound is given by:
\begin{equation}
E_M
>
E_{\mathrm{ref}}
\left(
\frac{K|\beta_{\mathrm{NL}}|}{\delta_{\mathrm{res}}}
\right)^{1/n}.
\label{eq:enhanced_EM_bound}
\end{equation}
Equations~\eqref{eq:direct_EM_bound} and \eqref{eq:enhanced_EM_bound} are the two formulas used below to get the first bounds from existing thorium--229 clock data. The direct bound uses the clock transition energy as $E_{\mathrm{ref}}$, and the enhanced bound uses the same clock data, but includes the known nuclear sensitivity enhancement as a phenomenological amplifier. The latter is stronger, but more model-dependent so should be taken with care.

\section{THE NUCLEUS AS A CLOCK}

The new thorium--229 nuclear clock uses the nucleus instead of the electron cloud as the frequency reference~\cite{Campbell2011,Campbell2012,Kazakov2012,VonDerWense2018,Thirolf2019,VonDerWenseSeiferle2020,Peik2021,BeeksReview2021,StrizhovTkalya1991,Tkalya1996,PeikTamm2003}. The important system is the two-level nuclear transition:
\begin{equation}
^{229}\mathrm{Th}_{g}
\longleftrightarrow
^{229m}\mathrm{Th}^+,
\label{eq:thorium_transition}
\end{equation}
where $^{229}\mathrm{Th}_{g}$ is the nuclear ground state and $^{229m}\mathrm{Th}^+$ is the long-lived low-energy isomeric state. The transition energy is given by:
\begin{equation}
E_{\mathrm{Th}}
=
E_m-E_g
=
h\nu_{\mathrm{Th}}
=
\hbar\omega_{\mathrm{Th}},
\label{eq:thorium_energy}
\end{equation}
where $E_g$ and $E_m$ are the nuclear energies of the ground and isomeric states, $\nu_{\mathrm{Th}}$ is the ordinary clock frequency, and $\omega_{\mathrm{Th}}=2\pi\nu_{\mathrm{Th}}$ is the angular frequency.

The interesting feature of $^{229}\mathrm{Th}$ is that this nuclear energy splitting lives in the optical or vacuum-ultraviolet range~\cite{VonDerWense2016,Thielking2018,Masuda2019,Seiferle2019,Yamaguchi2019,Sikorsky2020,Kraemer2023,Tiedau2024,Elwell2024,Yamaguchi2024,Zhang2024Frequency,Elwell2025}:
\begin{equation}
E_{\mathrm{Th}}\simeq 8.36\,\mathrm{eV},
\qquad
\nu_{\mathrm{Th}}\simeq 2.02\times 10^{15}\,\mathrm{Hz}.
\label{eq:thorium_numbers}
\end{equation}
The corresponding wavelength is then given by:
\begin{equation}
\lambda_{\mathrm{Th}}
=
\frac{c}{\nu_{\mathrm{Th}}}
=
\frac{hc}{E_{\mathrm{Th}}}
\simeq 148\,\mathrm{nm},
\label{eq:thorium_wavelength}
\end{equation}
where $c$ is the speed of light, so the nuclear transition can be driven directly by a narrow-band vacuum-ultraviolet laser~\cite{PeikTamm2003,Tiedau2024,Elwell2024}.

In the simplest clock description the thorium nucleus is treated as a two-level quantum system with basis states:
\begin{equation}
|g\rangle,
\qquad
|m\rangle,
\label{eq:clock_basis}
\end{equation}
where $|g\rangle$ is the ground nuclear state and $|m\rangle$ is the isomeric nuclear state. The unperturbed Hamiltonian can be written as:
\begin{equation}
H_0
=
E_g |g\rangle\langle g|
+
E_m |m\rangle\langle m| ,
\label{eq:clock_hamiltonian}
\end{equation}
where a laser of angular frequency $\omega_L$ drives the transition. The detuning is:
\begin{equation}
\delta
=
\omega_L-\omega_{\mathrm{Th}},
\label{eq:detuning}
\end{equation}
where $\omega_L$ is the laser angular frequency, when $\delta=0$, the laser is resonant with the nuclear transition.

For a coherent interrogation pulse of duration time $T$, the excitation probability has the standard Rabi form~\cite{Rabi1937,ScullyZubairy1997}:
\begin{equation}
P_m(\delta,T)
=
\frac{\Omega^2}{\Omega^2+\delta^2}
\sin^2
\left[
\frac{T}{2}
\sqrt{\Omega^2+\delta^2}
\right],
\label{eq:rabi_probability}
\end{equation}
where $P_m$ is the probability of finding the nucleus in the isomeric state and $\Omega$ is the Rabi frequency. The Rabi frequency measures the strength of the coupling between the nuclear transition and the laser field. On resonance, $\delta=0$, Eq.~\eqref{eq:rabi_probability} becomes:
\begin{equation}
P_m(0,T)
=
\sin^2\left(\frac{\Omega T}{2}\right).
\label{eq:on_resonance_rabi}
\end{equation}

The clock is built by using the nuclear resonance to stabilize the laser, if the laser frequency is too low or too high then the excitation probability changes. So the experiment measures this change through absorption, fluorescence, conversion-electron detection, or another readout channel. A feedback loop then corrects the laser frequency until it remains locked to the nuclear resonance, the clock output is therefore the stabilized laser frequency~\cite{ToscaniDeCol2026,Huang2026}:
\begin{equation}
\nu_{\mathrm{clock}}
=
\nu_{\mathrm{Th}} .
\label{eq:clock_output}
\end{equation}

In the solid-state implementation, the $^{229}\mathrm{Th}$ nuclei are embedded in a calcium fluoride crystal:
\begin{equation}
^{229}\mathrm{Th}:\mathrm{CaF}_2 ,
\label{eq:th_caf2}
\end{equation}
the crystal is transparent near the required vacuum-ultraviolet wavelength and helps suppress unwanted electronic decay channels~\cite{Rellergert2010,Kazakov2012,Dessovic2014,Nickerson2020,Nickerson2021,Pineda2025,Zhao2026Materials}. Chemically, thorium enters primarily as a tetravalent ion, $\mathrm{Th}^{4+}$, replacing a divalent calcium ion, $\mathrm{Ca}^{2+}$, this substitution requires local charge compensation so the nuclear clock transition is affected by crystal defects, strain, electric-field gradients, temperature shifts, and inhomogeneous broadening. The measured frequency is thus not simply the bare nuclear frequency, but is:
\begin{equation}
\nu_{\mathrm{obs}}
=
\nu_{\mathrm{Th}}
+
\Delta\nu_{\mathrm{crystal}}
+
\Delta\nu_{\mathrm{temp}}
+
\Delta\nu_{\mathrm{EM}}
+
\Delta\nu_{\mathrm{laser}}
+
\Delta\nu_{\mathrm{stat}},
\label{eq:observed_clock_frequency}
\end{equation}
where $\Delta\nu_{\mathrm{crystal}}$ is the crystal-field and defect shifts, $\Delta\nu_{\mathrm{temp}}$ is the thermal shifts, $\Delta\nu_{\mathrm{EM}}$ is the electromagnetic shifts such as Zeeman and Stark shifts, $\Delta\nu_{\mathrm{laser}}$ is laser noise, and $\Delta\nu_{\mathrm{stat}}$ is the statistical measurement uncertainty.

For our search for nonlocal time--energy structure, we allow one additional contribution, a nonlocal contribution:
\begin{equation}
\nu_{\mathrm{obs}}
=
\nu_{\mathrm{Th}}
+
\Delta\nu_{\mathrm{known}}
+
\Delta\nu_{\mathrm{NL}},
\label{eq:clock_with_nonlocal_shift}
\end{equation}
where $\Delta\nu_{\mathrm{known}}$ contains the ordinary local clock shifts and $\Delta\nu_{\mathrm{NL}}$ is the possible nonlocal contribution. If no unexplained residual is observed, then we say that:
\begin{equation}
\left|
\frac{\Delta\nu_{\mathrm{NL}}}{\nu_{\mathrm{Th}}}
\right|
<
\delta_{\mathrm{res}},
\label{eq:clock_residual_bound}
\end{equation}
where $\delta_{\mathrm{res}}$ is the experimentally allowed residual fractional frequency shift.

We would like to reiterate that existing $^{229}$Th clock measurements were not designed as a test of the nonlocal time--energy broadening of the uncertainty principle. Nevertheless, they already constrain how large an unexplained nonlocal contribution could be, so the logic is simple; if a nonlocal correction shifted or broadened the thorium clock transition by a fractional amount larger than the reported residual experimental uncertainty, it would already have appeared as an unexplained clock anomaly.

We will parameterize the leading nonlocal contribution by:
\begin{equation}
\left|
\frac{\Delta\nu_{\mathrm{NL}}}{\nu_{\mathrm{Th}}}
\right|
=
|\beta_{\mathrm{NL}}|
\left(
\frac{E_{\mathrm{ref}}}{E_M}
\right)^n ,
\label{eq:leading_clock_shift}
\end{equation}
where $\Delta\nu_{\mathrm{NL}}$ is the possible nonlocal frequency contribution, $\nu_{\mathrm{Th}}$ is the thorium--229 clock frequency, $E_{\mathrm{ref}}$ is the energy scale to which the clock is assumed to be sensitive, $E_M$ is the nonlocality scale, $n$ is the leading power in the low-energy expansion, and $\beta_{\mathrm{NL}}$ is a dimensionless coefficient. In the numerical bounds below we use the minimal choice:
\begin{equation}
n=2,
\qquad
|\beta_{\mathrm{NL}}|=1 .
\label{eq:minimal_model_values}
\end{equation}

If the largest allowed unexplained fractional residual is $\delta_{\mathrm{res}}$, then:
\begin{equation}
\left|
\frac{\Delta\nu_{\mathrm{NL}}}{\nu_{\mathrm{Th}}}
\right|
<
\delta_{\mathrm{res}} .
\label{eq:residual_condition}
\end{equation}
Combining Eqs.~\eqref{eq:leading_clock_shift} and \eqref{eq:residual_condition} gives us:
\begin{equation}
E_M
>
E_{\mathrm{ref}}
\left(
\frac{|\beta_{\mathrm{NL}}|}{\delta_{\mathrm{res}}}
\right)^{1/n}.
\label{eq:direct_bound_repeated}
\end{equation}
For the direct clock-energy channel, we take:
\begin{equation}
E_{\mathrm{ref}}=E_{\mathrm{Th}}=h\nu_{\mathrm{Th}} .
\label{eq:direct_reference_energy}
\end{equation}
Using:
\begin{equation}
\nu_{\mathrm{Th}}
=
2.020407384335\times 10^{15}\,\mathrm{Hz},
\label{eq:thorium_frequency_used}
\end{equation}
and:
\begin{equation}
h
=
4.135667696\times 10^{-15}\,\mathrm{eV\,s},
\label{eq:planck_constant_ev}
\end{equation}
we obtain:
\begin{equation}
E_{\mathrm{Th}}
=
h\nu_{\mathrm{Th}}
=
8.355733552\,\mathrm{eV}.
\label{eq:transition_energy_used}
\end{equation}

The direct frequency-uncertainty channel uses the reported uncertainty in the measured transition frequency as the residual scale~\cite{ToscaniDeCol2026,Huang2026}, this is:
\begin{equation}
\delta_{\mathrm{res}}
=
2.0\times 10^{-13}.
\label{eq:frequency_uncertainty_residual}
\end{equation}
Equation~\eqref{eq:direct_bound_repeated} then gives us:
\begin{equation}
E_M
>
8.355733552\,\mathrm{eV}
\left(
\frac{1}{2.0\times10^{-13}}
\right)^{1/2}
=
1.87\times10^{7}\,\mathrm{eV}.
\label{eq:direct_frequency_bound}
\end{equation}
Thus:
\begin{equation}
E_M>18.7\,\mathrm{MeV}
\label{eq:direct_frequency_bound_box}
\end{equation}
in the direct frequency-uncertainty channel.

A more conservative bound uses the inter-crystal reproducibility or systematic reproducibility scale~\cite{Ooi2025Reproducibility} of:
\begin{equation}
\delta_{\mathrm{res}}
=
1.4\times10^{-13}.
\label{eq:conservative_residual}
\end{equation}
This gives us:
\begin{equation}
E_M
>
8.355733552\,\mathrm{eV}
\left(
\frac{1}{1.4\times10^{-13}}
\right)^{1/2}
=
2.23\times10^7\,\mathrm{eV},
\label{eq:conservative_direct_bound}
\end{equation}
or, can be written in terms of MeV:
\begin{equation}
E_M>22.3\,\mathrm{MeV}.
\label{eq:conservative_direct_bound_box}
\end{equation}
This is our preferred conservative direct bound, because it is tied to reproducibility rather than only statistical precision.

An optimistic present-data estimate uses the day-scale stability inferred from the reported clock instability~\cite{ToscaniDeCol2026,Huang2026}:
\begin{equation}
\delta_{\mathrm{res}}
=
6.80\times10^{-15}.
\label{eq:optimistic_residual}
\end{equation}
The corresponding bound is then:
\begin{equation}
E_M
>
8.355733552\,\mathrm{eV}
\left(
\frac{1}{6.80\times10^{-15}}
\right)^{1/2}
=
1.01\times10^8\,\mathrm{eV},
\label{eq:optimistic_direct_bound}
\end{equation}
or again in MeV:
\begin{equation}
E_M>101\,\mathrm{MeV}.
\label{eq:optimistic_direct_bound_box}
\end{equation}
This bound is stronger, but it is also more dependent on the assumption that the day-scale stability is a reliable proxy for the largest allowed unexplained residual.

The thorium--229 transition may also provide an amplified sensitivity channel, we will discuss why later on in the paper. The observed transition energy is only of the order of electron-volts, but it arises from cancellations among much larger nuclear and electromagnetic contributions, so we therefore also consider a phenomenological enhancement factor $K$:
\begin{equation}
\left|
\frac{\Delta\nu_{\mathrm{NL}}}{\nu_{\mathrm{Th}}}
\right|
=
K|\beta_{\mathrm{NL}}|
\left(
\frac{E_{\mathrm{ref}}}{E_M}
\right)^n .
\label{eq:enhanced_shift_section}
\end{equation}
Then the corresponding lower bound is:
\begin{equation}
E_M
>
E_{\mathrm{ref}}
\left(
\frac{K|\beta_{\mathrm{NL}}|}{\delta_{\mathrm{res}}}
\right)^{1/n}.
\label{eq:enhanced_bound_section}
\end{equation}
Now using the reported thorium sensitivity estimate~\cite{Flambaum2006,Fadeev2020,Beeks2025}:
\begin{equation}
K=5900,
\label{eq:enhancement_value}
\end{equation}
with the same direct clock reference energy $E_{\mathrm{ref}}=E_{\mathrm{Th}}$, the conservative enhanced channel gives us:
\begin{equation}
E_M
>
8.355733552\,\mathrm{eV}
\left(
\frac{5900}{1.4\times10^{-13}}
\right)^{1/2}
=
1.72\times10^9\,\mathrm{eV},
\label{eq:conservative_enhanced_bound}
\end{equation}
or in GeV:
\begin{equation}
E_M>1.72\,\mathrm{GeV}.
\label{eq:conservative_enhanced_bound_box}
\end{equation}
The optimistic enhanced channel gives us:
\begin{equation}
E_M
>
8.355733552\,\mathrm{eV}
\left(
\frac{5900}{6.80\times10^{-15}}
\right)^{1/2}
=
7.78\times10^9\,\mathrm{eV},
\label{eq:optimistic_enhanced_bound}
\end{equation}
or in GeV:
\begin{equation}
E_M>7.78\,\mathrm{GeV}.
\label{eq:optimistic_enhanced_bound_box}
\end{equation}

Finally we can consider a more model-dependent nuclear-scale reference energy where if the nonlocal regulator couples not merely to the observed optical transition energy but to internal nuclear Coulomb or binding-energy scales, then $E_{\mathrm{ref}}$ may be much larger than $E_{\mathrm{Th}}$. By using the conservative residual $\delta_{\mathrm{res}}=1.4\times10^{-13}$, we find that:
\begin{equation}
E_{\mathrm{ref}}=1\,\mathrm{MeV}
\quad\Longrightarrow\quad
E_M>2.67\,\mathrm{TeV},
\label{eq:one_mev_model_bound}
\end{equation}
and:
\begin{equation}
E_{\mathrm{ref}}=10\,\mathrm{MeV}
\quad\Longrightarrow\quad
E_M>26.7\,\mathrm{TeV}.
\label{eq:ten_mev_model_bound}
\end{equation}
These nuclear-scale bounds are not used as the headline result but we included them only to show how strongly the reach can increase if the relevant nonlocal energy scale is set by internal nuclear dynamics rather than by the observed optical transition energy.

The resulting bounds are summarized in Table~\ref{tab:em_bounds}, the direct bounds are the most conservative because they use only the measured thorium clock transition energy, the enhanced bounds are stronger but require the additional assumption that the known nuclear sensitivity of thorium--229 amplifies the nonlocal response, and the nuclear-scale bounds are the most speculative and require a detailed model of how the nonlocal regulator enters the nuclear Hamiltonian.

\begin{widetext}
\begin{table*}[t]
\centering
\begin{tabular}{lccc}
\hline
Channel & $\delta_{\mathrm{res}}$ & Assumption & Bound on $E_M$ \\
\hline
Direct frequency uncertainty
& $2.0\times10^{-13}$
& $E_{\mathrm{ref}}=E_{\mathrm{Th}}$
& $18.7\,\mathrm{MeV}$ \\

Conservative direct
& $1.4\times10^{-13}$
& $E_{\mathrm{ref}}=E_{\mathrm{Th}}$
& $22.3\,\mathrm{MeV}$ \\

Optimistic direct
& $6.80\times10^{-15}$
& $E_{\mathrm{ref}}=E_{\mathrm{Th}}$
& $101\,\mathrm{MeV}$ \\

Conservative enhanced
& $1.4\times10^{-13}$
& $K=5900$
& $1.72\,\mathrm{GeV}$ \\

Optimistic enhanced
& $6.80\times10^{-15}$
& $K=5900$
& $7.78\,\mathrm{GeV}$ \\

Nuclear-scale model
& $1.4\times10^{-13}$
& $E_{\mathrm{ref}}=1\,\mathrm{MeV}$
& $2.67\,\mathrm{TeV}$ \\

Nuclear-scale model
& $1.4\times10^{-13}$
& $E_{\mathrm{ref}}=10\,\mathrm{MeV}$
& $26.7\,\mathrm{TeV}$ \\
\hline
\end{tabular}
\caption{The first phenomenological bounds on the nonlocality scale $E_M$ from existing thorium--229 nuclear clock data. The direct bounds use the measured clock transition energy. The enhanced bounds include the phenomenological thorium nuclear sensitivity factor. The nuclear-scale bounds are illustrative and model-dependent.}
\label{tab:em_bounds}
\end{table*}
\end{widetext}

So the main result to take away from this section is thus:
\begin{equation}
E_M>22.3\,\mathrm{MeV}.
\label{eq:main_conservative_bound}
\end{equation}
as the conservative direct lower bound from present thorium--229 clock data, and:
\begin{equation}
E_M>1.72\,\mathrm{GeV}.
\label{eq:main_enhanced_bound}
\end{equation}
as the corresponding phenomenological nuclear-enhanced bound. These values should be interpreted as first bounds from existing data, not as the result of a dedicated nonlocality experiment.

\section{THE NUCLEUS AS AN AMPLIFIER}

The direct bounds in the previous section used the most conservative reference energy:
\begin{equation}
E_{\mathrm{ref}}=E_{\mathrm{Th}},
\label{eq:direct_ref_recall}
\end{equation}
where $E_{\mathrm{Th}}$ is the observed thorium--229 clock transition energy. This is the cleanest assumption because it only uses the measured clock frequency, however it is not necessarily the most physically sensitive channel. The thorium--229 transition is a nuclear transition, and its small observed energy is not typical of nuclear physics. It is the result of a near cancellation among much larger nuclear and electromagnetic contributions~\cite{Flambaum2006,HayesFriar2007,FlambaumWiringa2009,Fadeev2020,Beeks2025}.

We can write the transition energy as:
\begin{equation}
E_{\mathrm{Th}}
=
E_m-E_g,
\label{eq:transition_difference2}
\end{equation}
where $E_m$ is the energy of the isomeric nuclear state and $E_g$ is the energy of the nuclear ground state. Each of these energies contains several contributions:
\begin{equation}
E_i
=
E_i^{\mathrm{strong}}
+
E_i^{\mathrm{Coul}}
+
E_i^{\mathrm{spin}}
+
E_i^{\mathrm{def}}
+
\cdots ,
\qquad
i=g,m \;\;,
\label{eq:nuclear_energy_decomposition}
\end{equation}
where $E_i^{\mathrm{strong}}$ are the strong-interaction contributions, the $E_i^{\mathrm{Coul}}$ are the electromagnetic Coulomb contributions, the $E_i^{\mathrm{spin}}$ are the spin-dependent nuclear structure contributions, the $E_i^{\mathrm{def}}$ are the deformation-dependent contributions, and the ellipsis indicates that additional nuclear-structure corrections are considered. The observed transition is therefore:
\begin{align}
E_{\mathrm{Th}}
&=
\left(E_m^{\mathrm{strong}}-E_g^{\mathrm{strong}}\right)
+
\left(E_m^{\mathrm{Coul}}-E_g^{\mathrm{Coul}}\right)
\nonumber\\
&\quad+
\left(E_m^{\mathrm{spin}}-E_g^{\mathrm{spin}}\right)
+
\left(E_m^{\mathrm{def}}-E_g^{\mathrm{def}}\right)
+\cdots .
\label{eq:transition_energy_decomposed}
\end{align}
The important point to note is that the final result is only of the order of electronvolts, while the individual nuclear contributions can be much larger. A small fractional modification of one of the larger internal contributions can therefore produce an enhanced fractional modification of the observed clock transition.

Suppose we introduce a parameter $X$ into the nuclear Hamiltonian, such as an electromagnetic coupling, an effective Coulomb contribution, or a parameter controlling a nonlocal correction to the nuclear interaction. The sensitivity of the clock transition to $X$ is given by:
\begin{equation}
\frac{\delta \nu_{\mathrm{Th}}}{\nu_{\mathrm{Th}}}
=
K_X
\frac{\delta X}{X},
\label{eq:sensitivity_definition}
\end{equation}
where $K_X$ is a dimensionless sensitivity coefficient. The coefficient $K_X$ can be large when the observed transition energy is small compared with the internal nuclear-energy differences responsible for the shift; this is the sense in which we mean that the thorium--229 nucleus may act as an amplifier~\cite{Flambaum2006,HayesFriar2007,FlambaumWiringa2009,Fadeev2020,Beeks2025,Caputo2025,Delaunay2025}.

For the purposes of the current phenomenological analysis we encode this amplification through a single factor $K$, so the nonlocal fractional shift is then written as:
\begin{equation}
\left|
\frac{\Delta\nu_{\mathrm{NL}}}{\nu_{\mathrm{Th}}}
\right|
=
K|\beta_{\mathrm{NL}}|
\left(
\frac{E_{\mathrm{ref}}}{E_M}
\right)^n ,
\label{eq:amplified_nonlocal_shift}
\end{equation}
here $K$ does not mean that the clock photon has higher energy, as the photon energy remains $E_{\mathrm{Th}}\simeq 8.36\,\mathrm{eV}$. But instead $K$ represents the fact that the transition frequency can respond strongly to small changes in the internal nuclear Hamiltonian.

A useful way to show this is to introduce an internal nuclear scale $E_{\mathrm{nuc}}$, where if a nonlocal correction perturbs the nuclear Hamiltonian by a fractional amount of order:
\begin{equation}
\frac{\delta H_{\mathrm{nuc}}}{H_{\mathrm{nuc}}}
\sim
\left(
\frac{E_{\mathrm{nuc}}}{E_M}
\right)^n ,
\label{eq:nuclear_hamiltonian_shift}
\end{equation}
then the induced clock-frequency shift may be estimated as:
\begin{equation}
\frac{\delta \nu_{\mathrm{Th}}}{\nu_{\mathrm{Th}}}
\sim
\frac{\delta E_{\mathrm{Th}}}{E_{\mathrm{Th}}}
\sim
\frac{\delta E_{\mathrm{nuc}}}{E_{\mathrm{Th}}}.
\label{eq:clock_from_nuclear_shift}
\end{equation}
Since $E_{\mathrm{Th}}$ is very small compared with ordinary nuclear-energy scales, the ratio:
\begin{equation}
\frac{E_{\mathrm{nuc}}}{E_{\mathrm{Th}}}
\label{eq:amplification_ratio}
\end{equation}
can be large; this is the qualitative origin of the enhanced channel.

However there is an important distinction between three possible choices of reference scale. The first is the direct clock-energy scale:
\begin{equation}
E_{\mathrm{ref}}=E_{\mathrm{Th}},
\label{eq:ref_direct_again}
\end{equation}
this gives us the most conservative and least model-dependent bound. The second is the enhanced clock-energy scale:
\begin{equation}
E_{\mathrm{ref}}=E_{\mathrm{Th}},
\qquad
K>1,
\label{eq:ref_enhanced}
\end{equation}
this uses the measured clock energy but allows the nuclear sensitivity of thorium--229 to amplify the observed fractional shift. The third is a nuclear-scale reference energy:
\begin{equation}
E_{\mathrm{ref}}=E_{\mathrm{nuc}},
\label{eq:ref_nuclear}
\end{equation}
where $E_{\mathrm{nuc}}$ may be a MeV-scale Coulomb, deformation, or binding-energy difference, this third choice gives much stronger bounds, but it is also the most model-dependent.

In this paper we therefore report the bounds separately as to not cause confusion or to over-claim our findings. The direct channel is our baseline result, the enhanced channel is a motivated phenomenological extension, and the nuclear-scale channel is included only as an illustration of what could be reached if the nonlocal regulator acts directly on internal nuclear structure.

This separation is necessary because a nonlocal regulator need not affect every part of the experiment in the same way, a correction to the propagation of the external clock photon is not the same as a correction to the nuclear Hamiltonian. For an on-shell photon we must be careful not to identify the effect simply with a power of the photon energy, the more relevant possibility is that the regulator modifies bound-state dynamics, off-shell exchange fields, nuclear Coulomb energy, or internal response functions~\cite{MoffatThompson2026Regulators,Thompson2026Covariance,Caputo2025,Delaunay2025}. In that case the thorium--229 clock can be sensitive to scales larger than the observed photon energy.

The amplification idea can be summarized simply: a conventional optical clock measures an electronic transition whose energy is directly set by electronic structure while a thorium--229 clock measures an optical-energy transition whose origin is nuclear, so therefore the clock combines optical controllability with nuclear sensitivity, so this is why it is a promising system for searching for non-Planckian nonlocality.

For the dedicated experiment that will be proposed in the future, the amplifier role of thorium--229 is not used just to shift the central frequency, but it is used to enhance the possibility that a small nonlocal modification of the nuclear response appears as an observable residual in the time--energy phase space of the clock transition. The desired observable is therefore not only $\Delta\nu_{\mathrm{NL}}$, but also a possible residual contribution to the measured covariance matrix:
\begin{equation}
\Sigma_F^{(TE)}
=
\begin{pmatrix}
\tau_F^2 & c_{TE}\\
c_{TE} & \epsilon_F^2
\end{pmatrix}.
\label{eq:amplified_covariance_recall}
\end{equation}
If the nuclear transition amplifies the response to internal nonlocal corrections then the elements of $\Sigma_F^{(TE)}$ may be more accessible in thorium--229 than in an ordinary electronic clock. This is the main physical motivation for moving from the first bound above to a dedicated thorium--229 time--energy experiment.

The main limitation of the present analysis in this paper is that the residual scale $\delta_{\mathrm{res}}$ is not a direct measurement of a nonlocal residual. It is inferred from clock uncertainty, reproducibility, and stability. In a solid-state thorium clock these quantities receive contributions from crystal-field shifts, local strain, temperature variation, defect structure, electric-field gradients, magnetic fields, laser noise, and statistical uncertainty. These effects are ordinary local physics and they must be controlled before any residual can be interpreted as fundamental.

So the present paper does not claim evidence for nonlocality and it also does not claim that the existing thorium--229 clock experiment was already an optimized test of nonlocal time--energy uncertainty. But instead it establishes that existing data are precise enough to place meaningful first bounds on a non-Planckian nonlocality scale and this is important because it shows that the scale $E_M$ can be constrained by laboratory clock data, without assuming from the outset that it lies near the Planck scale.

A dedicated Ramsey, linewidth, and time--energy phase-space experiment~\cite{Ramsey1950,Ramsey1985} will be developed separately, that experiment will be designed from the beginning to isolate a possible nonlocal covariance contribution, compare multiple crystals and platforms, and distinguish a fundamental residual from ordinary solid-state physics. The role of the present paper is only to give a first phenomenological bound from existing thorium--229 clock data and identifies the parameter range that a dedicated experiment should target.

\section{THE NUMERICAL EXTRACTION}

The numerical bounds that we reported above were obtained by applying a single bound formula to several experimentally motivated residual scales. The purpose of this section is to make the extraction transparent, no fit to a new experiment is performed. The calculation is a reanalysis of published thorium--229 clock quantities, interpreted as upper limits on an allowed nonlocal residual~\cite{Zhang2024Frequency,Ooi2025Reproducibility,ToscaniDeCol2026,Huang2026}.

The starting point is the measured thorium clock frequency~\cite{Zhang2024Frequency,ToscaniDeCol2026,Huang2026}:
\begin{equation}
\nu_{\mathrm{Th}}
=
2.020407384335\times10^{15}\,\mathrm{Hz}.
\label{eq:numerical_thorium_frequency}
\end{equation}
The corresponding transition energy is then:
\begin{equation}
E_{\mathrm{Th}}
=
h\nu_{\mathrm{Th}},
\label{eq:numerical_transition_energy_def}
\end{equation}
with $h=4.135667696\times10^{-15}\,\mathrm{eV\,s}$, therefore:
\begin{equation}
E_{\mathrm{Th}}
=
8.355733552\,\mathrm{eV}.
\label{eq:numerical_transition_energy}
\end{equation}

For each experimental scenario, we assign a residual fractional sensitivity:
\begin{equation}
\delta_{\mathrm{res}}^{(i)},
\label{eq:numerical_delta_i}
\end{equation}
here the label $i$ means the particular interpretation of the data. In the present analysis the three direct residual choices are:
\begin{align}
\delta_{\mathrm{freq}}
&=
2.0\times10^{-13},
\label{eq:delta_freq_num}
\\
\delta_{\mathrm{rep}}
&=
1.4\times10^{-13},
\label{eq:delta_rep_num}
\\
\delta_{\mathrm{day}}
&=
6.80\times10^{-15}.
\label{eq:delta_day_num}
\end{align}
where $\delta_{\mathrm{freq}}$ is the direct frequency-uncertainty channel, $\delta_{\mathrm{rep}}$ is the conservative reproducibility channel, and $\delta_{\mathrm{day}}$ is the optimistic day-stability channel~\cite{Ooi2025Reproducibility,ToscaniDeCol2026,Huang2026}.

The direct bound is computed from:
\begin{equation}
E_M^{(i)}
=
E_{\mathrm{Th}}
\left(
\frac{1}{\delta_{\mathrm{res}}^{(i)}}
\right)^{1/2},
\label{eq:numerical_direct_formula}
\end{equation}
where we have set $n=2$ and $|\beta_{\mathrm{NL}}|=1$. This gives us:
\begin{align}
E_M^{(\mathrm{freq})}
&=
8.355733552\,\mathrm{eV}
\left(
\frac{1}{2.0\times10^{-13}}
\right)^{1/2}
\nonumber\\
&=
1.87\times10^7\,\mathrm{eV}
=
18.7\,\mathrm{MeV},
\label{eq:numerical_freq_result}
\\[1ex]
E_M^{(\mathrm{rep})}
&=
8.355733552\,\mathrm{eV}
\left(
\frac{1}{1.4\times10^{-13}}
\right)^{1/2}
\nonumber\\
&=
2.23\times10^7\,\mathrm{eV}
=
22.3\,\mathrm{MeV},
\label{eq:numerical_rep_result}
\\[1ex]
E_M^{(\mathrm{day})}
&=
8.355733552\,\mathrm{eV}
\left(
\frac{1}{6.80\times10^{-15}}
\right)^{1/2}
\nonumber\\
&=
1.01\times10^8\,\mathrm{eV}
=
101\,\mathrm{MeV}.
\label{eq:numerical_day_result}
\end{align}

For the enhanced channel the same residual scales are used, but the bound formula includes the nuclear sensitivity factor $K=5900$~\cite{Beeks2025}. So the enhanced bound is:
\begin{equation}
E_{M,K}^{(i)}
=
E_{\mathrm{Th}}
\left(
\frac{K}{\delta_{\mathrm{res}}^{(i)}}
\right)^{1/2}.
\label{eq:numerical_enhanced_formula}
\end{equation}
Then using the conservative reproducibility residual gives us:
\begin{align}
E_{M,K}^{(\mathrm{rep})}
&=
8.355733552\,\mathrm{eV}
\left(
\frac{5900}{1.4\times10^{-13}}
\right)^{1/2}
\nonumber\\
&=
1.72\times10^9\,\mathrm{eV}
=
1.72\,\mathrm{GeV}.
\label{eq:numerical_enhanced_rep_result}
\end{align}
Using the optimistic day-stability residual gives:
\begin{align}
E_{M,K}^{(\mathrm{day})}
&=
8.355733552\,\mathrm{eV}
\left(
\frac{5900}{6.80\times10^{-15}}
\right)^{1/2}
\nonumber\\
&=
7.78\times10^9\,\mathrm{eV}
=
7.78\,\mathrm{GeV}.
\label{eq:numerical_enhanced_day_result}
\end{align}

The model-dependent nuclear-scale estimates are obtained by replacing the reference energy $E_{\mathrm{Th}}$ with an assumed internal nuclear reference scale, as motivated by nuclear-structure sensitivity analyses \cite{Flambaum2006,HayesFriar2007,FlambaumWiringa2009,Fadeev2020,Caputo2025,Delaunay2025}:
\begin{equation}
E_{\mathrm{ref}}=E_{\mathrm{nuc}}.
\label{eq:numerical_nuclear_reference}
\end{equation}
By using the conservative reproducibility residual and no additional enhancement factor gives us:
\begin{equation}
E_M
=
E_{\mathrm{nuc}}
\left(
\frac{1}{1.4\times10^{-13}}
\right)^{1/2}.
\label{eq:numerical_nuclear_formula}
\end{equation}
For $E_{\mathrm{nuc}}=1\,\mathrm{MeV}$, we obtain:
\begin{equation}
E_M
=
2.67\,\mathrm{TeV}.
\label{eq:one_mev_extraction}
\end{equation}
For:
\begin{equation}
E_{\mathrm{nuc}}=10\,\mathrm{MeV},
\label{eq:ten_mev_ref}
\end{equation}
we obtain:
\begin{equation}
E_M
=
26.7\,\mathrm{TeV}.
\label{eq:ten_mev_extraction}
\end{equation}

So our numerical extraction can be summarized as:
\begin{equation}
E_M
\propto
E_{\mathrm{ref}},
\qquad
E_M
\propto
\delta_{\mathrm{res}}^{-1/2},
\qquad
E_M
\propto
K^{1/2},
\label{eq:bound_scalings_summary}
\end{equation}
with $n=2$, and thus the bound improves linearly with the assumed reference energy, as the square root of the nuclear enhancement factor, and as the inverse square root of the residual clock precision.

\section{WHY A LOW-ENERGY CLOCK CAN BOUND A HIGHER SCALE}

At first sight it may seem surprising that an optical or vacuum-ultraviolet clock transition can bound a nonlocality scale that may be potentially orders of magnitude above the clock energy. The reason is that the experiment is not producing particles at the scale $E_M$ but it is instead testing whether low-energy observables contain suppressed remnants of the nonlocal regulator. This is the same logic by which precision measurements constrain heavy new physics without directly producing it~\cite{Safronova2018,Peik2021,Caputo2025,Delaunay2025}.

The general form of the bound is given by:
\begin{equation}
\frac{|\Delta\nu_{\mathrm{NL}}|}{\nu_{\mathrm{Th}}}
=
A
\left(
\frac{E_{\mathrm{ref}}}{E_M}
\right)^n ,
\label{eq:precision_logic}
\end{equation}
where $A$ is a dimensionless amplitude. In the direct channel we have that:
\begin{equation}
A=|\beta_{\mathrm{NL}}|,
\label{eq:direct_A}
\end{equation}
while in the enhanced channel we have that:
\begin{equation}
A=K|\beta_{\mathrm{NL}}|,
\label{eq:enhanced_A}
\end{equation}
so the experiment does not need to reach the energy $E_M$ but it only needs to measure the left-hand side with enough precision to exclude corrections of a given size.

If the experiment constrains the fractional residual to be smaller than $\delta_{\mathrm{res}}$, then we have:
\begin{equation}
A
\left(
\frac{E_{\mathrm{ref}}}{E_M}
\right)^n
<
\delta_{\mathrm{res}}.
\label{eq:precision_bound_logic}
\end{equation}
Solving gives us:
\begin{equation}
E_M
>
E_{\mathrm{ref}}
\left(
\frac{A}{\delta_{\mathrm{res}}}
\right)^{1/n},
\label{eq:precision_scale_bound}
\end{equation}
so a small reference energy can still bound a larger scale if the fractional precision is sufficiently high.

For the quadratic case used in this paper we have $n=2$, the bound becomes:
\begin{equation}
E_M
>
E_{\mathrm{ref}}
\sqrt{
\frac{A}{\delta_{\mathrm{res}}}
}.
\label{eq:quadratic_precision_bound}
\end{equation}
So this equation shows us the central reason clocks are powerful, a clock residual at the level of $10^{-13}$ multiplies the reference energy by approximately:
\begin{equation}
\delta_{\mathrm{res}}^{-1/2}
\sim
10^{6.5}.
\label{eq:clock_precision_multiplier}
\end{equation}
Therefore an electronvolt-scale transition can constrain MeV-scale nonlocality even without any enhancement, if a nuclear sensitivity factor is present, the reach increases further.

This should be distinguished from a collider bound as a collider searches for direct high-energy effects in scattering, but a clock searches for tiny low-energy distortions in a sharply measured quantum transition. The two approaches are complementary as a collider bound is most direct when the new scale modifies high-energy scattering amplitudes, while a clock bound is most direct when the new scale modifies precision observables, bound-state energies, phases, linewidths, or response functions.

In the present case we explore the clock observable is the thorium--229 nuclear transition frequency. The measured frequency can be written as:
\begin{equation}
\nu_{\mathrm{obs}}
=
\nu_{\mathrm{local}}
+
\Delta\nu_{\mathrm{NL}},
\label{eq:clock_low_energy_observable}
\end{equation}
where $\nu_{\mathrm{local}}$ is the prediction after ordinary local effects are included, and $\Delta\nu_{\mathrm{NL}}$ is the possible nonlocal contribution. The absence of an unexplained residual implies that:
\begin{equation}
|\Delta\nu_{\mathrm{NL}}|
<
\nu_{\mathrm{Th}}\delta_{\mathrm{res}},
\label{eq:allowed_absolute_shift}
\end{equation}
this is why the result is a clock bound. It is not a direct observation of nonlocality, but rather it is a limit on how large a nonlocal contribution could be without modifying the observed clock behavior.

The thorium--229 clock is especially useful because it has two advantages for us, first is that the transition can be measured with optical-clock precision~\cite{Ludlow2015,Zhang2024Frequency,Ooi2025Reproducibility,ToscaniDeCol2026,Huang2026}. The second is that the transition is nuclear. The first property gives a small $\delta_{\mathrm{res}}$ and the second property motivates a possible enhancement factor $K$. Together, these two features make the system sensitive to nonlocality effects occurring at energy scales much larger than the observed photon energy.

The conservative direct bound does not require the enhancement, it follows only from the precision of the measured transition and the assumption of a quadratic leading correction. The enhanced bound is stronger, since it assumes that the nuclear structure of thorium--229 amplifies the effect. The nuclear-scale estimates are stronger still, because they assume that the relevant reference energy is an internal nuclear energy rather than the observed transition energy.

So our result is best viewed as a low-energy precision bound on a nonlocality scale, it does not prove that $E_M$ lies near the MeV or GeV scale, but it says that under the stated leading-correction model existing thorium--229 clock data already exclude values of $E_M$ below the reported bounds. Future clock improvements will push this exclusion upward.

\section{CONCLUDING REMARKS}

With the new thorium--229 nuclear clocks, we can now connect the effects of nonlocal quantum field theory with a precision laboratory measurement. Our idea says that if nonlocality introduces a finite time--energy response width, then an ultra-narrow nuclear clock transition may retain a residual frequency shift, linewidth contribution, dephasing channel, or response broadening that is absent in ordinary local quantum field theory. The existing thorium--229 data were not collected for this purpose, but they are already precise enough to give a first phenomenological lower bound on the nonlocality energy scale $E_M$~\cite{Zhang2024Frequency,Ooi2025Reproducibility,ToscaniDeCol2026,Huang2026}.

We have found that the most conservative result is the direct clock-energy bound, which used only the measured thorium--229 transition energy and the reported reproducibility scale of the clock data and this gives a lower bound of $E_M>22.3\,\mathrm{MeV}$. We believe that this is the cleanest result, because it does not require a detailed model of how the nonlocal regulator affects nuclear binding, Coulomb energy, or the internal nuclear Hamiltonian. This result should be regarded as the baseline conclusion of the present work.

The enhanced result of $E_M>1.72\,\mathrm{GeV}$ is also physically motivated since thorium--229 is not an ordinary optical clock, but rather its low-energy transition is nuclear and thus involves length scales five orders of manitude smaller. Moreover, its narrow energy gap comes from cancellations among much larger nuclear and electromagnetic contributions. This makes it plausible that thorium--229 can act as an amplifier for high-energy-sector effects, but this interpretation requires an additional sensitivity assumption, so it should be treated separately from the direct clock-energy bound.

The TeV-scale values we obtained by inserting MeV-scale nuclear reference energies are best understood as reach estimates and not as model-independent bounds. They show us that if the relevant nonlocal correction acts directly on internal nuclear scales, then thorium--229 could become a powerful probe of nonlocality far above the optical transition energy. A dedicated nuclear-structure calculation would be required before those values could be promoted to direct empirical constraints.

How might these bounds relate to collider physics? If a universal unsuppressed nonlocality scale as low as tens of MeV or a few GeV modified all Standard Model amplitudes in the same way, it would have almost certainly appeared already in particle physics as such a scale would affect precision QED, hadronic physics, electroweak observables, flavor data, and collider scattering long before it became detectable via a thorium clock. Likewise the TeV-scale nonlocality with unsuppressed couplings to quarks, leptons, gauge bosons, or Higgs processes would be expected to show up in high-energy tails, contact-interaction searches, precision cross sections, or form-factor deviations at modern colliders.

So for that reason the present result should not be interpreted as a global lower bound on every possible form of nonlocal quantum field theory, but it is instead a clock-channel bound where it constrains the size of a nonlocal correction in the thorium--229 nuclear-clock observable under the leading low-energy correction model used here. A universal Standard Model nonlocality scale would have to be compared with collider, electroweak, flavor, and low-energy precision data. A sector-dependent or nuclear-response-enhanced nonlocality scale can be meaningfully constrained by the clock analysis developed in this paper, but that is not our goal in nonlocal quantum field theory.

We want to note that collider experiments test high-energy scattering while nuclear clocks test ultra-small residual shifts in sharply defined quantum transitions. These are complementary probes as colliders are powerful when the new scale modifies hard scattering directly, and clocks are powerful when the new scale leaves suppressed traces in phases, line centers, linewidths, coherence times, or bound-state energies.

So the takeaway we want to present from this paper is not that $E_M$ has been measured, nor that the fundamental nonlocality scale is near the MeV or GeV range. But the realistic conclusion is that existing thorium--229 nuclear clock data already exclude nonlocal time--energy corrections below the tens-of-MeV scale in the direct clock channel, and motivate a GeV-scale sensitivity if the known thorium nuclear enhancement is applicable.

We regard the most realistic summary bound for this paper as the conservative direct bound of $E_M>22.3\,\mathrm{MeV}$. This is the number least vulnerable to assumptions about nuclear amplification as the enhanced GeV-scale bound should be emphasized as a motivated secondary result, and the TeV-scale estimates should be presented only as model-dependent future reach.

The main result we want to focus on is that we have shown that nonlocal time--energy uncertainty can be constrained with real nuclear-clock data. The present bound is only a first step, but it establishes the method. A future dedicated experiment, designed specifically to isolate residual linewidth, dephasing, and time--energy covariance contributions, can turn this phenomenological constraint into a sharper test of nonlocal quantum field theory.

\section*{Acknowledgments}

We thank Martin. A. Green for insightful and stimulating discussions. Research at the Perimeter Institute for Theoretical Physics is supported by the Government of Canada through Industry Canada and by the Province of Ontario through the Ministry of Research and Innovation (MRI).

\clearpage

\section{A Proposed Experiment for the use of Nuclear Entanglement and Squeezed States to Test Nonlocal Quantum Field Theory with a Thorium–229 Clock}

\section{abstract}

As a follow-up to a previous paper on the topic of using nuclear clocks to test nonlocal quantum field theories, we will explore the idea of using a squeezed-state experiment with the $^{229}\mathrm{Th}$ nuclear clock to test the time--energy structure of nonlocal quantum field theory. The idea is reasonably obtainable within the near future of nuclear clock experiments. One would prepare an ensemble of thorium nuclei in a coherent superposition of the nuclear ground state and the low-lying isomeric clock state, entangle the participating nuclei through a collective interaction, and generate a family of spin-squeezed states with a tunable squeezing parameter $r$. Then a phase-controlled analysis pulse will rotate the selected collective nuclear quadrature into a measurable ground--isomer population difference. Near a strongly polarized collective state the normalized operators $J_y/\sqrt{S}$ and $J_z/\sqrt{S}$ obey the same approximate canonical algebra as the phase and amplitude quadratures of a squeezed optical mode. We use this experiment to either probe or bound the nonlocal energy scale $E_M$. In this proposal the standard nuclear-clock equipment is used for the clock transition and frequency reference, but the full experiment requires coherent pulsed VUV control, a collective entangling interaction, quantum-noise-resolved population detection, and phase-resolved quadrature tomography. This is all within reach of current nuclear clocks but requires some extra equipment. The experiment would be able to probe other proposals of fundamental physics such as ultralight dark matter, violations of Lorentz invariance and the Einstein equivalence principle, and other aspects of fundamental physics.

\section{Introductory Considerations}

In June of 2026 there entered a new precision measurement tool for physics, the $^{229}$Th:CaF$_2$ clock~\cite{Kraemer2023,Tiedau2024,Elwell2024,Yamaguchi2024,Zhang2024Frequency,Hiraki2024,Ooi2026,ToscaniDeCol2026,Huang2026,Derevianko2026Colloquium,Morawetz2026}. While traditional atomic clocks have been used since the 1950s, they measure the vibrations of electrons, so they are sensitive to outside interference. Nuclear clocks measure the energy shifts inside the atom's nucleus, providing a significantly more stable and robust time signal~\cite{Ramsey1985,Ludlow2015,Diddams2020,VanierAudoin1989,Riehle2004,Allan1966,Riley2008,BIPM2019}.

Thorium--229 is the key system we will use~\cite{KrogerReich1976,ReichHelmer1990,Burke1990,HelmerReich1994,Ruchowska2006,Beck2007,PeikTamm2003,Rellergert2010,Campbell2011,Campbell2012,Kazakov2012,VonDerWense2016,Thielking2018,Masuda2019,Seiferle2019,Yamaguchi2019,Sikorsky2020,VonDerWense2018Review,Thirolf2019,VonDerWenseSeiferle2020,BeeksReview2021,Peik2021,Thirolf2024Review}. Its first isomeric state lies at an unusually low excitation energy, so vacuum-ultraviolet light can address a genuinely nuclear transition with optical precision. The observed energy is small, but it emerges from much larger nuclear and electromagnetic contributions. A tiny modification of that internal structure could therefore produce a measurable change in the clock signal.

This is the motivation for us to ask if this is sensitive enough to test time--energy structure of nonlocal quantum field theory.

Our proposal uses squeezed states, where we can use precision devices to reduce the uncertainty of one observable in one quadrature, subsequently broadening the uncertainty in the other, this is called anti-squeezing. In the experiment one would prepare many thorium nuclei in a coherent superposition of the nuclear ground state and the isomeric clock state, then apply a collective interaction that entangles the participating nuclei. This squeezes the collective clock phase while increasing the uncertainty in the conjugate population-energy variable. The experiment does not remove quantum uncertainty but it instead redistributes it~\cite{Caves1981,KitagawaUeda1993,Wineland1992,Wineland1994,Ma2011,Gross2012,Pezze2018,Wang2025NuclearQubits}.

This idea comes from squeezed-light experiments where a coherent optical state begins with equal noise in two quadratures, then a nonlinear interaction compresses one quadrature and stretches the other. Then balanced homodyne detection then scans the resulting uncertainty ellipse by changing the phase of a local oscillator~\cite{Slusher1985,Wu1986,Smithey1993,LvovskyRaymer2009,BraunsteinVanLoock2005,Weedbrook2012,Leonhardt1997,WallsMilburn2008,BachorRalph2019}.

This nuclear clock experiment follows that same logic where the collective nuclear phase replaces one optical quadrature. The ground-versus-isomer population replaces the other quadrature. A phase-controlled analysis pulse plays the role of the homodyne setting by rotating the chosen nuclear quadrature into a measurable population difference. Repeating the experiment for several analysis settings (the $r$ in standard squeezed state equations) reconstructs the full nuclear covariance ellipse.

In the local quantum theory it predicts that stronger squeezing will continue to reduce the clock-phase variance, provided decoherence and technical noise remain under control. But a nonlocal observable structure would alter that scaling. In our framework, it was shown that physical measurements inherit a finite response from the nonlocal regulator that defines the ultraviolet theory. If that response extends into the time--energy variables of a clock energy Hamiltonian, then the squeezed variance may stop decreasing and approach a residual floor.

Now we do know that any real experiments generate apparent floors through optical loss, imperfect pulse control, finite detection efficiency, host-crystal disorder, inhomogeneous broadening, laser noise, decoherence, etc. So the experiment must vary the squeezing strength, Ramsey time, analysis angle, effective particle number, readout gain, and material environment so that we can reduce noise and get a cleaner detection our bound. If there is a fundamental contribution, then it should remain consistent across those changes where the technical one usually will not.

Here the nucleus and the squeezing protocol serve different purposes. The nuclear structure may amplify sensitivity to high-energy physics and squeezing suppresses ordinary projection noise~\cite{Flambaum2006,Litvinova2009,Berengut2009,Fadeev2020,BeeksFineStructure2025,Safronova2018,DereviankoPospelov2014,Arvanitaki2015,Hees2016,Fuchs2025,Caputo2025,KosteleckyRussell2011,Will2014,KouroshniaMoffat2026,Arvanitaki2018}.

We will consider ensembles of $^{229}\mathrm{Th}$ nuclei embedded in fluoride hosts, with particular attention to calcium fluoride and strontium fluoride as to compare what would be better for the experiment~\cite{Zhang2024Films,Higgins2025,Pineda2025,Schaden2025,Terhune2025,Morgan2025,Elwell2025,Nickerson2020,Dessovic2014,Nickerson2021,BeeksThesis2022,BeeksGrowth2023,BeeksTransmission2024,Gong2025,Rehn2026,Abragam1961,Slichter1990,Stoneham1975,MorganSpinless2025,Wu2026PhotonuclearSrF2,Wu2024Isomer}. Existing nuclear-clock platforms already provide the transition, the host material, the frequency reference, and a basis for spectroscopy. Now our proposed test requires more as it needs coherent pulsed vacuum-ultraviolet control, a collective entangling interaction, quantum-noise-resolved population detection, and phase-sensitive covariance tomography. This is all within current reach.

This paper develops that experiment. We describe the thorium clock as a collective two-level nuclear system, construct coherent and squeezed nuclear states, map the protocol to optical homodyne detection, and derive the measurement sequence needed to reconstruct the time--energy covariance. We then introduce the nonlocal response model, separate frequency shifts from variance and linewidth effects, identify the main false positives, and show how either a null result or a residual covariance can be translated into information about the Moffat energy.

The central question is simple yet elegant: 
\\
\\
\emph{How far can we squeeze a nuclear clock before ordinary local quantum theory stops describing what we measure?}

\section{The $^{229}$Th nuclear clock as a two-level quantum system}

To be able to construct the quantum-mechanical description required for the squeezing experiment we will propose, we first identify the effective clock degree of freedom of a single $^{229}$Th nucleus. While the complete nuclear spectrum contains many states and magnetic substates, the clock interrogation can be restricted to a single spectroscopically resolved transition between the nuclear ground state and the exceptionally low-lying isomeric state. Within this resolved subspace, the nucleus can be treated as an effective two-level quantum system, providing the elementary nuclear qubit from which the collective clock dynamics will later be constructed later in the paper.

The physical oscillator of the thorium nuclear clock is the transition between the ground state and its low-lying isomeric state:
\begin{equation}
    ^{229}\text{Th}_g\longleftrightarrow^{229e}\text{Th}.
\end{equation}
For a $^{229}$Th nuclear clock, the nuclei has two states we will need for this experiment, the ground state $|g\rangle$, and the isomeric or excited state $|e\rangle$\footnote{Sometimes denoted as $|m\rangle$, like in our last paper.}:
\begin{equation}
    |g\rangle=\left|I_g^\pi=\frac{5^+}{2}\right\rangle, \qquad |e\rangle=\left| I_e^\pi\frac{3^+}{2}\right\rangle,
\end{equation}
where $I$ is the total nuclear angular momentum (spin), and $\pi$ is the parity. These represent a nuclear two-level system which dynamically couple through $\text{M}1$ and $\text{E}2$ transitions and form the basis for the thorium-229 optical nuclear clock~\cite{BohrMottelson1969,BohrMottelson1975,RingSchuck1980,Krane1987}.

The transition is predominantly magnetic dipole, with a smaller electric-quadrupole contribution. Its energy separation lies in the vacuum-ultraviolate range:
\begin{equation}
    E_\text{Th}=E_e-E_g=\hbar\omega_0=h\nu_0\simeq 8.36\;\text{eV},
\end{equation}
where:
\begin{equation}
    \nu_0=\frac{\omega_0}{2\pi}\simeq2.024\times10^{15}\;\text{Hz},\qquad \lambda_0=\frac{c}{\nu_0}\simeq148.38\;\text{nm},
\end{equation}
this unusually small nuclear excitation energy allows the transition to be interrogated directly using a narrow band vacuum-ultraviolet field. Each of these energies contains several contributions from:
\begin{equation}
E_i
=
E_i^{\mathrm{strong}}
+
E_i^{\mathrm{Coul}}
+
E_i^{\mathrm{spin}}
+
E_i^{\mathrm{def}}
+
\cdots ,
\end{equation}
with $i=g,m$, where $E_i^{\mathrm{strong}}$ is the strong-interaction contribution, $E_i^{\mathrm{Coul}}$ is the electromagnetic Coulomb contribution, $E_i^{\mathrm{spin}}$ is the spin-dependent nuclear structure contribution, and $E_i^{\mathrm{def}}$ is the deformation-dependent contribution, for each $i$. The ellipsis denotes additional nuclear-structure corrections. The observed transition is therefore given by:
\begin{align}
E_{\mathrm{Th}}
&=
\left(E_m^{\mathrm{strong}}-E_g^{\mathrm{strong}}\right)
+
\left(E_m^{\mathrm{Coul}}-E_g^{\mathrm{Coul}}\right)
\nonumber\\
&\quad+
\left(E_m^{\mathrm{spin}}-E_g^{\mathrm{spin}}\right)
+
\left(E_m^{\mathrm{def}}-E_g^{\mathrm{def}}\right)
+\cdots .
\end{align}
The important point to note is that the final result is only of the order of electronvolts, while the individual nuclear contributions can be much larger. A small fractional modification of one of the larger internal contributions may therefore produce an enhanced fractional modification of the observed clock transition. This is why the nucleus may act as an amplifier of high energy effects.

The labels $|g\rangle$ and $|e\rangle$ should be understood as effective clock-state labels rather than as the complete nuclear manifolds. Since the two nuclear states have nonzero angular momentum, each one contains several magnetic substates. In a solid-state host, electric-field gradients, local magnetic fields, lattice strain, and other environmental interactions further split and shift these substates.

The effective nuclear Hamiltonian before selecting a clock component may be written as:
\begin{widetext}
\begin{equation}
\begin{aligned}
\begin{split}
    H_{\mathrm{nuc}}^{\mathrm{eff}}
    =
    H_{\mathrm{nuc}}^{(0)}
    +
    H_Q(V_{ij})
    +
    H_{\mathrm{mag}}(\mathbf B_{\mathrm{loc}})
    +
    H_{\mathrm{iso}}[\rho_e(0)]
    +
    H_{\mathrm{phonon}}(T)
    +\cdots ,
    \label{eq:full_effective_nuclear_hamiltonian}
\end{split}
\end{aligned}
\end{equation}
\end{widetext}
where $H_Q$ is the electric-quadrupole interaction and the remaining terms represent magnetic, isomer-shift, and phonon-dependent contributions. The host crystal therefore determines the detailed spectrum that is observed, even though the underlying oscillator remains the same nuclear transition.

We assume that one component has been spectroscopically isolated by using a highly stable, narrowband laser, which selectively interacts with exactly one specific transition:
\begin{equation}
    |g,\alpha\rangle
    \longleftrightarrow
    |e,\beta\rangle
    \label{eq:resolved_clock_component}
\end{equation}
which allows us to selectively address this transition. We can do this because the laser is strongly detuned away from any other transitions, and so the rest of the nuclear Hilbert space can be safely ignored. We now introduce the projector:
\begin{equation}
    P
    =
    |g,\alpha\rangle\langle g,\alpha|
    +
    |e,\beta\rangle\langle e,\beta|,
    \label{eq:clock_subspace_projector}
\end{equation}
the projected Hamiltonian is then:
\begin{equation}
    H_{\mathrm{2L}}
    =
    P H_{\mathrm{nuc}}^{\mathrm{eff}}P ,
    \label{eq:projected_two_level_hamiltonian}
\end{equation}
this is useful because it strips away all irrelevant energy levels, and leaves us with a simplified $2 \times 2$ matrix representation ($H_{\mathrm{2L}}$) that governs only our chosen clock states. We subsequently suppress the sublevel labels and write:
\begin{equation}
    |g,\alpha\rangle\rightarrow |g\rangle,
    \qquad
    |e,\beta\rangle\rightarrow |e\rangle .
\end{equation}
The frequency $\omega_0$ below is thus the frequency of the selected, resolved clock component under the specified environmental conditions.

Once trapped inside this two-level subspace, the environmental sublevel labels ($\alpha, \beta$) become redundant. Dropping them simplifies the states to a universal two-level basis:
\begin{equation}
|g\rangle =\left(\begin{matrix}0\\ 1\end{matrix}\right),\qquad |e\rangle =\left(\begin{matrix}1\\ 0\end{matrix}\right)
\end{equation}
The transition frequency between them is $\omega _{0}$.

Now we move on to the Pauli representation of the clock states. We define the projectors:
\begin{align}
    P_g&=|g\rangle\langle g|,
    \\
    P_e&=|e\rangle\langle e|,
    \\
    P_g+P_e&=\mathbbm{1},
    \label{eq:clock_projectors}
\end{align}
together with the Pauli operators:
\begin{align}
    \sigma_z
    &=
    |e\rangle\langle e|
    -
    |g\rangle\langle g|,
    \label{eq:sigma_z_clock}
    \\
    \sigma_x
    &=
    |e\rangle\langle g|
    +
    |g\rangle\langle e|,
    \label{eq:sigma_x_clock}
    \\
    \sigma_y
    &=
    -i|e\rangle\langle g|
    +
    i|g\rangle\langle e|.
    \label{eq:sigma_y_clock}
\end{align}
Now it is also useful to introduce the raising and lowering operators:
\begin{equation}
    \sigma_+
    =
    |m\rangle\langle g|
    =
    \frac{\sigma_x+i\sigma_y}{2},
    \qquad
    \sigma_-
    =
    |g\rangle\langle m|
    =
    \frac{\sigma_x-i\sigma_y}{2}.
    \label{eq:clock_ladder_operators}
\end{equation}
Then level projectors can be written as:
\begin{equation}
    P_m=\frac{\mathbbm{1}+\sigma_z}{2},
    \qquad
    P_g=\frac{\mathbbm{1}-\sigma_z}{2}.
    \label{eq:clock_projectors_pauli}
\end{equation}

Then the exact Hamiltonian restricted to the selected two-dimensional subspace is given by:
\begin{equation}
    H_0
    =
    E_g|g\rangle\langle g|
    +
    E_m|m\rangle\langle m|.
    \label{eq:exact_two_level_hamiltonian}
\end{equation}
By substituting Eq.~\eqref{eq:clock_projectors_pauli} gives us:
\begin{align}
    H_0
    &=
    E_g\frac{\mathbbm{1}-\sigma_z}{2}
    +
    E_m\frac{\mathbbm{1}+\sigma_z}{2}
    \nonumber\\
    &=
    \frac{E_m+E_g}{2}\mathbbm{1}
    +
    \frac{E_m-E_g}{2}\sigma_z .
    \label{eq:two_level_hamiltonian_decomposition}
\end{align}
Now defining:
\begin{equation}
    \overline{E}
    =
    \frac{E_m+E_g}{2},
    \qquad
    E_m-E_g=\hbar\omega_0,
    \label{eq:mean_and_splitting}
\end{equation}
we obtain:
\begin{equation}
    H_0
    =
    \overline{E}\,\mathbbm{1}
    +
    \frac{\hbar\omega_0}{2}\sigma_z .
    \label{eq:two_level_hamiltonian_pauli}
\end{equation}

The first term generates only a global phase. So we have:
\begin{equation}
    e^{-iH_0t/\hbar}
    =
    e^{-i\overline{E}t/\hbar}
    e^{-i\omega_0t\sigma_z/2},
    \label{eq:free_two_level_propagator}
\end{equation}
where the factor $e^{-i\overline{E}t/\hbar}$ multiplies every state by the same phase and cannot affect any expectation value. Adding an identity term also leaves the energy variance unchanged:
\begin{align}
    \left[\Delta\!\left(H+c\mathbbm{1}\right)\right]^2
    &=
    \left\langle
    \left[
    H+c\mathbbm{1}
    -
    \langle H+c\mathbbm{1}\rangle
    \right]^2
    \right\rangle
    \nonumber\\
    &=
    \left\langle
    \left(H-\langle H\rangle\right)^2
    \right\rangle
    =
    (\Delta H)^2 .
    \label{eq:identity_does_not_change_variance}
\end{align}
We can this use the dynamically equivalent clock Hamiltonian:
\begin{equation}
    H_{\mathrm{clock}}
    =
    \frac{\hbar\omega_0}{2}\sigma_z .
    \label{eq:reduced_clock_hamiltonian}
\end{equation}

If we consider an arbitrary pure state in the clock subspace:
\begin{equation}
    |\psi(0)\rangle
    =
    c_g|g\rangle+c_m|m\rangle,
    \qquad
    |c_g|^2+|c_m|^2=1.
    \label{eq:general_clock_state}
\end{equation}
then under the free Hamiltonian:
\begin{align}
    |\psi(t)\rangle
    &=
    e^{-i\overline{E}t/\hbar}
    \left(
    c_g e^{+i\omega_0t/2}|g\rangle
    +
    c_m e^{-i\omega_0t/2}|m\rangle
    \right)
    \nonumber\\
    &\doteq
    c_g|g\rangle
    +
    c_m e^{-i\omega_0t}|m\rangle ,
    \label{eq:clock_relative_phase_evolution}
\end{align}
where $\doteq$ denotes equality up to an overall phase. The physical clock signal is then encoded in the relative phase between the ground and isomeric amplitudes. Equivalently, the off-diagonal density-matrix element evolves according to:
\begin{equation}
    \rho_{mg}(t)
    =
    e^{-i\omega_0t}\rho_{mg}(0).
    \label{eq:clock_coherence_evolution}
\end{equation}
Population in either energy eigenstate alone does not provide a running clock phase as a clock requires a coherent superposition of the two levels.

To find the energy expectation value and variance we first let:
\begin{equation}
    p_m=\langle P_m\rangle,
    \qquad
    p_g=\langle P_g\rangle,
    \qquad
    p_m+p_g=1,
\end{equation}
and since $\langle\sigma_z\rangle=p_m-p_g$, the expectation value of the full Hamiltonian is given by:
\begin{equation}
    \langle H_0\rangle
    =
    \overline{E}
    +
    \frac{\hbar\omega_0}{2}\langle\sigma_z\rangle .
    \label{eq:single_clock_mean_energy}
\end{equation}
now using $\sigma_z^2=\mathbbm{1}$, its variance is:
\begin{align}
    (\Delta H_0)^2
    &=
    \left(\frac{\hbar\omega_0}{2}\right)^2
    \left(
    \langle\sigma_z^2\rangle
    -
    \langle\sigma_z\rangle^2
    \right)
    \nonumber\\
    &=
    \left(\frac{\hbar\omega_0}{2}\right)^2
    \left[
    1-(p_m-p_g)^2
    \right]
    \nonumber\\
    &=
    (\hbar\omega_0)^2p_mp_g .
    \label{eq:single_clock_energy_variance}
\end{align}
So an energy eigenstate has zero intrinsic energy variance, whereas an equal superposition $p_g=p_m=\frac{1}{2}$, has thee following:
\begin{equation}
    \Delta H_0
    =
    \frac{\hbar\omega_0}{2}.
    \label{eq:equal_superposition_energy_variance}
\end{equation}
The energy uncertainty discussed later in the paper is therefore the quantum variance of the prepared clock state. It must not be identified with the spectroscopic linewidth or with the uncertainty of an estimated transition frequency for reasons discussed later.

To explain the coherent vacuum-ultraviolet driving we let a phase-coherent interrogation field have angular frequency $\omega_L$, phase $\phi_L$, and a slowly varying amplitude. After projection onto the selected clock transition and removal of counter-rotating terms, the interaction Hamiltonian can be written as:
\begin{equation}
    H_{\mathrm{int}}^{\mathrm{RWA}}(t)
    =
    \frac{\hbar\Omega(t)}{2}
    \left[
    e^{-i(\omega_Lt+\phi_L)}\sigma_+
    +
    e^{+i(\omega_Lt+\phi_L)}\sigma_-
    \right],
    \label{eq:clock_drive_lab_frame}
\end{equation}
where $\Omega(t)$ is the Rabi angular frequency. Microscopically, $\Omega(t)$ is determined by the matrix element of the interrogation field between the selected ground and isomeric states, including the dominant magnetic-dipole and smaller electric-quadrupole contributions~\cite{Rabi1937,Ramsey1950,ScullyZubairy1997,Foot2005,SakuraiNapolitano2020,CohenTannoudji1977}.

We can introduce the rotating-frame transformation:
\begin{equation}
    R(t)
    =
    \exp\left(
    \frac{i\omega_Lt}{2}\sigma_z
    \right).
    \label{eq:clock_rotating_transformation}
\end{equation}
The Hamiltonian in this frame is given by:
\begin{equation}
    H_{\mathrm{rot}}
    =
    R\left(H_{\mathrm{clock}}+H_{\mathrm{int}}^{\mathrm{RWA}}\right)R^\dagger
    +
    i\hbar\dot R R^\dagger .
    \label{eq:rotating_frame_definition}
\end{equation}
Using:
\begin{equation}
    R\sigma_{\pm}R^\dagger
    =
    e^{\pm i\omega_Lt}\sigma_{\pm},
\end{equation}
we find that:
\begin{equation}
    H_{\mathrm{rot}}
    =
    \frac{\hbar\delta}{2}\sigma_z
    +
    \frac{\hbar\Omega(t)}{2}
    \left(
    e^{-i\phi_L}\sigma_+
    +
    e^{+i\phi_L}\sigma_-
    \right),
    \label{eq:clock_rotating_hamiltonian_ladder}
\end{equation}
where $\delta=\omega_0-\omega_L$ is the laser de-tuning. In terms of the Cartesian Pauli operators we have:
\begin{equation}
    H_{\mathrm{rot}}
    =
    \frac{\hbar\delta}{2}\sigma_z
    +
    \frac{\hbar\Omega(t)}{2}
    \left(
    \cos\phi_L\,\sigma_x
    +
    \sin\phi_L\,\sigma_y
    \right).
    \label{eq:clock_rotating_hamiltonian}
\end{equation}
Consequently then the effective Bloch-sphere rotation vector is written as:
\begin{equation}
    \boldsymbol{\Omega}_{\mathrm{eff}}
    =
    \left(
    \Omega\cos\phi_L,
    \Omega\sin\phi_L,
    \delta
    \right).
    \label{eq:effective_bloch_rotation_vector}
\end{equation}
For a square pulse, the generalized Rabi angular frequency is given by:
\begin{equation}
    \Omega_{\mathrm{R}}
    =
    \sqrt{\Omega^2+\delta^2}.
    \label{eq:generalized_rabi_frequency}
\end{equation}

On resonance where $\delta=0$, and for a pulse with fixed phase the Hamiltonian commutes with itself at different times. We define the pulse area:
\begin{equation}
    \Theta
    =
    \int_{\mathrm{pulse}}dt\,\Omega(t),
    \label{eq:clock_pulse_area}
\end{equation}
the pulse propagator becomes:
\begin{widetext}
\begin{equation}
\begin{aligned}
\begin{split}
    U_{\mathrm{p}}(\Theta,\phi_L)
    &=
    \exp\left[
    -\frac{i\Theta}{2}
    \left(
    \cos\phi_L\,\sigma_x
    +
    \sin\phi_L\,\sigma_y
    \right)
    \right]
    \nonumber\\
    &=
    \cos\left(\frac{\Theta}{2}\right)\mathbbm{1}
    -
    i\sin\left(\frac{\Theta}{2}\right)
    \left(
    \cos\phi_L\,\sigma_x
    +
    \sin\phi_L\,\sigma_y
    \right).
    \label{eq:resonant_pulse_unitary}
\end{split}
\end{aligned}
\end{equation}
\end{widetext}
Acting on the ground state gives us:
\begin{equation}
    U_{\mathrm{p}}(\Theta,\phi_L)|g\rangle
    =
    \cos\left(\frac{\Theta}{2}\right)|g\rangle
    -
    i e^{-i\phi_L}
    \sin\left(\frac{\Theta}{2}\right)|m\rangle .
    \label{eq:pulse_action_on_ground_state}
\end{equation}
A $\pi/2$ pulse therefore prepares the equal superposition:
\begin{equation}
    |g\rangle
    \longrightarrow
    \frac{1}{\sqrt{2}}
    \left(
    |g\rangle
    -
    i e^{-i\phi_L}|m\rangle
    \right),
    \label{eq:pi_over_two_clock_pulse}
\end{equation}
whereas a $\pi$ pulse transfers the population:
\begin{equation}
    |g\rangle
    \longrightarrow
    -i e^{-i\phi_L}|m\rangle .
    \label{eq:pi_clock_pulse}
\end{equation}

This two-level description is valid so long as the selected transition is spectroscopically isolated, the drive bandwidth and Rabi frequency are small compared with the separation from unwanted components, and the rotating-wave conditions:
\begin{equation}
    |\delta|,\Omega\ll\omega_L\simeq\omega_0
\end{equation}
are satisfied. Then under these assumptions the $^{229}\mathrm{Th}$ clock is a controllable nuclear qubit whose free evolution accumulates a relative phase and whose state can be rotated by phase-coherent vacuum-ultraviolet pulses. This structure supplies the starting point for Ramsey spectroscopy and, for an ensemble of nuclei, the collective clock-phase and energy-population observables that will be developed below.

Before we move on to the next section we want to distinguish between superpositions and mixture.

A resonant $\pi/2$ pulse prepares a coherent superposition of the two nuclear clock states. Up to the phase convention of the interrogation field, this state can be written as:
\begin{equation}
    |\psi_{\phi}\rangle
    =
    \frac{1}{\sqrt{2}}
    \left(
    |g\rangle
    +
    e^{i\phi}|m\rangle
    \right),
    \label{eq:clock_coherent_superposition}
\end{equation}
where $\phi$ is the relative phase established by the coherent driving field. The corresponding density operator is given by:
\begin{align}
    \rho_{\mathrm{sup}}
    &=
    |\psi_{\phi}\rangle\langle\psi_{\phi}|
    \nonumber\\
    &=
    \frac{1}{2}
    \left[
    |g\rangle\langle g|
    +
    |m\rangle\langle m|
    +
    e^{-i\phi}|g\rangle\langle m|
    +
    e^{i\phi}|m\rangle\langle g|
    \right].
    \label{eq:clock_superposition_density_operator}
\end{align}
In the ordered basis $(|g\rangle,|m\rangle)$, then:
\begin{equation}
    \rho_{\mathrm{sup}}
    =
    \frac{1}{2}
    \begin{pmatrix}
        1 & e^{-i\phi}\\
        e^{i\phi} & 1
    \end{pmatrix}.
    \label{eq:clock_superposition_density_matrix}
\end{equation}
This state must be distinguished from an incoherent statistical mixture containing equal ground- and isomer-state populations:
\begin{equation}
    \rho_{\mathrm{mix}}
    =
    \frac{1}{2}
    |g\rangle\langle g|
    +
    \frac{1}{2}
    |m\rangle\langle m|
    =
    \frac{1}{2}
    \begin{pmatrix}
        1 & 0\\
        0 & 1
    \end{pmatrix}.
    \label{eq:clock_incoherent_mixture}
\end{equation}
Both states satisfy:
\begin{equation}
    \langle P_g\rangle
    =
    \langle P_m\rangle
    =
    \frac{1}{2},
    \qquad
    \langle\sigma_z\rangle=0,
    \label{eq:equal_populations_superposition_mixture}
\end{equation}
and both possess the same energy variance:
\begin{equation}
    \Delta H_{\mathrm{clock}}
    =
    \frac{\hbar\omega_0}{2}.
    \label{eq:same_energy_variance_superposition_mixture}
\end{equation}
Energy populations and energy variance alone therefore cannot distinguish a coherent clock state from an incoherent ensemble.

So the distinction lies in the off-diagonal matrix elements. For the superposition we have:
\begin{equation}
    \langle g|\rho_{\mathrm{sup}}|m\rangle
    =
    \frac{1}{2}e^{-i\phi}\neq0,
\end{equation}
whereas for incoherent statistical mixture we have:
\begin{equation}
    \langle g|\rho_{\mathrm{mix}}|m\rangle=0.
\end{equation}
These off-diagonal elements encode the nuclear coherence and the relative phase that evolves under the clock Hamiltonian. Equivalently we can write:
\begin{equation}
    \langle\sigma_x\rangle_{\mathrm{sup}}=\cos\phi,
    \qquad
    \langle\sigma_y\rangle_{\mathrm{sup}}=\sin\phi,
\end{equation}
while:
\begin{equation}
    \langle\sigma_x\rangle_{\mathrm{mix}}
    =
    \langle\sigma_y\rangle_{\mathrm{mix}}
    =
    0.
\end{equation}
The coherent superposition can consequently produce Ramsey interference, whereas the incoherent mixture contains no phase information and produces no Ramsey fringe. A working nuclear clock therefore requires not just simultaneous populations in
$|g\rangle$ and $|m\rangle$, but a phase-coherent superposition of them.

\section{Minimal Moffat–Thompson nonlocal dynamics of the clock
transition}
\label{sec:minimal_MT_clock}

We will now incorporate the effective two-level nuclear clock into the Moffat--Thompson nonlocal quantum field theory framework~\cite{Yukawa1950I,Yukawa1950II,PaisUhlenbeck1950,Efimov1967,AlebastrovEfimov1973,Efimov1977,Krasnikov1987,Moffat1990,Evens1991,KleppeWoodard1992,Biswas2012,Tomboulis2015,ModestoPivaRachwal2016,ModestoRachwal2017,MoffatThompson2026,Thompson2026Localization,KouroshniaMoffat2026,Thompson2026PhaseSpace,Thompson2026Covariance,MoffatKouroshnia2026Doubling}. We explain how a nonlocal field theory can modify the energy difference between the $^{229}\mathrm{Th}$ ground and isomeric states, and thus the phase accumulated by the nuclear clock.

The minimal construction has four assumptions. First, the theory contains one nonlocality scale $E_M>0$, which we call the Moffat energy. Second, the local kinetic operators are dressed by an entire function of an appropriate Lorentz- and gauge-covariant differential operator. Third, the entire function is normalized in the infrared and has no finite zeros. Finally, the canonical quantum algebra is not deformed by hand. The clock correction is instead obtained by matching the regulated field theory onto an effective nuclear Hamiltonian and projecting that Hamiltonian onto the selected ground--isomer subspace.

To explore the nonlocal deformation of the field dynamics we first let $\Phi$ collectively represent the fields whose interactions generate the nuclear Hamiltonian. Their local quadratic action may be written as:
\begin{equation}
    S_0^{(2)}
    =
    \frac{1}{2}
    \langle \Phi,K_0\Phi\rangle ,
    \label{eq:local_quadratic_action_clock}
\end{equation}
where $K_0$ is the local kinetic operator and the bracket includes the
spacetime integral and all internal-index contractions.

In the minimal nonlocal theory, the quadratic action is replaced by:
\begin{equation}
    S_M^{(2)}
    =
    \frac{1}{2}
    \left\langle
    \Phi,
    \mathcal{F}^{-1}
    \left(
        \frac{\mathcal{D}}{E_M^2}
    \right)
    K_0\Phi
    \right\rangle ,
    \label{eq:nonlocal_quadratic_action_clock}
\end{equation}
where $\mathcal{D}$ is an appropriate covariant second-order operator. We note that depending on the field sector, it may be constructed from the d'Alembertian, covariant Laplacian, or a gauge-covariant kinetic operator. The regulator $\mathcal{F}(z)$ is assumed to satisfy:
\begin{align}
    \mathcal{F}(0)&=1,
    \label{eq:F_IR_normalization}
    \\
    \mathcal{F}(z^*)&=\mathcal{F}(z)^*,
    \label{eq:F_reality_condition}
    \\
    \mathcal{F}(z)&\neq 0
    \qquad \forall\;\;
    \text{finite }z\in\mathbb{C}.
    \label{eq:F_zero_free_condition}
\end{align}
The first condition makes sure that we recover the local theory at energies much smaller than $E_M$. The second ensures the appropriate reality or Hermiticity property on physical configurations. The third makes the regulator invertible and prevents it from introducing additional finite-energy zeros into the free kinetic operator~\cite{Moffat1990,Evens1991,KleppeWoodard1992,MoffatThompson2026}.

The free nonlocal field equation is written as:
\begin{equation}
    \mathcal{F}^{-1}
    \left(
        \frac{\mathcal{D}}{E_M^2}
    \right)
    K_0\Phi=0.
    \label{eq:nonlocal_free_field_equation}
\end{equation}
Since $\mathcal{F}$ has no finite zeros, the operator $\mathcal{F}^{-1}$ is invertible on the physical domain. Acting on Eq.~\eqref{eq:nonlocal_free_field_equation} with $\mathcal{F}$ therefore gives us:
\begin{equation}
    K_0\Phi=0.
    \label{eq:local_free_equation_recovered}
\end{equation}
Consequently we have:
\begin{equation}
    \ker K_M=\ker K_0,
    \qquad
    K_M
    \equiv
    \mathcal{F}^{-1}
    \left(
        \frac{\mathcal{D}}{E_M^2}
    \right)K_0.
    \label{eq:equal_free_kernels}
\end{equation}

The same point is transparent in a basis that diagonalizes the free kinetic operator. In momentum space:
\begin{equation}
    K_M(p)
    =
    \mathcal{F}^{-1}
    \left(
        \frac{\mathcal{D}(p)}{E_M^2}
    \right)
    K_0(p),
    \label{eq:nonlocal_momentum_kinetic_operator}
\end{equation}
and thus the propagator is:
\begin{align}
    G_M(p)
    &=
    K_M^{-1}(p)
    \nonumber\\
    &=
    \mathcal{F}
    \left(
        \frac{\mathcal{D}(p)}{E_M^2}
    \right)
    K_0^{-1}(p)
    \nonumber\\
    &=
    \mathcal{F}
    \left(
        \frac{\mathcal{D}(p)}{E_M^2}
    \right)
    G_0(p).
    \label{eq:nonlocal_propagator_clock}
\end{align}
Since $\mathcal{F}$ contains no finite poles or zeros, the finite pole locations of $G_M$ are the same as those of $G_0$. The regulator modifies off-shell propagation, ultraviolet behavior, and interaction amplitudes, but it does not create a new free particle pole or move an existing free pole through multiplication by an invertible function.

This observation is important for the nuclear clock, it makes sure that one cannot obtain a shift of the $^{229}\mathrm{Th}$ transition simply by replacing the free two-level equation with an invertibly dressed version of itself. The physical clock correction must come through the regulated interactions that determine the nuclear binding energies.

The full nonlocal action contains interaction terms in addition to Eq.~\eqref{eq:nonlocal_quadratic_action_clock}. At energies
$E\ll E_M$, analyticity permits the regulated interaction functional to be expanded as:
\begin{equation}
    \mathcal{I}_M
    =
    \mathcal{I}_0
    +
    \frac{1}{E_M^2}\mathcal{I}_2
    +
    \frac{1}{E_M^4}\mathcal{I}_4
    +\mathcal{O}(E_M^{-6}).
    \label{eq:nonlocal_interaction_expansion}
\end{equation}
Only even inverse powers appear here because we assume that the relevant entire-function dependence is analytic in a second-order covariant operator. Different symmetries or regulator choices may eliminate some of these terms, but Eq.~\eqref{eq:nonlocal_interaction_expansion} is the minimal generic expansion for the theroy we work with.

After the field degrees of freedom are matched onto a low-energy nuclear description, the nuclear Hamiltonian takes the form of:
\begin{equation}
    H_{\mathrm{nuc},M}
    =
    H_{\mathrm{nuc}}
    +
    \frac{1}{E_M^2}Q_3
    +
    \frac{1}{E_M^4}Q_5
    +\mathcal{O}(E_M^{-6}),
    \label{eq:matched_nonlocal_nuclear_hamiltonian}
\end{equation}
where subscripts on $Q_3$ and $Q_5$ indicate their energy dimensions:
\begin{equation}
    [Q_3]=(\mathrm{energy})^3,
    \qquad
    [Q_5]=(\mathrm{energy})^5.
\end{equation}
They represent the effective nuclear operators generated by the regulated strong, electromagnetic, spin-dependent, and other interactions after the high-energy field degrees of freedom have been integrated out.

We let $|g\rangle$ and $|m\rangle$ be the resolved ground and isomeric clock states introduced in the previus section. We define the projector onto the clock subspace:
\begin{equation}
    P_{\mathrm{c}}
    =
    |g\rangle\langle g|
    +
    |m\rangle\langle m|.
    \label{eq:clock_subspace_projector_MT}
\end{equation}
The effective two-level Hamiltonian is then given by:
\begin{equation}
    H_{\mathrm{c},M}
    =
    P_{\mathrm{c}}H_{\mathrm{nuc},M}P_{\mathrm{c}}.
    \label{eq:projected_nonlocal_clock_hamiltonian}
\end{equation}

For a nondegenerate resolved clock component, ordinary perturbation theory gives the corrected energy of either state $|i\rangle$, with
$i\in\{g,m\}$, as:
\begin{align}
    E_{i,M}
    &=
    E_i
    +
    \frac{1}{E_M^2}
    \langle i|Q_3|i\rangle
    \nonumber\\
    &\quad
    +
    \frac{1}{E_M^4}
    \left[
        \langle i|Q_5|i\rangle
        +
        \sum_{n\neq i}
        \frac{
            |\langle n|Q_3|i\rangle|^2
        }{
            E_i-E_n
        }
    \right]
    +
    \mathcal{O}(E_M^{-6}).
    \label{eq:nonlocal_energy_level_perturbation}
\end{align}
The second term is the first-order correction generated by $Q_3$. The sum in the third line is the usual second-order perturbative correction from virtual mixing with nuclear states outside the selected clock subspace. If the nonlocal correction mixes the two selected levels, we must first diagonalize $P_{\mathrm{c}}H_{\mathrm{nuc},M}P_{\mathrm{c}}$; the symbols $|g\rangle$ and $|m\rangle$ then denote the resulting corrected eigenstates.

We define the quantum expectation value of an operator:
\begin{equation}
    q_{3,i}
    \equiv
    \langle i|Q_3|i\rangle,
    \qquad
    \Delta q_3
    \equiv
    q_{3,m}-q_{3,g},
    \label{eq:q3_definitions}
\end{equation}
where $\Delta q_3$ is the physical difference between this expectation value in a metastable or excited state ($m$) versus the ground state ($g$), and $q_{3,i}$ is the expectation value (or matrix element) of an operator $Q_{3}$ for state $i$. The local transition energy is again:
\begin{equation}
\notag    E_{\mathrm{Th}}
    =
    E_m-E_g
    =
    \hbar\omega_0.
    \label{eq:local_thorium_transition_again}
\end{equation}
Subtracting the two corrected levels gives us:
\begin{align}
    E_{\mathrm{Th},M}
    &=
    E_{m,M}-E_{g,M}
    \nonumber\\
    &=
    E_{\mathrm{Th}}
    +
    \frac{\Delta q_3}{E_M^2}
    +
    \mathcal{O}(E_M^{-4}).
    \label{eq:corrected_thorium_transition_energy}
\end{align}
Thus the leading deterministic transition shift is:
\\
\begin{equation}
    \delta E_{\mathrm{Th}}
    \equiv
    E_{\mathrm{Th},M}-E_{\mathrm{Th}}
    =
    \frac{
        \langle m|Q_3|m\rangle
        -
        \langle g|Q_3|g\rangle
    }{E_M^2}
    +
    \mathcal{O}(E_M^{-4}).
    \label{eq:leading_MT_clock_shift}
\end{equation}
A correction common to both nuclear levels cancels. The clock is sensitive only to the difference between the nonlocal corrections to the ground and isomeric states.

A simple model will help illustrate why the leading correction scales as a nuclear energy cubed divided by $E_M^2$. We first let $\rho_i(\mathbf{x})$ be the static charge density of nuclear state $|i\rangle$, with Fourier transform $\rho_i(\mathbf{k})$. Its local Coulomb self-energy is given by:
\begin{equation}
    E_{\mathrm{C}}^{(i)}
    =
    \frac{1}{2}
    \int\frac{d^3\mathbf{k}}{(2\pi)^3}
    |\rho_i(\mathbf{k})|^2
    \frac{4\pi\alpha}{\mathbf{k}^2}.
    \label{eq:local_coulomb_self_energy}
\end{equation}
Now, suppose the static propagator is modified according to:
\begin{equation}
    \frac{4\pi\alpha}{\mathbf{k}^2}
    \longrightarrow
    \frac{4\pi\alpha}{\mathbf{k}^2}
    \mathcal{F}
    \left(
        \frac{\mathbf{k}^2}{E_M^2}
    \right),
    \label{eq:nonlocal_coulomb_propagator}
\end{equation}
and expand the entire function at small momentum:
\begin{equation}
    \mathcal{F}(u)
    =
    1-\gamma u+\mathcal{O}(u^2),
    \label{eq:F_small_u_expansion}
\end{equation}
where coefficient $\gamma$ is dimensionless and depends on the regulator convention. Substitution into Eq.~\eqref{eq:local_coulomb_self_energy} gives us:
\begin{align}
    E_{\mathrm{C},M}^{(i)}
    &=
    E_{\mathrm{C}}^{(i)}
    -
    \frac{2\pi\alpha\gamma}{E_M^2}
    \int\frac{d^3\mathbf{k}}{(2\pi)^3}
    |\rho_i(\mathbf{k})|^2
    +
    \mathcal{O}(E_M^{-4}).
    \label{eq:nonlocal_coulomb_energy_momentum}
\end{align}
Using Parseval's identity we have:
\begin{equation}
    \int\frac{d^3\mathbf{k}}{(2\pi)^3}
    |\rho_i(\mathbf{k})|^2
    =
    \int d^3\mathbf{x}\,
    \rho_i^2(\mathbf{x}),
    \label{eq:parseval_charge_density}
\end{equation}
the correction becomes:
\begin{equation}
    \delta E_{\mathrm{C}}^{(i)}
    =
    -
    \frac{2\pi\alpha\gamma}{E_M^2}
    \int d^3\mathbf{x}\,
    \rho_i^2(\mathbf{x})
    +
    \mathcal{O}(E_M^{-4}).
    \label{eq:nonlocal_coulomb_energy_position}
\end{equation}
The corresponding contribution to the clock transition is therefore given by:
\begin{equation}
    \delta E_{\mathrm{Th}}^{\mathrm{C}}
    =
    -
    \frac{2\pi\alpha\gamma}{E_M^2}
    \int d^3\mathbf{x}
    \left[
        \rho_m^2(\mathbf{x})
        -
        \rho_g^2(\mathbf{x})
    \right]
    +
    \mathcal{O}(E_M^{-4}).
    \label{eq:nonlocal_coulomb_clock_shift}
\end{equation}

For a nuclear density localized over a radius $R_{\mathrm{nuc}}$, dimensional analysis gives us:
\begin{equation}
    \int d^3\mathbf{x}\,\rho_i^2(\mathbf{x})
    \sim
    \frac{1}{R_{\mathrm{nuc}}^3}
    \sim
    E_{\mathrm{nuc}}^3,
    \label{eq:nuclear_density_scaling}
\end{equation}
in units with $\hbar=c=1$. Thus we have:
\begin{equation}
    \delta E_{\mathrm{Th}}^{\mathrm{C}}
    \sim
    \alpha\,
    \frac{E_{\mathrm{nuc}}^3}{E_M^2}
    \times
    \left(
        \text{difference of nuclear structures}
    \right).
    \label{eq:coulomb_clock_scaling}
\end{equation}
This calculation is illustrative rather than a complete nuclear-structure prediction. A quantitative coefficient requires the actual ground- and isomer-state charge distributions and the regulator appropriate to the electromagnetic sector. The same matching logic applies to regulated strong and spin-dependent nuclear interactions.

The thorium transition is only a few electronvolts, but each nuclear level is determined by much larger internal contributions. We can show this again as:
\begin{equation}
    \notag E_i
    =
    E_i^{\mathrm{strong}}
    +
    E_i^{\mathrm{Coul}}
    +
    E_i^{\mathrm{spin}}
    +
    E_i^{\mathrm{def}}
    +\cdots .
\end{equation}
The small transition energy is the difference between the corresponding ground- and isomer-state energies. A small fractional modification of a large internal contribution may therefore produce a comparatively large fractional change in $E_{\mathrm{Th}}$.

We let $X_a$ be a nuclear parameter or energy contribution on which the transition depends. To first order we have:
\begin{equation}
    \delta E_{\mathrm{Th}}
    =
    \sum_a
    \frac{\partial E_{\mathrm{Th}}}{\partial X_a}
    \delta X_a.
    \label{eq:clock_sensitivity_taylor}
\end{equation}
From this we define the dimensionless sensitivity coefficient:
\begin{equation}
    K_a
    \equiv
    \frac{X_a}{E_{\mathrm{Th}}}
    \frac{\partial E_{\mathrm{Th}}}{\partial X_a}
    =
    \frac{\partial\ln E_{\mathrm{Th}}}
         {\partial\ln X_a}.
    \label{eq:nuclear_sensitivity_coefficient}
\end{equation}
Dividing Eq.~\eqref{eq:clock_sensitivity_taylor} by $E_{\mathrm{Th}}$ gives us:
\begin{equation}
    \frac{\delta E_{\mathrm{Th}}}{E_{\mathrm{Th}}}
    =
    \sum_a
    K_a
    \frac{\delta X_a}{X_a}.
    \label{eq:fractional_clock_sensitivity}
\end{equation}
If the low-energy matching gives:
\begin{equation}
    \frac{\delta X_a}{X_a}
    =
    \beta_a
    \left(
        \frac{E_a}{E_M}
    \right)^2
    +
    \mathcal{O}(E_M^{-4}),
    \label{eq:parameter_MT_correction}
\end{equation}
where $E_a$ is the characteristic scale entering that contribution, then:
\begin{equation}
    \frac{\delta E_{\mathrm{Th}}}{E_{\mathrm{Th}}}
    =
    \sum_a
    K_a\beta_a
    \left(
        \frac{E_a}{E_M}
    \right)^2
    +
    \mathcal{O}(E_M^{-4}).
    \label{eq:enhanced_MT_clock_sensitivity}
\end{equation}
The enhancement coefficients $K_a$ do not mean that the clock laser excites the nucleus to a MeV-scale state, the laser still drives the approximately $8.36\,\mathrm{eV}$ ground--isomer transition. But the enhancement arises because that small energy difference depends on much larger internal nuclear contributions.

The corrected angular frequency is given by:
\begin{align}
    \omega_M
    &=
    \frac{E_{\mathrm{Th},M}}{\hbar}
    \nonumber\\
    &=
    \omega_0
    +
    \frac{\Delta q_3}{\hbar E_M^2}
    +
    \mathcal{O}(E_M^{-4}),
    \label{eq:corrected_MT_clock_frequency}
\end{align}
so that we have:
\begin{equation}
    \delta\omega_M
    \equiv
    \omega_M-\omega_0
    =
    \frac{\Delta q_3}{\hbar E_M^2}
    +
    \mathcal{O}(E_M^{-4}).
    \label{eq:MT_clock_frequency_shift}
\end{equation}

Within the two-level clock subspace we have:
\begin{equation}
    H_{\mathrm{c},M}
    =
    E_{g,M}|g\rangle\langle g|
    +
    E_{m,M}|m\rangle\langle m|.
    \label{eq:full_corrected_two_level_hamiltonian}
\end{equation}
Defining the nonlocal mean energy:
\begin{equation}
    \overline{E}_M
    =
    \frac{E_{m,M}+E_{g,M}}{2},
\end{equation}
we obtain:
\begin{equation}
    H_{\mathrm{c},M}
    =
    \overline{E}_M\mathbb{I}
    +
    \frac{\hbar\omega_M}{2}\sigma_z.
    \label{eq:corrected_two_level_pauli_hamiltonian}
\end{equation}
The identity term again produces only a common phase. The dynamically relevant one-nucleus Hamiltonian is:
\begin{equation}
    H_{2,M}
    =
    \frac{\hbar\omega_M}{2}\sigma_z.
    \label{eq:reduced_MT_two_level_hamiltonian}
\end{equation}

For free evolution over a Ramsey time $T_R$, relative to an interrogation field of angular frequency $\omega_L$, the rotating-frame Hamiltonian is:
\begin{equation}
    H_{\mathrm{free},M}
    =
    \frac{\hbar}{2}
    \left(
        \omega_M-\omega_L
    \right)\sigma_z.
    \label{eq:MT_rotating_free_hamiltonian}
\end{equation}
The corresponding propagator is then:
\begin{equation}
    U_{\mathrm{free},M}(T_R)
    =
    \exp
    \left[
        -\frac{i}{2}
        \left(
            \omega_M-\omega_L
        \right)
        T_R\sigma_z
    \right].
    \label{eq:MT_Ramsey_propagator}
\end{equation}
The accumulated clock phase is thus given by:
\begin{equation}
    \phi_M(T_R)
    =
    \left(
        \omega_M-\omega_L
    \right)T_R.
    \label{eq:MT_Ramsey_phase}
\end{equation}
Relative to the local prediction:
\begin{align}
    \delta\phi_M(T_R)
    &=
    \left(
        \omega_M-\omega_0
    \right)T_R
    \nonumber\\
    &=
    \frac{\Delta q_3}{\hbar E_M^2}T_R
    +
    \mathcal{O}(E_M^{-4}).
    \label{eq:nonlocal_Ramsey_phase_shift}
\end{align}
A constant $\Delta q_3$ therefore displaces the centre of the Ramsey fringe. It does not by itself broaden the fringe or create a random phase variance.

Before we move on we want to note that three energy quantities must always remain distinct:
\begin{equation}
    \delta E_{\mathrm{Th}}
    \neq
    \Delta H_{\mathrm{state}}
    \neq
    \epsilon_{\mathrm{line}},
    \label{eq:three_distinct_clock_energies}
\end{equation}
where $\delta E_{\mathrm{Th}}$ is the deterministic displacement of the nuclear transition caused by the matched nonlocal interactions. $\Delta H_{\mathrm{state}}$ is the intrinsic quantum variance of the clock Hamiltonian in a prepared superposition or squeezed state. And $\epsilon_{\mathrm{line}}$ denotes a stochastic or spectroscopic width produced by dephasing, inhomogeneity, decay, technical noise, or some specified nonlocal response process.

The minimal Moffat--Thompson dynamics derived here directly
provides the corrected generator for the system:
\begin{align}
    H_{2,M}
    &=
    \frac{\hbar\omega_M}{2}\sigma_z,
    \\
    \omega_M
    &=
    \omega_0
    +
    \frac{\Delta q_3}{\hbar E_M^2}
    +\mathcal{O}(E_M^{-4}),
    \label{eq:minimal_MT_final_generator}
\end{align}
and therefore a deterministic Ramsey-phase shift. It does not without an additional dynamical or measurement model imply an independent stochastic energy width, a nonzero phase-noise floor, or an additive time--energy covariance.

\section{Time–energy observables of a collective nuclear clock}

The two-level description developed above identifies the internal states and Hamiltonian of a single $^{229}\mathrm{Th}$ nucleus. A clock measurement, however, is generally performed on an ensemble of nuclei that are prepared, interrogated, and read out collectively. We now will construct the corresponding collective observables and show how the nuclear phase defines an operational clock-time variable conjugate, in the local tangent-plane sense, to the collective clock energy.

The main distinction is that the variable called ``time'' below is not the external laboratory parameter appearing in the Schr\"odinger equation~\cite{Heisenberg1927,Kennard1927,Robertson1929,Schrodinger1930,Pauli1933,MandelstamTamm1945,AharonovBohm1961,Busch1990I,Busch1990II,Busch2008,DeffnerCampbell2017}. It is an estimator obtained from the phase of the nuclear ensemble. So we will distinguish three quantities $T_R$, this is the externally selected Ramsey interrogation duration, $T$ is the temporal offset inferred from the collective clock phase, and $\Delta t_{\mathrm{env}}$, this is the duration of a pulse, coherence envelope, or decay signal.

Only $T$ will be treated as the operational clock observable~\cite{Wigner1932,Hillery1984,BraunsteinCaves1994,Paris2009,Giovannetti2006,Giovannetti2011,Helstrom1976,Holevo1982,TothApellaniz2014}. Similarly, the intrinsic quantum-state energy uncertainty $\Delta H$ must not be identified with either an estimated frequency uncertainty $\hbar\Delta\omega$ or a spectroscopic linewidth $\hbar\sigma_{\omega,\mathrm{line}}$.

We first consider $N_{\mathrm{eff}}$ nuclei that are coherently addressed on the same resolved ground--isomer transition. The number $N_{\mathrm{eff}}$ is the effective participating ensemble and need not equal the total number of thorium nuclei in the crystal. Nuclei outside the selected spectral component, nuclei that do not remain coherent, and nuclei that are not coupled to the interrogation field do not contribute fully to $N_{\mathrm{eff}}$.

For the $a$th nucleus, we define the Pauli operators as:
\begin{align}
    \sigma_x^{(a)}
    &=
    |m\rangle_a\langle g|
    +
    |g\rangle_a\langle m|,
    \\
    \sigma_y^{(a)}
    &=
    -i|m\rangle_a\langle g|
    +
    i|g\rangle_a\langle m|,
    \\
    \sigma_z^{(a)}
    &=
    |m\rangle_a\langle m|
    -
    |g\rangle_a\langle g|.
\end{align}
The superscript $(a)$ just indicates that the operator acts on the $a$th nuclear two-level system and as the identity on all remaining nuclei.

The dimensionless collective angular-momentum operators are given by:
\begin{equation}
    J_k
    =
    \frac{1}{2}
    \sum_{a=1}^{N_{\mathrm{eff}}}
    \sigma_k^{(a)},
    \qquad
    k=x,y,z.
    \label{eq:collective_spin_definition}
\end{equation}
Since Pauli operators belonging to distinct nuclei commute, these collective operators satisfy the usual angular-momentum algebra:
\begin{align}
    [J_x,J_y]&=iJ_z,
    \\
    [J_y,J_z]&=iJ_x,
    \\
    [J_z,J_x]&=iJ_y,
    \label{eq:collective_spin_algebra}
\end{align}
where the operators $J_k$ are dimensionless. Here the factor of $\hbar$ therefore enters the physical Hamiltonian rather than the commutation relations.

The isomeric-state number operator is written as:
\begin{align}
    N_m
    &=
    \sum_{a=1}^{N_{\mathrm{eff}}}
    |m\rangle_a\langle m|
    \nonumber\\
    &=
    \frac{1}{2}
    \sum_{a=1}^{N_{\mathrm{eff}}}
    \left(
        \mathbbm{1}+\sigma_z^{(a)}
    \right)
    \nonumber\\
    &=
    J_z+\frac{N_{\mathrm{eff}}}{2}.
    \label{eq:isomer_number_operator}
\end{align}
So $J_z$ measures the population difference between the isomeric and ground states:
\begin{equation}
    J_z
    =
    \frac{N_m-N_g}{2},
    \qquad
    N_m+N_g=N_{\mathrm{eff}}.
\end{equation}

Now by using the nonlocally corrected transition frequency $\omega_M$ derived in the
last section, the physical internal energy of the ensemble is given by:
\begin{align}
    H_{\mathrm{phys},M}
    &=
    \hbar\omega_M N_m
    \nonumber\\
    &=
    \hbar\omega_M
    \left(
        J_z+\frac{N_{\mathrm{eff}}}{2}
    \right),
    \label{eq:physical_collective_clock_hamiltonian}
\end{align}
here the second term is proportional to the identity and contributes only a common phase. It has no effect on any variance, commutator, or Ramsey signal. Consequently, the dynamically relevant collective clock Hamiltonian is given by:
\begin{equation}
    H_{\mathrm{clock},M}
    =
    \hbar\omega_M J_z.
    \label{eq:collective_clock_hamiltonian}
\end{equation}

For any operator $A$, we define its centered form and variance by:
\begin{equation}
    \delta A
    =
    A-\langle A\rangle,
    \qquad
    (\Delta A)^2
    =
    \langle(\delta A)^2\rangle.
    \label{eq:centered_operator_variance}
\end{equation}
The centered collective energy-fluctuation operator is thus:
\begin{align}
    \delta H
    &=
    H_{\mathrm{clock},M}
    -
    \langle H_{\mathrm{clock},M}\rangle
    \nonumber\\
    &=
    \hbar\omega_M
    \left(
        J_z-\langle J_z\rangle
    \right).
    \label{eq:centered_collective_energy}
\end{align}
Its variance is given by:
\begin{equation}
    (\Delta H)^2
    =
    \hbar^2\omega_M^2(\Delta J_z)^2,
    \label{eq:collective_energy_variance}
\end{equation}
or equivalently:
\begin{equation}
    \Delta H
    =
    \hbar\omega_M\Delta J_z.
    \label{eq:collective_energy_standard_deviation}
\end{equation}
So the collective energy uncertainty is therefore directly determined by the fluctuation of the ground--isomer population difference.

We choose a Ramsey operating state whose mean collective spin points along the positive $x$ direction:
\begin{align}
    \langle J_x\rangle&=S>0,
    \\
    \langle J_y\rangle_0&=0,
    \\
    \langle J_z\rangle_0&\simeq0,
    \label{eq:x_polarized_operating_state}
\end{align}
where:
\begin{equation}
    S\equiv\langle J_x\rangle
    \label{eq:mean_spin_length}
\end{equation}
is the mean collective-spin length at the operating point. For an ideal, fully coherent ensemble, $S=N_{\mathrm{eff}}/2$. Decoherence, incomplete state preparation, and reduced Ramsey contrast generally produce
$S<N_{\mathrm{eff}}/2$.

A clock phase $\phi$ is generated by a rotation around the $z$ axis:
\begin{equation}
    U_\phi
    =
    e^{-i\phi J_z}.
    \label{eq:collective_phase_rotation}
\end{equation}
The transformed $y$ component in the Heisenberg picture is:
\begin{equation}
    J_y(\phi)
    =
    U_\phi^\dagger J_yU_\phi.
    \label{eq:transformed_Jy_definition}
\end{equation}
To derive its form, we first define:
\begin{equation}
    F(\phi)
    =
    U_\phi^\dagger J_yU_\phi.
\end{equation}
Differentiating gives us:
\begin{align}
    \frac{dF}{d\phi}
    &=
    iU_\phi^\dagger[J_z,J_y]U_\phi
    \nonumber\\
    &=
    U_\phi^\dagger J_xU_\phi.
\end{align}
Together with the corresponding equation for $J_x$, this gives us the standard rotation relation:
\begin{equation}
    J_y(\phi)
    =
    J_y\cos\phi+J_x\sin\phi.
    \label{eq:Jy_rotation}
\end{equation}
For a small phase displacement:
\begin{equation}
    |\phi|\ll1,
    \qquad
    \cos\phi\simeq1,
    \qquad
    \sin\phi\simeq\phi,
\end{equation}
and hence:
\begin{equation}
    J_y(\phi)
    \simeq
    J_y+\phi J_x.
    \label{eq:linearized_Jy_rotation}
\end{equation}
Taking the expectation value in the operating state gives us:
\begin{align}
    \langle J_y(\phi)\rangle
    &\simeq
    \langle J_y\rangle_0
    +
    \phi\langle J_x\rangle
    \nonumber\\
    &=
    \phi S.
    \label{eq:mean_Jy_phase_response}
\end{align}
The observable $J_y$ thus provides a linear signal for a small collective phase.

A locally unbiased phase estimator is written as:
\begin{equation}
    \widehat{\phi}
    =
    \frac{J_y-\langle J_y\rangle_0}{S}.
    \label{eq:operational_phase_estimator}
\end{equation}
At the chosen operating point, $\langle J_y\rangle_0=0$, so we have:
\begin{equation}
    \widehat{\phi}
    =
    \frac{J_y}{S}.
    \label{eq:simple_phase_estimator}
\end{equation}
It is locally unbiased because:
\begin{equation}
    \langle\widehat{\phi}\rangle_\phi
    \simeq
    \phi
\end{equation}
to first order in $\phi$. Its variance is given by:
\begin{equation}
    (\Delta\phi)^2
    =
    \frac{(\Delta J_y)^2}{S^2},
    \label{eq:phase_estimator_variance}
\end{equation}
and therefore:
\begin{equation}
    \Delta\phi
    =
    \frac{\Delta J_y}{S}.
    \label{eq:phase_estimator_uncertainty}
\end{equation}
The transverse fluctuation $\Delta J_y$ is thus the phase noise of the
collective clock, while $S$ is the phase-to-signal conversion slope.

A temporal displacement $T$ of the nuclear clock relative to an ideal phase reference produces the carrier phase is written as $\phi = \omega_M T$. This relation differs from $\phi=\delta T_R$, which is used to infer a detuning $\delta$ from phase accumulated during an externally chosen Ramsey duration $T_R$. In the present construction $T=\phi/\omega_M$ is instead the time offset encoded by the phase of the nuclear carrier itself.

Promoting the phase estimator to a clock-time estimator gives us:
\begin{equation}
    \widehat{T}
    =
    \frac{\widehat{\phi}}{\omega_M}
    =
    \frac{J_y-\langle J_y\rangle_0}
         {\omega_M S}.
    \label{eq:operational_clock_time_estimator}
\end{equation}
At the operating point we have:
\begin{equation}
    \widehat{T}
    =
    \frac{J_y}{\omega_M S}.
    \label{eq:simple_clock_time_estimator}
\end{equation}
Its variance is given by:
\begin{equation}
    (\Delta T)^2
    =
    \frac{(\Delta J_y)^2}
         {\omega_M^2S^2},
    \label{eq:clock_time_variance}
\end{equation}
so that:
\begin{equation}
    \Delta T
    =
    \frac{\Delta J_y}{\omega_M S}.
    \label{eq:clock_time_uncertainty}
\end{equation}

Equation~\eqref{eq:operational_clock_time_estimator} does not define a universal self-adjoint time operator conjugate to every semibounded Hamiltonian. It defines a locally valid estimator of a small temporal displacement of this particular nuclear clock state. Its meaning depends on the prepared mean spin, the selected operating point, and the linear response of $J_y$ to the clock phase.

We now can compare the operational time estimator with the centered energy fluctuation. By using:
\begin{equation}
    \widehat{T}
    =
    \frac{J_y-\langle J_y\rangle_0}
         {\omega_M S}
\end{equation}
and:
\begin{equation}
    \delta H
    =
    \hbar\omega_M
    \left(
        J_z-\langle J_z\rangle
    \right),
\end{equation}
their commutator is:
\begin{align}
    [\widehat{T},\delta H]
    &=
    \left[
        \frac{J_y}{\omega_M S},
        \hbar\omega_M J_z
    \right]
    \nonumber\\
    &=
    \frac{\hbar}{S}[J_y,J_z]
    \nonumber\\
    &=
    i\hbar\frac{J_x}{S}.
    \label{eq:effective_clock_commutator}
\end{align}
This is not the exact operator identity $[\widehat{T},\delta H]=i\hbar\mathbbm{1}$ because the right-hand side still contains $J_x$. Nevertheless, taking its expectation value in the operating
state gives us:
\begin{align}
    \left\langle
        [\widehat{T},\delta H]
    \right\rangle
    &=
    i\hbar\frac{\langle J_x\rangle}{S}
    \nonumber\\
    &=
    i\hbar.
    \label{eq:expected_clock_commutator}
\end{align}
So the operational clock time and collective energy behave as a canonical pair in the tangent plane around the polarized clock state. In the strongly polarized regime, where fluctuations of $J_x$ are small relative to $S$, can can write:
\begin{equation}
    J_x\simeq S,
    \qquad
    [\widehat{T},\delta H]\simeq i\hbar.
    \label{eq:tangent_plane_canonical_pair}
\end{equation}

Robertson's inequality for two self-adjoint observables $A$ and $B$ is:
\begin{equation}
    \Delta A\,\Delta B
    \geq
    \frac{1}{2}
    \left|
        \langle[A,B]\rangle
    \right|.
    \label{eq:Robertson_general}
\end{equation}
Applying it to $\widehat{T}$ and $\delta H$, and using Eq.~\eqref{eq:expected_clock_commutator}, gives us:
\begin{equation}
    \Delta T\,\Delta H
    \geq
    \frac{\hbar}{2}.
    \label{eq:operational_time_energy_uncertainty}
\end{equation}

The same result follows directly from the collective-spin algebra. From:
\begin{equation}
    [J_y,J_z]=iJ_x,
\end{equation}
Robertson's relation gives us:
\begin{equation}
    \Delta J_y\,\Delta J_z
    \geq
    \frac{1}{2}
    |\langle J_x\rangle|
    =
    \frac{S}{2}.
    \label{eq:collective_spin_uncertainty}
\end{equation}
Now using:
\begin{equation}
    \Delta T
    =
    \frac{\Delta J_y}{\omega_M S},
    \qquad
    \Delta H
    =
    \hbar\omega_M\Delta J_z,
\end{equation}
their product is then:
\begin{align}
    \Delta T\,\Delta H
    &=
    \frac{\Delta J_y}{\omega_M S}
    \left(
        \hbar\omega_M\Delta J_z
    \right)
    \nonumber\\
    &=
    \frac{\hbar}{S}
    \Delta J_y\Delta J_z \;\;\geq
    \frac{\hbar}{2}.
    \label{eq:spin_derivation_time_energy}
\end{align}
The factors of $\omega_M$ cancel. The uncertainty relation therefore does not depend on the numerical value of the nuclear carrier frequency, although the separate time and energy widths do.

For correlated states, the stronger Schr\"odinger--Robertson relation is:
\begin{equation}
    (\Delta T)^2(\Delta H)^2
    -
    \operatorname{Cov}(T,H)^2
    \geq
    \frac{1}{4}
    \left|
        \left\langle
            [\widehat{T},\delta H]
        \right\rangle
    \right|^2,
    \label{eq:Schrodinger_Robertson_TE}
\end{equation}
where the symmetrized covariance is:
\begin{equation}
    \operatorname{Cov}(T,H)
    =
    \frac{1}{2}
    \left\langle
        \delta\widehat{T}\,\delta H
        +
        \delta H\,\delta\widehat{T}
    \right\rangle.
    \label{eq:time_energy_covariance}
\end{equation}
At the polarized operating point:
\begin{equation}
    (\Delta T)^2(\Delta H)^2
    -
    \operatorname{Cov}(T,H)^2
    \geq
    \frac{\hbar^2}{4}.
    \label{eq:full_time_energy_covariance_bound}
\end{equation}
Equivalently, defining the time--energy covariance matrix gives us:
\begin{equation}
    \Gamma_{TE}
    =
    \begin{pmatrix}
        (\Delta T)^2
        &
        \operatorname{Cov}(T,H)
        \\
        \operatorname{Cov}(T,H)
        &
        (\Delta H)^2
    \end{pmatrix},
    \label{eq:time_energy_covariance_matrix_early}
\end{equation}
we obtain:
\begin{equation}
    \det\Gamma_{TE}
    \geq
    \frac{\hbar^2}{4}.
    \label{eq:time_energy_covariance_determinant}
\end{equation}
This determinant form will be especially useful when the transverse nuclear state is squeezed and its uncertainty ellipse is rotated away from the $J_y$ and $J_z$ axes.

The operational construction can also be expressed as a Mandelstam--Tamm relation. For an observable $A$ with no explicit time dependence we write:
\begin{equation}
    \frac{d\langle A\rangle}{dt}
    =
    \frac{i}{\hbar}
    \langle[H,A]\rangle.
    \label{eq:Heisenberg_expectation_equation}
\end{equation}
Now taking:
\begin{equation}
    A=J_y,
    \qquad
    H=H_{\mathrm{clock},M}
    =
    \hbar\omega_MJ_z,
\end{equation}
gives us:
\begin{align}
    \frac{d\langle J_y\rangle}{dt}
    &=
    \frac{i}{\hbar}
    \left\langle
        [\hbar\omega_MJ_z,J_y]
    \right\rangle
    \nonumber\\
    &=
    \omega_M\langle J_x\rangle
    \nonumber\\
    &=
    \omega_M S.
    \label{eq:Jy_clock_rate}
\end{align}
The characteristic time associated with the change of $J_y$ is therefore given by:
\begin{equation}
    T_{J_y}
    =
    \frac{\Delta J_y}
         {|d\langle J_y\rangle/dt|}
    =
    \frac{\Delta J_y}{\omega_M S}.
    \label{eq:Mandelstam_Tamm_clock_time}
\end{equation}
This is exactly the uncertainty $\Delta T$ of the operational estimator. Robertson's inequality between $J_y$ and $H$ then gives us:
\begin{equation}
    T_{J_y}\Delta H
    \geq
    \frac{\hbar}{2}.
    \label{eq:Mandelstam_Tamm_clock_result}
\end{equation}

Finally, we let a temporal displacement be encoded in a pure collective state as:
\begin{equation}
    |\psi(T)\rangle
    =
    e^{-iH_{\mathrm{clock},M}T/\hbar}
    |\psi_0\rangle.
    \label{eq:time_parameterized_clock_state}
\end{equation}
The quantum Fisher information for estimating $T$ is given by:
\begin{equation}
    F_Q(T)
    =
    \frac{4}{\hbar^2}
    (\Delta H)^2.
    \label{eq:clock_quantum_Fisher_information}
\end{equation}
For one preparation, the quantum Cram\'er--Rao bound states:
\begin{equation}
    (\Delta\widehat{T})^2
    \geq
    \frac{1}{F_Q}.
    \label{eq:quantum_Cramer_Rao_clock}
\end{equation}
Substitution gives:
\begin{equation}
    (\Delta\widehat{T})^2
    \geq
    \frac{\hbar^2}
         {4(\Delta H)^2},
\end{equation}
and therefore:
\begin{equation}
    \Delta\widehat{T}\,\Delta H
    \geq
    \frac{\hbar}{2}.
    \label{eq:QFI_time_energy_result}
\end{equation}
The collective-spin, effective-commutator, Mandelstam--Tamm, and quantum Fisher-information derivations therefore agree. The nuclear clock supplies an operational time variable through its measurable collective phase, while the conjugate energy variable is the quantum fluctuation of the ground--isomer population.

\section{The coherent nuclear ensemble and its quantum projection-noise scale}

The time--energy observables that we derived above apply to an arbitrary collective nuclear state. We now establish the unsqueezed reference state against which squeezing and excess noise will later be measured. This reference is a coherent spin state where every participating nucleus occupies the same coherent ground--isomer superposition, but the nuclei remain mutually unentangled~\cite{Dicke1954,TavisCummings1968,Radcliffe1971,Arecchi1972}.

We start by supposing that $N$ nuclei are initially prepared in the ground state:
\begin{equation}
    |\Psi_0\rangle
    =
    |g\rangle^{\otimes N}.
    \label{eq:all_ground_initial_state}
\end{equation}
A spatially uniform resonant $\pi/2$ pulse with phase $\phi_L$ acts identically on every nucleus. With an appropriate phase convention, the single-nucleus state after the pulse is:
\begin{equation}
    |\psi_{\phi_L}\rangle
    =
    \frac{1}{\sqrt{2}}
    \left(
        |g\rangle
        +
        e^{i\phi_L}|m\rangle
    \right).
    \label{eq:single_nucleus_equatorial_state}
\end{equation}
The full ensemble is therefore represented by:
\begin{equation}
    |\mathrm{CSS}_{\phi_L}\rangle
    =
    \left[
        \frac{
            |g\rangle+e^{i\phi_L}|m\rangle
        }{\sqrt{2}}
    \right]^{\otimes N}.
    \label{eq:general_coherent_spin_state}
\end{equation}
For $\phi_L=0$, the mean collective spin points along the positive $x$
direction, as shown by:
\begin{equation}
    |\mathrm{CSS}_{x}\rangle
    =
    \left(
        \frac{|g\rangle+|m\rangle}{\sqrt{2}}
    \right)^{\otimes N}.
    \label{eq:x_polarized_CSS}
\end{equation}

Because the state factorizes into a product of single-nucleus states, it is separable, as shown by:
\begin{equation}
    \rho_{\mathrm{CSS}}
    =
    \bigotimes_{i=1}^{N}
    \rho_i.
    \label{eq:CSS_separable_density_matrix}
\end{equation}
So all nuclei share the same phase, but no entanglement has yet been created.

For one nucleus in the state of equal superposition state of two quantum levels, $(|g\rangle+|m\rangle)/\sqrt{2}$:
\begin{equation}
    \langle\sigma_x\rangle=1,
    \qquad
    \langle\sigma_y\rangle
    =
    \langle\sigma_z\rangle=0.
\end{equation}
By using:
\begin{equation}
    J_k
    =
    \frac{1}{2}\sum_{i=1}^{N}\sigma_k^{(i)},
\end{equation}
the collective mean spin is then:
\begin{equation}
    S
    \equiv
    \langle J_x\rangle
    =
    \frac{N}{2}.
    \label{eq:CSS_mean_spin}
\end{equation}
Since distinct nuclei are statistically independent in a product state, their variances add. For either transverse component:
\begin{align}
    (\Delta J_y)^2
    &=
    \frac{1}{4}
    \sum_{i=1}^{N}
    (\Delta\sigma_y^{(i)})^2
    =
    \frac{N}{4},
    \label{eq:CSS_Jy_variance}
    \\
    (\Delta J_z)^2
    &=
    \frac{1}{4}
    \sum_{i=1}^{N}
    (\Delta\sigma_z^{(i)})^2
    =
    \frac{N}{4}.
    \label{eq:CSS_Jz_variance}
\end{align}
The coherent spin state therefore has equal transverse uncertainties:
\begin{equation}
    \Delta J_y
    =
    \Delta J_z
    =
    \frac{\sqrt{N}}{2}.
    \label{eq:CSS_transverse_standard_deviations}
\end{equation}

The variance in Eq.~\eqref{eq:CSS_Jz_variance} is quantum projection noise, it arises because each nucleus in the coherent superposition yields either $|g\rangle$ or $|m\rangle$ when its energy population is measured.

We let $N_m$ be the number of nuclei found in the isomeric state. For the equatorial coherent spin state, each nucleus has probability $p_m=\frac{1}{2}$ of being measured in $|m\rangle$. The probability of obtaining exactly $n$ isomeric nuclei is therefore binomial:
\begin{equation}
    P(N_m=n)
    =
    \binom{N}{n}
    \left(\frac{1}{2}\right)^n
    \left(\frac{1}{2}\right)^{N-n}
    =
    2^{-N}\binom{N}{n}.
    \label{eq:CSS_binomial_distribution}
\end{equation}
Its mean and variance are given by:
\begin{equation}
    \langle N_m\rangle=\frac{N}{2},
    \qquad
    (\Delta N_m)^2=\frac{N}{4}.
    \label{eq:binomial_population_statistics}
\end{equation}
Since:
\begin{equation}
    J_z=N_m-\frac{N}{2},
\end{equation}
we obtain:
\begin{equation}
    (\Delta J_z)^2
    =
    (\Delta N_m)^2
    =
    \frac{N}{4}.
\end{equation}
Projection noise is this not detector noise or classical uncertainty about how many nuclei were prepared. It is the irreducible shot-to-shot fluctuation produced when a coherent quantum superposition is projected onto the energy basis.

Using the phase estimator derived previously:
\begin{equation}
    \Delta\phi
    =
    \frac{\Delta J_y}{|\langle J_x\rangle|},
\end{equation}
the coherent-state phase uncertainty is:
\begin{align}
    \Delta\phi_{\mathrm{CSS}}
    &=
    \frac{\sqrt{N}/2}{N/2}
    \nonumber\\
    &=
    \frac{1}{\sqrt{N}}.
    \label{eq:CSS_standard_quantum_limit}
\end{align}
This is the standard quantum limit for a single interrogation of $N$ independent nuclei.

The corresponding operational clock-time uncertainty is given by:
\begin{equation}
    \Delta T_{\mathrm{CSS}}
    =
    \frac{\Delta\phi_{\mathrm{CSS}}}{\omega_M}
    =
    \frac{1}{\omega_M\sqrt{N}},
    \label{eq:CSS_clock_time_uncertainty}
\end{equation}
whereas the collective energy uncertainty is:
\begin{equation}
    \Delta H_{\mathrm{CSS}}
    =
    \hbar\omega_M\Delta J_z
    =
    \frac{\hbar\omega_M\sqrt{N}}{2}.
    \label{eq:CSS_energy_uncertainty}
\end{equation}
Their product is given by:
\begin{equation}
    \Delta T_{\mathrm{CSS}}
    \Delta H_{\mathrm{CSS}}
    =
    \frac{\hbar}{2}.
    \label{eq:CSS_minimum_TE_product}
\end{equation}
So the unsqueezed coherent nuclear ensemble is thus already a minimum-uncertainty state. Squeezing will not reduce this phase-space area, but it will redistribute the noise between the phase and energy-population quadratures.

Although Eq.~\eqref{eq:general_coherent_spin_state} is a product state, it can
also be expanded in the symmetric Dicke basis:
\begin{equation}
    |\mathrm{CSS}_{\phi_L}\rangle
    =
    2^{-N/2}
    \sum_{n=0}^{N}
    \sqrt{\binom{N}{n}}\,
    e^{in\phi_L}
    |D_n^{(N)}\rangle,
    \label{eq:CSS_Dicke_expansion}
\end{equation}
where $|D_n^{(N)}\rangle$ is the normalized symmetric state containing exactly $n$ nuclei in $|m\rangle$ and $N-n$ nuclei in $|g\rangle$. The binomial coefficients reproduce the population distribution in Eq.~\eqref{eq:CSS_binomial_distribution}.

This expansion does not make the coherent spin state a GHZ or cat state. A GHZ state would have the form:
\begin{equation}
    |\mathrm{GHZ}\rangle
    =
    \frac{
        |g\rangle^{\otimes N}
        +
        e^{i\varphi}|m\rangle^{\otimes N}
    }{\sqrt{2}},
    \label{eq:GHZ_comparison}
\end{equation}
and would contain only the two extreme Dicke sectors $n=0$ and $n=N$. The coherent spin state instead contains the full binomial distribution and requires only one collective pulse, not individual control of the nuclei.

A macroscopic crystal may contain a very large total number of thorium nuclei, but they need not contribute equally to the measured collective mode. We let $w_i$ be the real coupling or detection weight of nucleus $i$. It may include the local probe intensity, cavity-mode amplitude, transition strength, spectral addressing efficiency, and readout efficiency. The measured weighted population operator is given by:
\begin{equation}
    J_z^{(w)}
    =
    \frac{1}{2}
    \sum_i w_i\sigma_z^{(i)}.
    \label{eq:weighted_collective_population}
\end{equation}
For independent nuclei, its projection-noise variance is:
\begin{equation}
    \left(\Delta J_z^{(w)}\right)^2
    =
    \frac{1}{4}\sum_i w_i^2.
    \label{eq:weighted_projection_noise}
\end{equation}
By contrast, the small-phase signal is proportional to $\sum_i w_i$. The resulting phase uncertainty is then given by:
\begin{equation}
    \Delta\phi
    =
    \frac{\sqrt{\sum_i w_i^2}}
         {\left|\sum_i w_i\right|}
    =
    \frac{1}{\sqrt{N_{\mathrm{eff}}}},
    \label{eq:weighted_phase_uncertainty}
\end{equation}
where:
\begin{equation}
    N_{\mathrm{eff}}
    =
    \frac{
        \left(\sum_i w_i\right)^2
    }{
        \sum_i w_i^2
    }
    \label{eq:effective_particle_number}
\end{equation}
is the effective coherent particle number. Uniform coupling, $w_i=w$, gives $N_{\mathrm{eff}}=N$. Strongly nonuniform coupling can instead give $N_{\mathrm{eff}}\ll N$, even when the crystal contains many nuclei.

If nucleus $i$ also has a detuning $\delta_i$, its coherence acquires the phase $e^{-i\delta_i t}$. The time-dependent coherent number becomes:
\begin{equation}
    N_{\mathrm{eff}}(t)
    =
    \frac{
        \left|
            \sum_i w_i e^{-i\delta_i t}
        \right|^2
    }{
        \sum_i w_i^2
    }.
    \label{eq:time_dependent_effective_number}
\end{equation}
Inhomogeneous broadening therefore reduces the coherent particle number as the individual nuclear phases separate, even though the physical number of thorium nuclei remains unchanged. In all subsequent clock-sensitivity and squeezing relations, $N$ should consequently be replaced by $N_{\mathrm{eff}}$, unless uniform coherent participation of the full ensemble has been established.

\section{Collective nonlinear generation of tunable nuclear spin squeezing}
\label{sec:tunable_nuclear_squeezing}

The coherent spin state introduced above contains a well-defined collective nuclear phase, but it is a separable product state and therefore remains limited by quantum projection noise. A resonant $\pi/2$ pulse alone cannot produce spin squeezing as it applies the same single-nucleus rotation to every member of the ensemble and does not correlate their fluctuations. Preparing an entangled squeezed state requires an additional nonlinear collective interaction~\cite{Appel2009,Leroux2010,SchleierSmith2010,LouchetChauvet2010,GrossNature2010,Riedel2010,Hosten2016,Cox2016,Davis2016,PolzikYe2016,Galanis2026,Boyers2025}.

Throughout this section, we write:
\begin{equation}
    N\equiv N_{\mathrm{eff}},
    \qquad
    S=\langle J_x\rangle\simeq\frac{N}{2},
    \label{eq:squeezing_effective_number}
\end{equation}
where $N_{\mathrm{eff}}$ is the effective number of nuclei that participate coherently in the selected collective mode.

A minimal theoretical mechanism is one-axis twisting, this is described by the Hamiltonian:
\begin{equation}
    H_{\mathrm{sq}}
    =
    \hbar\chi J_z^2,
    \label{eq:one_axis_twisting_hamiltonian}
\end{equation}
where $\chi$ has units of angular frequency and determines the strength of the nonlinear collective interaction. The corresponding evolution over a squeezing time $t_s$ is given by:
\begin{equation}
    U_{\mathrm{sq}}(t_s)
    =
    \exp\left(
        -i\chi t_sJ_z^2
    \right).
    \label{eq:one_axis_twisting_unitary}
\end{equation}
By starting from the $x$-polarized coherent spin state, the state after the entangling evolution is:
\begin{equation}
    |\psi(t_s)\rangle
    =
    U_{\mathrm{sq}}(t_s)
    |\mathrm{CSS}_x\rangle.
    \label{eq:twisted_nuclear_state}
\end{equation}

The operator $J_z^2$ is genuinely collective. By using:
\begin{equation}
    J_z
    =
    \frac{1}{2}
    \sum_{i=1}^{N}\sigma_z^{(i)},
\end{equation}
we find that:
\begin{align}
    J_z^2
    &=
    \frac{1}{4}
    \sum_{i,j=1}^{N}
    \sigma_z^{(i)}\sigma_z^{(j)}
    \nonumber\\
    &=
    \frac{N}{4}\mathbbm{1}
    +
    \frac{1}{2}
    \sum_{i<j}
    \sigma_z^{(i)}\sigma_z^{(j)}.
    \label{eq:Jz_squared_pairwise}
\end{align}
The first term contributes only a global phase. The second contains effective pairwise interactions between different nuclei. Consequently, $U_{\mathrm{sq}}$ cannot generally be written as a product of independent single-nucleus operations. For nonzero interaction strength and generic $t_s$, it correlates the nuclear populations and produces an entangled collective state.

Physically, different eigencomponents of $J_z$ acquire different phases proportional to the square of their population imbalance as shwon by:
\begin{equation}
    |J,m_z\rangle
    \longrightarrow
    e^{-i\chi t_sm_z^2}
    |J,m_z\rangle.
    \label{eq:dicke_twisting_phase}
\end{equation}
The phase therefore varies nonlinearly across the Dicke-state distribution. This distorts the initially circular projection-noise distribution into a tilted ellipse.

Near the strongly $x$-polarized state, we define the dimensionless transverse quadratures to be:
\begin{equation}
    X
    =
    \frac{J_y}{\sqrt{S}},
    \qquad
    P
    =
    \frac{J_z}{\sqrt{S}}.
    \label{eq:nuclear_squeezing_quadratures}
\end{equation}
Their commutator is given by:
\begin{equation}
    [X,P]
    =
    \frac{iJ_x}{S}
    \simeq i,
    \label{eq:squeezing_quadrature_commutator}
\end{equation}
so the transverse collective-spin plane behaves approximately as a canonical continuous-variable phase space. For the initial coherent spin state we write:
\begin{equation}
    (\Delta X)^2
    =
    (\Delta P)^2
    =
    \frac{1}{2},
    \qquad
    \operatorname{Cov}(X,P)=0.
    \label{eq:initial_CSS_quadratures}
\end{equation}

The Heisenberg equations generated by
Eq.~\eqref{eq:one_axis_twisting_hamiltonian} are:
\begin{align}
    \frac{dJ_z}{dt}
    &=
    \frac{i}{\hbar}
    [H_{\mathrm{sq}},J_z]
    =
    0,
    \label{eq:OAT_Jz_constant}
    \\
    \frac{dJ_y}{dt}
    &=
    \frac{i}{\hbar}
    [H_{\mathrm{sq}},J_y]
    \nonumber\\
    &=
    \chi
    \left(
        J_zJ_x+J_xJ_z
    \right).
    \label{eq:OAT_Jy_exact}
\end{align}
In the polarized tangent-plane regime, $J_x\simeq S$, and thus:
\begin{equation}
    \frac{dJ_y}{dt}
    \simeq
    2\chi S J_z.
    \label{eq:OAT_Jy_linearized}
\end{equation}
Integrating over $t_s$ gives us:
\begin{equation}
    J_y(t_s)
    \simeq
    J_y(0)+\kappa J_z(0),
    \qquad
    J_z(t_s)=J_z(0),
    \label{eq:OAT_spin_shear}
\end{equation}
where $\kappa = 2\chi S t_s$ is the dimensionless shearing strength. In quadrature form we write this as:
\begin{equation}
    \begin{pmatrix}
        X(t_s)\\
        P(t_s)
    \end{pmatrix}
    =
    \begin{pmatrix}
        1 & \kappa\\
        0 & 1
    \end{pmatrix}
    \begin{pmatrix}
        X(0)\\
        P(0)
    \end{pmatrix}.
    \label{eq:quadrature_shear_matrix}
\end{equation}
One-axis twisting therefore shears the circular coherent-state distribution rather than immediately compressing it along $J_y$ or $J_z$.

Applying this transformation to the initial covariance matrix
$\Gamma_0=\mathbbm{1}/2$ gives us:
\begin{equation}
    \Gamma(t_s)
    =
    \frac{1}{2}
    \begin{pmatrix}
        1+\kappa^2 & \kappa\\
        \kappa & 1
    \end{pmatrix}.
    \label{eq:OAT_covariance_matrix}
\end{equation}
Its eigenvalues, which are the minimum and maximum quadrature variances, are:
\begin{equation}
    V_{\mp}
    =
    \frac{1}{4}
    \left[
        2+\kappa^2
        \mp
        |\kappa|\sqrt{\kappa^2+4}
    \right].
    \label{eq:OAT_principal_variances}
\end{equation}
They can be written in the canonical squeezed-state form:
\begin{equation}
    V_-
    =
    \frac{1}{2}e^{-2r},
    \qquad
    V_+
    =
    \frac{1}{2}e^{2r},
    \label{eq:principal_variances_r}
\end{equation}
so long as:
\begin{equation}
    r
    =
    \operatorname{arsinh}
    \left(
        \frac{|\kappa|}{2}
    \right)
    =
    \operatorname{arsinh}
    \left(
        |\chi|St_s
    \right)
    .
    \label{eq:r_from_OAT}
\end{equation}
So $r=0$ when $t_s=0$, and increasing the interaction time or nonlinear coupling increases the squeezing magnitude within the Gaussian, high-polarization regime.

The principal squeezed direction produced by one-axis twisting is generally tilted relative to the clock-phase axis $J_y$. A collective rotation about the mean-spin direction:
\begin{equation}
    R_x(\alpha)
    =
    e^{-i\alpha J_x},
    \label{eq:squeezing_alignment_rotation}
\end{equation}
transforms the measured quadrature according to:
\begin{equation}
    J_y(\alpha)
    =
    J_y\cos\alpha-J_z\sin\alpha.
    \label{eq:rotated_phase_quadrature}
\end{equation}
By choosing $\alpha$ to coincide with the minimum-variance principal direction aligns the squeezed quadrature with $J_y$, which is the phase quadrature used by the Ramsey clock.

The complete preparation sequence is therefore:
\begin{equation}
    |g\rangle^{\otimes N}
    \xrightarrow{\;\pi/2\;}
    |\mathrm{CSS}_x\rangle
    \xrightarrow{\;e^{-i\chi t_sJ_z^2}\;}
    |\psi_{\mathrm{tilt}}\rangle
    \xrightarrow{\;R_x(\alpha_s)\;}
    |\psi_r\rangle.
    \label{eq:complete_squeezing_sequence}
\end{equation}
A tunable family is generated by selecting:
\begin{equation}
    t_s=0,t_1,t_2,t_3,\ldots,
\end{equation}
or equivalently by changing $\chi$. In an effective cavity- or measurement-mediated implementation, $\chi$ may depend on probe intensity, detuning, cavity coupling, or measurement strength. The microscopic realization of this interaction is separate from the state-level theory developed here.

The value of $r$ should ultimately be determined from the measured principal variances rather than inferred only from the nominal product $\chi t_s$:
\begin{equation}
    r_{\mathrm{exp}}
    =
    \frac{1}{4}
    \ln\left(
        \frac{V_+}{V_-}
    \right).
    \label{eq:experimental_r}
\end{equation}
For every chosen value of $r$, a fresh ensemble must be initialized, squeezed, rotated, and measured repeatedly. The same quantum state cannot generally be reused at successively larger $r$, because population detection and measurement-mediated squeezing condition or destroy the prepared state.

\section{Local Gaussian benchmark for the squeezed nuclear clock}
\label{sec:local_squeezed_clock_prediction}

Before we move on, we will explain the prediction of ordinary local quantum mechanics for the entangled nuclear squeezed states prepared in the previous section. This prediction supplies the reference curve against which any additional nonlocal broadening must be tested.

Throughout this section, we have $N_{\mathrm{eff}}$ as the effective number of coherently participating nuclei:
\begin{equation}
    S
    \equiv
    \langle J_x\rangle
    \simeq
    \frac{N_{\mathrm{eff}}}{2},
    \label{eq:local_squeezed_mean_spin}
\end{equation}
and $\omega_M$ is the corrected nuclear transition frequency derived from the minimal Moffat--Thompson Hamiltonian. At this stage, however, the state fluctuations are assumed to obey the ordinary local collective-spin algebra.

Near the $x$-polarized operating state, we define the normalized transverse
quadratures as:
\begin{equation}
    X=\frac{J_y}{\sqrt{S}},
    \qquad
    P=\frac{J_z}{\sqrt{S}},
    \label{eq:local_clock_quadratures}
\end{equation}
for which:
\begin{equation}
    [X,P]
    =
    i\frac{J_x}{S}
    \simeq i.
    \label{eq:local_quadrature_commutator}
\end{equation}
The variable $X$ is the collective phase quadrature, whereas $P$ is the population-difference, and hence energy, quadrature. An ideal Gaussian squeezing operation with squeezing magnitude $r\geq0$, aligned so that the phase quadrature is squeezed, produces:
\begin{equation}
    (\Delta X)^2_r
    =
    \frac{1}{2}e^{-2r},
    \qquad
    (\Delta P)^2_r
    =
    \frac{1}{2}e^{2r}.
    \label{eq:local_squeezed_quadrature_variances}
\end{equation}
Transforming back to collective-spin variables gives us:
\begin{equation}
    (\Delta J_y)^2_r
    =
    \frac{S}{2}e^{-2r},
    \qquad
    (\Delta J_z)^2_r
    =
    \frac{S}{2}e^{2r}.
    \label{eq:local_squeezed_spin_variances}
\end{equation}
For $S=N_{\mathrm{eff}}/2$, we have:
\begin{equation}
    \Delta J_y(r)
    =
    \frac{\sqrt{N_{\mathrm{eff}}}}{2}e^{-r},
    \qquad
    \Delta J_z(r)
    =
    \frac{\sqrt{N_{\mathrm{eff}}}}{2}e^{r}.
    \label{eq:local_squeezed_spin_widths}
\end{equation}
Increasing $r$ suppresses the phase fluctuation while increasing the conjugate ground--isomer population fluctuation.

The local phase estimator is:
\begin{equation}
    \widehat{\phi}
    =
    \frac{J_y}{S},
\end{equation}
so its uncertainty becomes:
\begin{align}
    \Delta\phi_{\mathrm{loc}}(r)
    &=
    \frac{\Delta J_y(r)}{S}
    \nonumber\\
    &=
    \frac{e^{-r}}{\sqrt{N_{\mathrm{eff}}}}.
    \label{eq:local_squeezed_phase_uncertainty}
\end{align}
A time offset of the nuclear carrier satisfies $\phi=\omega_M T$. The operational clock-time uncertainty is thus given by:
\begin{equation}
    \Delta T_{\mathrm{loc}}(r)
    =
    \frac{e^{-r}}
         {\omega_M\sqrt{N_{\mathrm{eff}}}}.
    \label{eq:local_squeezed_time_uncertainty}
\end{equation}
This quantity is distinct from the uncertainty in a detuning measured during a Ramsey interval $T_R$. Since $\phi=\delta T_R$, the corresponding single-shot detuning uncertainty is:
\begin{equation}
    \Delta\delta_{\mathrm{loc}}(r)
    =
    \frac{e^{-r}}
         {T_R\sqrt{N_{\mathrm{eff}}}}.
    \label{eq:local_squeezed_detuning_uncertainty}
\end{equation}

The collective clock Hamiltonian is:
\begin{equation}
    H_{\mathrm{clock},M}
    =
    \hbar\omega_MJ_z.
\end{equation}
Its quantum-state energy uncertainty is consequently:
\begin{align}
    \Delta H_{\mathrm{loc}}(r)
    &=
    \hbar\omega_M\Delta J_z(r)
    \nonumber\\
    &=
    \frac{
        \hbar\omega_M\sqrt{N_{\mathrm{eff}}}
    }{2}
    e^{r}.
    \label{eq:local_squeezed_energy_uncertainty}
\end{align}
By combining Eqs.~\eqref{eq:local_squeezed_time_uncertainty} and \eqref{eq:local_squeezed_energy_uncertainty} we find:
\begin{align}
    \Delta T_{\mathrm{loc}}(r)
    \Delta H_{\mathrm{loc}}(r)
    &=
    \frac{e^{-r}}
         {\omega_M\sqrt{N_{\mathrm{eff}}}}
    \frac{
        \hbar\omega_M\sqrt{N_{\mathrm{eff}}}
    }{2}
    e^r
    \nonumber\\
    &=
    \frac{\hbar}{2}.
    \label{eq:local_squeezed_minimum_product}
\end{align}
Ideal squeezing therefore does not violate the operational time--energy uncertainty relation. It redistributes the fixed minimum uncertainty:
\begin{equation}
    \Delta T_{\mathrm{loc}}
    \longrightarrow
    e^{-r}\Delta T_{\mathrm{CSS}},
    \qquad
    \Delta H_{\mathrm{loc}}
    \longrightarrow
    e^r\Delta H_{\mathrm{CSS}}.
    \label{eq:local_uncertainty_redistribution}
\end{equation}

For an imperfect state, reduced phase noise must be compared with the remaining mean-spin coherence. The appropriate metrological parameter is:
\begin{equation}
    \xi_R^2
    =
    \frac{
        N_{\mathrm{eff}}(\Delta J_y)^2
    }{
        |\langle J_x\rangle|^2
    }.
    \label{eq:local_Wineland_parameter}
\end{equation}
A coherent spin state has $\xi_R^2=1$, while a metrologically useful squeezed
state has $\xi_R^2<1$. Its phase and clock-time uncertainties are given by:
\begin{equation}
    \Delta\phi_{\mathrm{loc}}
    =
    \frac{\xi_R}{\sqrt{N_{\mathrm{eff}}}},
    \qquad
    \Delta T_{\mathrm{loc}}
    =
    \frac{\xi_R}
         {\omega_M\sqrt{N_{\mathrm{eff}}}}.
    \label{eq:local_Wineland_clock_prediction}
\end{equation}
For ideal Gaussian squeezing with negligible loss of polarization, $\xi_R=e^{-r}$. Equations \eqref{eq:local_squeezed_time_uncertainty}--\eqref{eq:local_squeezed_minimum_product} constitute the local squeezed-state clock prediction, that the temporal width decreases exponentially with $r$, the state-energy width increases exponentially, and no independent squeezing floor appears in the ideal local theory.

\section{A Residual covariance and squeezing saturation as the nonlocal clock signature}
\label{sec:nonlocal_covariance_signature}

The local theory predicts that the clock-phase variance can be reduced continuously by increasing the squeezing parameter $r$. In the ideal principal-axis configuration we have:
\begin{equation}
    \left(\Delta T_q\right)^2(r)
    =
    T_0^2e^{-2r},
    \qquad
    \left(\Delta H_q\right)^2(r)
    =
    H_0^2e^{2r},
    \label{eq:local_principal_widths}
\end{equation}
where:
\begin{equation}
    T_0
    =
    \frac{1}
         {\omega_M\sqrt{N_{\mathrm{eff}}}},
    \qquad
    H_0
    =
    \frac{
        \hbar\omega_M\sqrt{N_{\mathrm{eff}}}
    }{2}.
    \label{eq:local_unsqueezed_widths}
\end{equation}
The subscript $q$ tells us that it is the ordinary quantum contribution. These widths satisfy:
\begin{equation}
    T_0H_0=\frac{\hbar}{2},
    \qquad
    \Delta T_q(r)\Delta H_q(r)=\frac{\hbar}{2}.
    \label{eq:local_minimum_area_again}
\end{equation}

For a realistic state with Ramsey contrast $C\leq1$ and Wineland parameter $\xi_R$, the measured phase contribution is more generally expressed as:
\begin{equation}
    \left(\Delta\phi_q\right)^2
    =
    \frac{\xi_R^2}
         {C^2N_{\mathrm{eff}}},
    \label{eq:realistic_local_phase_variance}
\end{equation}
and thus:
\begin{equation}
    \left(\Delta T_q\right)^2
    =
    \frac{\xi_R^2}
         {C^2\omega_M^2N_{\mathrm{eff}}}.
    \label{eq:realistic_local_time_variance}
\end{equation}
The ideal result in Eq.~\eqref{eq:local_principal_widths} is recovered for $C=1$ and $\xi_R=e^{-r}$.

To formulate a possible nonlocal residual response, we let $W_q(T,H;r)$ be the ideal time--energy phase-space distribution of the prepared nuclear state at squeezing strength $r$. Here $T$ is the operational clock-time estimator and $H$ is the corresponding collective clock energy. In the Gaussian tangent-plane regime, their covariance matrix is:
\begin{equation}
    \Gamma_q^{(TE)}(r)
    =
    \begin{pmatrix}
        (\Delta T_q)^2
        &
        \operatorname{Cov}_q(T,H)
        \\
        \operatorname{Cov}_q(T,H)
        &
        (\Delta H_q)^2
    \end{pmatrix}.
    \label{eq:local_TE_covariance_again}
\end{equation}

We then introduce a joint nonlocal response kernel as $K_M(u,\varepsilon)$, where $u$ is a temporal response displacement and $\varepsilon$ is an energy response displacement. The kernel is normalized as:
\begin{equation}
    \int du\,d\varepsilon\,
    K_M(u,\varepsilon)=1,
    \label{eq:response_kernel_normalization}
\end{equation}
and is taken to be centered:
\begin{equation}
    \int du\,d\varepsilon\,
    uK_M(u,\varepsilon)=0,
    \qquad
    \int du\,d\varepsilon\,
    \varepsilon K_M(u,\varepsilon)=0.
    \label{eq:response_kernel_centering}
\end{equation}
Centering means that the kernel broadens the measured distribution without producing an additional mean clock shift. A deterministic shift of the transition frequency $\omega_M-\omega_0$ has already been included in the corrected Hamiltonian and must not be counted as a covariance contribution.

The response covariance matrix is defined by:
\begin{equation}
    \Sigma_M^{(TE)}
    =
    \begin{pmatrix}
        \tau_M^2 & c_{TE}\\
        c_{TE} & \epsilon_M^2
    \end{pmatrix},
    \label{eq:nonlocal_response_covariance}
\end{equation}
where:
\begin{align}
    \tau_M^2
    &=
    \int du\,d\varepsilon\,
    u^2K_M(u,\varepsilon),
    \label{eq:nonlocal_time_width}
    \\
    \epsilon_M^2
    &=
    \int du\,d\varepsilon\,
    \varepsilon^2K_M(u,\varepsilon),
    \label{eq:nonlocal_energy_width}
    \\
    c_{TE}
    &=
    \int du\,d\varepsilon\,
    u\varepsilon K_M(u,\varepsilon).
    \label{eq:nonlocal_cross_covariance}
\end{align}
So $\tau_M$ has dimensions of time, $\epsilon_M$ has dimensions of energy, and $c_{TE}$ has dimensions of action. Positivity of the covariance matrix
requires that:
\begin{equation}
    \tau_M^2\epsilon_M^2-c_{TE}^2\geq0,
\end{equation}
or that:
\begin{equation}
    |c_{TE}|\leq\tau_M\epsilon_M.
    \label{eq:nonlocal_covariance_positivity}
\end{equation}

The measured distribution is defined by the convolution:
\begin{equation}
    W_M(T,H;r)
    =
    \int du\,d\varepsilon\,
    K_M(u,\varepsilon)
    W_q(T-u,H-\varepsilon;r).
    \label{eq:nonlocal_phase_space_convolution}
\end{equation}
This equation says that an ideal outcome $(T_q,H_q)$ is observed with the additional response displacement $(u,\varepsilon)$:
\begin{equation}
    T_M=T_q+u,
    \qquad
    H_M=H_q+\varepsilon.
    \label{eq:additive_response_variables}
\end{equation}
Assuming that the response displacement is independent of the prepared quantum fluctuations, the mixed terms vanish:
\begin{equation}
    \operatorname{Cov}(T_q,u)
    =
    \operatorname{Cov}(T_q,\varepsilon)
    =
    \operatorname{Cov}(H_q,u)
    =
    \operatorname{Cov}(H_q,\varepsilon)
    =
    0.
\end{equation}
Consequently we see that:
\begin{align}
    \operatorname{Var}(T_M)
    &=
    \operatorname{Var}(T_q)+\operatorname{Var}(u),
    \\
    \operatorname{Var}(H_M)
    &=
    \operatorname{Var}(H_q)+\operatorname{Var}(\varepsilon),
    \\
    \operatorname{Cov}(T_M,H_M)
    &=
    \operatorname{Cov}(T_q,H_q)
    +
    \operatorname{Cov}(u,\varepsilon).
\end{align}
The full covariance matrices therefore add:
\begin{equation}
    \Gamma_M^{(TE)}(r)
    =
    \Gamma_q^{(TE)}(r)
    +
    \Sigma_M^{(TE)}.
    \label{eq:nonlocal_covariance_addition}
\end{equation}

We can first consider an aligned squeezed state with:
\begin{equation}
    \operatorname{Cov}_q(T,H)=0
\end{equation}
and a diagonal response kernel $c_{TE}=0$. Now using Eq.~\eqref{eq:local_principal_widths}, covariance addition gives us:
\begin{equation}
    \left(\Delta T_M\right)^2(r)
    =
    \frac{e^{-2r}}
         {\omega_M^2N_{\mathrm{eff}}}
    +
    \tau_M^2,
    \label{eq:nonlocal_time_squeezing_curve}
\end{equation}
and:
\begin{equation}
    \left(\Delta H_M\right)^2(r)
    =
    \frac{
        \hbar^2\omega_M^2N_{\mathrm{eff}}
    }{4}
    e^{2r}
    +
    \epsilon_M^2.
    \label{eq:nonlocal_energy_squeezing_curve}
\end{equation}

The local contribution to the temporal variance decreases exponentially, but the response term does not depend on $r$. Therefore:
\begin{equation}
    \lim_{r\rightarrow+\infty}
    \Delta T_M(r)
    =
    \tau_M.
    \label{eq:nonlocal_time_floor}
\end{equation}
The clock-time uncertainty no longer approaches zero, it asymptotically approaches a nonzero squeezing floor.

\begin{widetext}
\begin{center}
  \includegraphics[width=\textwidth]{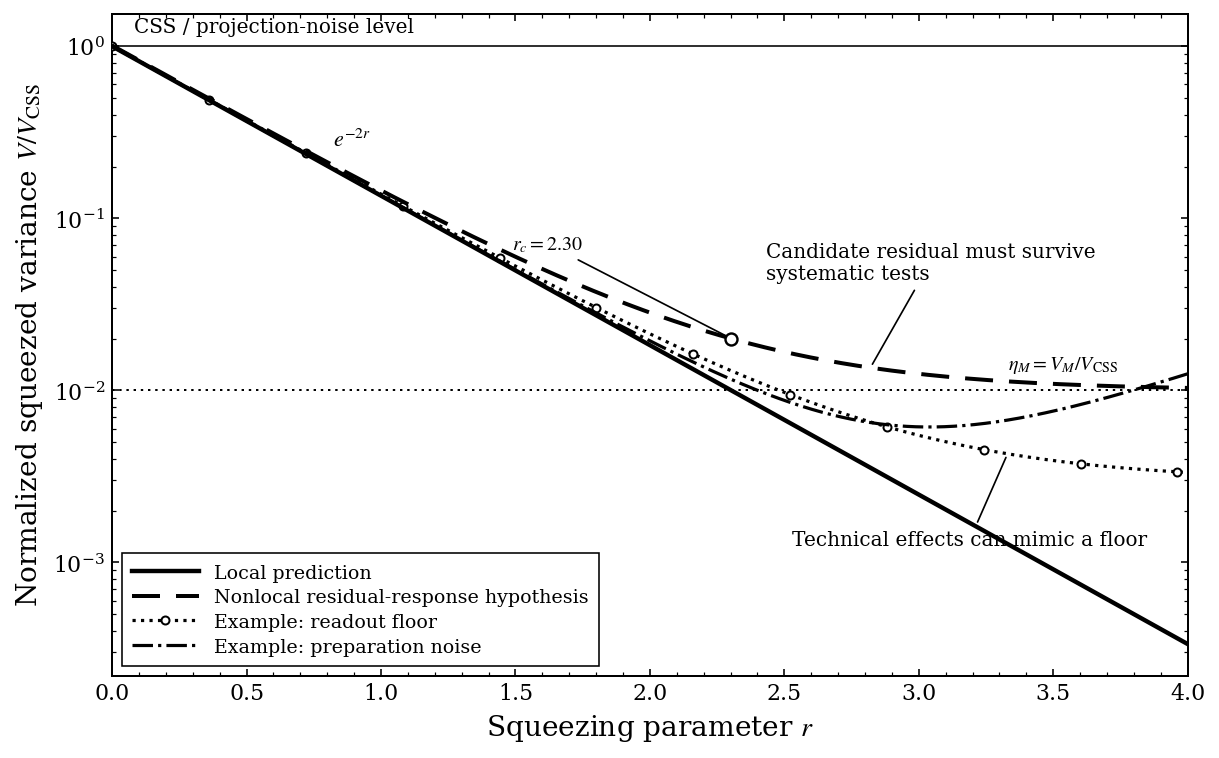}
  \captionof{figure}{The normalized squeezed-quadrature variance $V/V_{\rm CSS}$ as a function of the squeezing parameter $r$, with $V_{\rm CSS}=1/2$ the coherent-spin-state projection-noise reference. The local Gaussian prediction decreases as $V_{\rm local}/V_{\rm CSS}=e^{-2r}$, whereas the phenomenological nonlocal residual-response model $V_{\rm NL}/V_{\rm CSS}=e^{-2r}+\eta_M$ approaches the residual variance $\eta_M=V_M/V_{\rm CSS}=\tau_M^2/T_0^2$. The crossover is defined by $e^{-2r_c}=\eta_M$, giving $r_c=-\frac{1}{2}\ln\eta_M$. This for the illustrative choice $\eta_M=10^{-2}$, $r_c\simeq2.30$. The dotted and dash-dotted curves show representative readout- and preparation-noise contributions, respectively, and are included only to illustrate how technical effects can mimic a variance floor. A measured saturation thus constitutes a candidate residual only after independently calibrated systematic contributions have been excluded.}
  \label{fig:local-nonlocal-squeezing}
\end{center}
\end{widetext}

As shown in Fig.~\ref{fig:local-nonlocal-squeezing}, the distinction is not simply an enlarged variance at a single squeezing setting. The local contribution continues to decrease exponentially with $r$, whereas an $r$-independent residual response produces asymptotic saturation. Technical noise can generate similar behavior, and that is why the residual must be tested against independently calibrated systematic contributions.

The conjugate energy floor is exposed by squeezing in the opposite direction. Equivalently, we may rotate the squeezing axis by $\pi/2$ or formally replace $r\rightarrow-r$, then:
\begin{equation}
    \lim_{r\rightarrow-\infty}
    \Delta H_M(r)
    =
    \epsilon_M.
    \label{eq:nonlocal_energy_floor}
\end{equation}
The experimental signature is therefore not jus an enlarged variance at one squeezing setting. It is a systematic departure from the local exponential scaling over a tunable family of states.

For the diagonal independent-kernel model we have:
\begin{align}
    \left[
        \Delta T_M(r)\Delta H_M(r)
    \right]^2
    &=
    \left(
        T_0^2e^{-2r}+\tau_M^2
    \right)
    \left(
        H_0^2e^{2r}+\epsilon_M^2
    \right)
    \nonumber\\
    &=
    T_0^2H_0^2
    +
    \tau_M^2\epsilon_M^2
    +
    T_0^2\epsilon_M^2e^{-2r}
    +
    \tau_M^2H_0^2e^{2r}.
    \label{eq:nonlocal_product_expansion}
\end{align}
Differentiating with respect to $r$ gives us:
\begin{equation}
    -2T_0^2\epsilon_M^2e^{-2r}
    +
    2\tau_M^2H_0^2e^{2r}
    =
    0.
\end{equation}
Hence the stationary squeezing strength satisfies:
\begin{equation}
    e^{4r_*}
    =
    \frac{
        T_0^2\epsilon_M^2
    }{
        \tau_M^2H_0^2
    },
\end{equation}
or:
\begin{equation}
    r_*
    =
    \frac{1}{2}
    \ln\left(
        \frac{T_0\epsilon_M}
             {\tau_MH_0}
    \right).
    \label{eq:optimal_nonlocal_squeezing}
\end{equation}
At this point the two $r$-dependent cross terms are equal, and:
\begin{equation}
    \min_r
    \left[
        \Delta T_M(r)\Delta H_M(r)
    \right]
    =
    T_0H_0+\tau_M\epsilon_M.
\end{equation}
Since $T_0H_0=\hbar/2$:
\begin{equation}
    \min_r
    \left[
        \Delta T_M(r)\Delta H_M(r)
    \right]
    =
    \frac{\hbar}{2}
    +
    \tau_M\epsilon_M.
    \label{eq:minimum_nonlocal_TE_product}
\end{equation}
This additive expression is valid only when the local and response covariance matrices are diagonal in the same basis and their fluctuations are independent.

For the general case, we may write:
\begin{equation}
    \Gamma_q^{(TE)}
    =
    \begin{pmatrix}
        a & c\\
        c & b
    \end{pmatrix},
    \qquad
    \Sigma_M^{(TE)}
    =
    \begin{pmatrix}
        \alpha & \gamma\\
        \gamma & \beta
    \end{pmatrix},
    \label{eq:general_covariance_components}
\end{equation}
with:
\begin{equation}
    \alpha=\tau_M^2,
    \qquad
    \beta=\epsilon_M^2,
    \qquad
    \gamma=c_{TE}.
\end{equation}
Then:
\begin{align}
    \det\Gamma_M^{(TE)}
    &=
    (a+\alpha)(b+\beta)
    -
    (c+\gamma)^2
    \nonumber\\
    &=
    \det\Gamma_q^{(TE)}
    +
    a\beta+b\alpha-2c\gamma
    +
    \det\Sigma_M^{(TE)}.
    \label{eq:general_nonlocal_determinant}
\end{align}
A correlated response therefore cannot, in general, be characterized by the simple product $\tau_M\epsilon_M$. The rigorous experimental target is the entire residual matrix, represented by
\begin{equation}
    \Sigma_M^{(TE)}
    =
    \Gamma_M^{(TE)}(r)
    -
    \Gamma_q^{(TE)}(r).
    \label{eq:experimentally_reconstructed_residual}
\end{equation}
Measurements at several squeezing magnitudes and quadrature angles should fit the three independent residual parameters
\begin{equation}
    \tau_M^2,
    \qquad
    \epsilon_M^2,
    \qquad
    c_{TE},
\end{equation}
rather than imposing $c_{TE}=0$ in advance. Under the local null hypothesis:
\begin{equation}
    \Sigma_M^{(TE)}=0,
\end{equation}
and the phase variance continues to scale as $e^{-2r}$. Under the operational nonlocal hypothesis, the reconstructed covariance contains an $r$-independent positive residual and the squeezed variance saturates.

The strict minimal Moffat--Thompson Hamiltonian derived above produces the corrected transition frequency:
\begin{equation}
    \omega_M
    =
    \omega_0
    +
    \frac{\Delta q_3}
         {\hbar E_M^2}
    +
    \mathcal{O}(E_M^{-4}).
\end{equation}
This is a deterministic line displacement, but it does not by itself imply $\tau_M$, $\epsilon_M$, or $c_{TE}$.

A temporal response scale may be parameterized dimensionally as:
\begin{equation}
    \tau_M
    =
    \alpha_t\frac{\hbar}{E_M},
    \label{eq:temporal_response_Moffat_scale}
\end{equation}
where $\alpha_t$ is a dimensionless coefficient fixed by the form factor, clock observable, and measurement prescription. An independent energy width $\epsilon_M$, however, cannot be derived from dimensional analysis alone. It requires a specified microscopic energy-response kernel, fluctuating nuclear matrix elements, open-system dynamics, or a joint time--energy measurement POVM.

Accordingly, Eq.~\eqref{eq:nonlocal_covariance_addition} is an nonlocal completion and a falsifiable experimental signal model. Its characteristic signature is a residual covariance matrix that remains after the known quantum-state covariance, contrast loss, finite particle number, and ordinary technical noise have been accounted for. The main question is whether the measured clock-phase variance continues to decrease according to the local $e^{-2r}$ law or approaches a reproducible, nonzero squeezing floor.

\section{Ramsey quadrature readout and reconstruction of the nuclear time--energy ellipse}
\label{sec:ramsey_covariance_tomography}

The clock state and its time--energy covariance cannot just be inferred from a single population measurement. Direct detection naturally measures the ground--isomer population difference, represented by $J_z$, whereas the operational clock-time observable is proportional to the transverse phase quadrature $J_y$. The complete covariance matrix is therefore reconstructed by combining Ramsey interrogation with controlled collective analysis rotations.

Now we look at Ramsey phase accumulation. To start let the squeezed nuclear ensemble be prepared with its mean collective spin oriented along the positive $x$ direction:
\begin{equation}
    \langle J_x\rangle=S>0,
    \qquad
    \langle J_y\rangle\simeq0,
    \qquad
    \langle J_z\rangle\simeq0,
    \label{eq:tomography_operating_point}
\end{equation}
where $S$ is the remaining collective coherence and equals $N_{\mathrm{eff}}/2$ for an ideal fully polarized ensemble.

During the Ramsey free-evolution interval $T_R$, the nuclear transition of angular frequency $\omega_M$ is compared with an interrogation field of angular frequency $\omega_L$. We define the detuning as:
\begin{equation}
    \delta=\omega_M-\omega_L,
    \label{eq:tomography_detuning}
\end{equation}
the rotating-frame Hamiltonian is given by:
\begin{equation}
    H_R
    =
    \hbar\delta J_z.
    \label{eq:collective_ramsey_hamiltonian}
\end{equation}
The corresponding collective propagator is:
\begin{equation}
    U_R(T_R)
    =
    \exp\left(
        -i\delta T_RJ_z
    \right)
    =
    e^{-i\phi J_z},
    \label{eq:collective_ramsey_propagator}
\end{equation}
where:
\begin{equation}
    \phi=\delta T_R
    \label{eq:ramsey_accumulated_phase_tomography}
\end{equation}
is the accumulated Ramsey phase.

Under this rotation:
\begin{equation}
    U_R^\dagger J_yU_R
    =
    J_y\cos\phi+J_x\sin\phi.
    \label{eq:ramsey_Jy_rotation_tomography}
\end{equation}
For operation near the centre of the Ramsey fringe, when $|\phi|\ll1$. We have:
\begin{equation}
    J_y(\phi)
    \simeq
    J_y+\phi J_x.
\end{equation}
Taking the expectation value gives us:
\begin{equation}
    \langle J_y(\phi)\rangle
    \simeq
    S\phi.
    \label{eq:ramsey_phase_to_collective_signal}
\end{equation}
The transverse collective spin thus records the phase accumulated relative to the interrogation field. The local phase estimator is given by:
\begin{equation}
    \widehat{\phi}
    =
    \frac{J_y}{S},
\end{equation}
while the corresponding detuning estimator is:
\begin{equation}
    \widehat{\delta}
    =
    \frac{\widehat{\phi}}{T_R}
    =
    \frac{J_y}{ST_R}.
    \label{eq:ramsey_detuning_estimator_tomography}
\end{equation}

The directly measured nuclear observable is the number of nuclei occupying
the isomeric state is:
\begin{equation}
    N_m
    =
    J_z+\frac{N_{\mathrm{eff}}}{2}.
    \label{eq:isomer_number_tomography}
\end{equation}
Consequently we have:
\begin{equation}
    \langle J_z\rangle
    =
    \langle N_m\rangle
    -
    \frac{N_{\mathrm{eff}}}{2},
    \qquad
    (\Delta J_z)^2
    =
    (\Delta N_m)^2.
    \label{eq:population_readout_relation}
\end{equation}
A population measurement thus gives the energy-population quadrature directly, because the centered clock energy is given by:
\begin{equation}
    \delta H
    =
    \hbar\omega_M
    \left(
        J_z-\langle J_z\rangle
    \right).
    \label{eq:energy_from_population_tomography}
\end{equation}
It does not, without an analysis rotation, directly measure the conjugate phase quadrature $J_y$.

Before population detection, one would apply a controlled collective rotation about the mean-spin axis:
\begin{equation}
    R_x(\theta)
    =
    e^{-i\theta J_x},
    \label{eq:tomography_analysis_rotation}
\end{equation}
where $\theta$ is the chosen analysis angle. In the Heisenberg picture, the population operator becomes:
\begin{align}
    J_z(\theta)
    &=
    R_x^\dagger(\theta)J_zR_x(\theta)
    \nonumber\\
    &=
    J_z\cos\theta-J_y\sin\theta.
    \label{eq:rotated_population_quadrature}
\end{align}
So the same physical population detector measures a different transverse
quadrature for each value of $\theta$, in particular:
\begin{equation}
    J_z(0)=J_z,
    \qquad
    J_z\left(\frac{\pi}{2}\right)=-J_y.
    \label{eq:direct_and_phase_quadratures}
\end{equation}
The sign at $\theta=\pi/2$ is fixed by the rotation convention and has no
effect on the measured variance.

We then define the transverse variances and symmetrized covariance by:
\begin{equation}
    V_y
    =
    \langle(\delta J_y)^2\rangle,
    \qquad
    V_z
    =
    \langle(\delta J_z)^2\rangle,
    \label{eq:tomography_spin_variances}
\end{equation}
and:
\begin{equation}
    C_{yz}
    =
    \frac{1}{2}
    \left\langle
        \delta J_y\delta J_z
        +
        \delta J_z\delta J_y
    \right\rangle.
    \label{eq:tomography_spin_covariance}
\end{equation}
The variance measured after an analysis rotation is:
\begin{align}
    V_\theta
    &\equiv
    \left[
        \Delta J_z(\theta)
    \right]^2
    \nonumber\\
    &=
    V_z\cos^2\theta
    +
    V_y\sin^2\theta
    -
    2C_{yz}\sin\theta\cos\theta.
    \label{eq:rotated_quadrature_variance_tomography}
\end{align}

Measurements at four analysis angles are sufficient to reconstruct the full symmetric covariance matrix:
\begin{align}
    V_0&=V_z,
    \label{eq:tomography_V0}
    \\
    V_{\pi/2}&=V_y,
    \label{eq:tomography_Vpi2}
    \\
    V_{\pi/4}
    &=
    \frac{V_y+V_z}{2}-C_{yz},
    \label{eq:tomography_Vpi4}
    \\
    V_{-\pi/4}
    &=
    \frac{V_y+V_z}{2}+C_{yz}.
    \label{eq:tomography_Vminuspi4}
\end{align}
It thus follows that:
\begin{equation}
    C_{yz}
    =
    \frac{
        V_{-\pi/4}-V_{\pi/4}
    }{2}.
    \label{eq:reconstructed_spin_covariance}
\end{equation}
Each $V_\theta$ is obtained from repeated preparations and population
measurements at fixed $\theta$. Because the readout is generally destructive, different analysis angles require separately prepared realizations of the same nominal clock state.

The reconstructed transverse-spin covariance matrix is:
\begin{equation}
    \Gamma_J
    =
    \begin{pmatrix}
        V_y & C_{yz}\\
        C_{yz} & V_z
    \end{pmatrix}.
    \label{eq:reconstructed_spin_covariance_matrix}
\end{equation}
Its eigenvalues give the squeezed and anti-squeezed principal variances, while its eigenvectors would determine the orientation of the nuclear uncertainty ellipse.

The operational clock variables are related linearly to the collective-spin
quadratures as:
\begin{equation}
    \delta T
    =
    \frac{\delta J_y}{\omega_MS},
    \qquad
    \delta H
    =
    \hbar\omega_M\delta J_z.
    \label{eq:spin_to_TE_tomography_map}
\end{equation}
The measured time--energy covariance matrix is thus:
\begin{equation}
    \Gamma_{TE}
    =
    \begin{pmatrix}
        \displaystyle
        \frac{V_y}{\omega_M^2S^2}
        &
        \displaystyle
        \frac{\hbar C_{yz}}{S}
        \\[8pt]
        \displaystyle
        \frac{\hbar C_{yz}}{S}
        &
        \displaystyle
        \hbar^2\omega_M^2V_z
    \end{pmatrix}.
    \label{eq:tomographically_reconstructed_TE_matrix}
\end{equation}
So then:
\begin{align}
    (\Delta T)^2
    &=
    \frac{V_y}{\omega_M^2S^2},
    \\
    (\Delta H)^2
    &=
    \hbar^2\omega_M^2V_z,
    \\
    \operatorname{Cov}(T,H)
    &=
    \frac{\hbar C_{yz}}{S}.
\end{align}
Its determinant is:
\begin{equation}
    \det\Gamma_{TE}
    =
    \frac{\hbar^2}{S^2}
    \left(
        V_yV_z-C_{yz}^2
    \right)
    \geq
    \frac{\hbar^2}{4}.
    \label{eq:tomographic_TE_determinant}
\end{equation}

Ramsey interrogation therefore does more than determine the mean clock
frequency. Combined with analysis rotations and repeated population
measurements, it can reconstruct the complete orientation, eccentricity, and area of the nuclear time--energy uncertainty ellipse. This reconstructed matrix is the main observable to be compared with the local squeezed-state prediction and searched for an additional residual nonlocal covariance.

\section{Systematic covariance budget and rejection of spurious squeezing floors}
\label{sec:false_positive_rejection}

A saturation of the squeezed clock variance is not by itself evidence for a nonlocal response. The measured time--energy covariance must be decomposed as:
\begin{equation}
    \Gamma_{\mathrm{meas}}^{(TE)}
    =
    \Gamma_q^{(TE)}
    +
    \Sigma_M^{(TE)}
    +
    \Sigma_{\mathrm{laser}}^{(TE)}
    +
    \Sigma_{\mathrm{host}}^{(TE)}
    +
    \Sigma_{\mathrm{ro}}^{(TE)}
    +
    \Sigma_{\mathrm{prep}}^{(TE)},
    \label{eq:complete_noise_covariance_budget}
\end{equation}
where $\Gamma_q^{(TE)}$ is the predicted quantum-state covariance, $\Sigma_M^{(TE)}$ is the putative nonlocal residual, and the remaining terms describe interrogation-laser noise, crystal-environment fluctuations, readout noise, and imperfect state preparation.

For the measured phase variance, the corresponding scalar model is given by:
\begin{equation}
    V_{\phi,\mathrm{meas}}
    =
    \frac{\xi_R^2}{C^2N_{\mathrm{eff}}}
    +
    \omega_M^2\tau_M^2
    +
    V_{\phi,\mathrm{tech}},
    \label{eq:measured_phase_noise_model}
\end{equation}
where:
\begin{equation}
    V_{\phi,\mathrm{tech}}
    =
    V_{\phi,\mathrm{laser}}
    +
    V_{\phi,\mathrm{host}}
    +
    V_{\phi,\mathrm{ro}}
    +
    V_{\phi,\mathrm{prep}},
\end{equation}
where the first term is the local squeezed-state prediction, including the Wineland parameter $\xi_R$, Ramsey contrast $C$, and effective coherent particle number $N_{\mathrm{eff}}$. The second is the candidate temporal response floor.

Each technical contribution has a distinguishable control dependence. The laser-frequency noise changes with Ramsey duration and the measured laser-noise spectrum. The host noise changes with temperature, magnetic field, strain, crystal orientation, and material; readout noise is determined from dark and unsqueezed calibration measurements. The preparation noise tracks pulse-area, phase, and squeezing-control errors~\cite{Lindblad1976,Gorini1976,BreuerPetruccione2002,GardinerZoller2004,Huelga1997,Escher2011,Demkowicz2012,Wang2025NuclearQubits}. A candidate nonlocal covariance may be assigned only to the residual:
\begin{equation}
    \widehat{\Sigma}_M^{(TE)}
    =
    \Gamma_{\mathrm{meas}}^{(TE)}
    -
    \Gamma_q^{(TE)}
    -
    \sum_j\widehat{\Sigma}_j^{(TE)},
    \label{eq:calibrated_nonlocal_residual}
\end{equation}
and only if it remains positive, reproducible, and independent of the control parameters that govern the calibrated technical noises. A full microscopic noise analysis and the corresponding null tests will be developed in a dedicated follow-up study.

\section{Likelihood-based inference of the residual covariance and nonlocality scale}
\label{sec:statistical_inference_EM}

The statistical analysis would need to be done in two stages. We first determine whether the data require a positive residual time--energy covariance after subtracting the local quantum prediction and calibrated technical noise. We then would translate the resulting confidence interval into a lower bound on the Moffat scale $E_M$. This ordering helps avoid assuming the nonlocal interpretation before a residual has been established.

Suppose that the experiment is repeated at squeezing settings:
\begin{equation}
    r_1,r_2,\ldots,r_K.
\end{equation}
At setting $j$, the measured phase variance is modeled as:
\begin{equation}
    V_{\phi,\mathrm{meas}}^{(j)}
    =
    \frac{\xi_{R,j}^{\,2}}
         {C_j^{\,2}N_{\mathrm{eff},j}}
    +
    \omega_M^2\tau_M^2
    +
    V_{\phi,\mathrm{tech}}^{(j)},
    \label{eq:phase_variance_statistical_model}
\end{equation}
where $\xi_{R,j}$ is the measured Wineland squeezing parameter, $C_j$ is the Ramsey contrast, $N_{\mathrm{eff},j}$ is the effective coherent particle number, and $V_{\phi,\mathrm{tech}}^{(j)}$ is the independently calibrated technical phase-noise contribution. The parameter $\tau_M^2\geq0$ is the possible nonlocal temporal-response variance. The factor $\omega_M^2$ converts it into phase variance because $\phi=\omega_MT$.

The more complete analysis uses the reconstructed time--energy covariance
matrix of:
\begin{equation}
    \Gamma_{\mathrm{meas}}^{(j)}
    =
    \Gamma_q(r_j)
    +
    \Sigma_M
    +
    \Sigma_{\mathrm{tech}}^{(j)},
    \label{eq:covariance_statistical_model}
\end{equation}
where:
\begin{equation}
    \Sigma_M
    =
    \begin{pmatrix}
        \tau_M^2 & c_{TE}\\
        c_{TE} & \epsilon_M^2
    \end{pmatrix}.
    \label{eq:fitted_nonlocal_covariance}
\end{equation}
The parameters $\epsilon_M^2$ and $c_{TE}$ are respectively the residual energy variance and time--energy cross covariance. Physical positivity requires that:
\begin{equation}
    \tau_M^2\geq0,
    \qquad
    \epsilon_M^2\geq0,
    \qquad
    c_{TE}^2\leq\tau_M^2\epsilon_M^2.
    \label{eq:fitted_covariance_constraints}
\end{equation}

We let $\widehat{\Gamma}_j$ be the sample covariance reconstructed from $n_j$ independent preparations at setting $r_j$~\cite{Wishart1928,Anderson2003}. In the Gaussian approximation, the covariance likelihood can be written as:
\begin{equation}
    -2\ln\mathcal{L}
    =
    \sum_{j=1}^{K}
    n_j
    \left[
        \ln\det\Gamma_j
        +
        \operatorname{tr}
        \left(
            \Gamma_j^{-1}\widehat{\Gamma}_j
        \right)
    \right]
    +\mathrm{constant},
    \label{eq:covariance_likelihood}
\end{equation}
where:
\begin{equation}
    \Gamma_j
    =
    \Gamma_q(r_j)
    +
    \Sigma_M
    +
    \Sigma_{\mathrm{tech}}^{(j)}.
\end{equation}
Uncertain calibration quantities, including $C_j$, $N_{\mathrm{eff},j}$, $\xi_{R,j}$, and the technical-noise parameters, are included as nuisance parameters and profiled or marginalized rather than fixed without uncertainty.

The local and nonlocal hypotheses are respectively:
\begin{align}
    H_0:&\qquad \Sigma_M=0,
    \label{eq:local_null_hypothesis}
    \\
    H_1:&\qquad \Sigma_M\succeq0,
    \label{eq:nonlocal_alternative_hypothesis}
\end{align}
where $\succeq0$ means a positive-semidefinite matrix. A likelihood-ratio statistic compares the best fit under these hypotheses~\cite{Wilks1938,Cowan2011}:
\begin{equation}
    q_0
    =
    -2\ln
    \frac{
        \mathcal{L}(\Sigma_M=0,\widehat{\widehat{\boldsymbol{\eta}}})
    }{
        \mathcal{L}(\widehat{\Sigma}_M,\widehat{\boldsymbol{\eta}})
    },
    \label{eq:null_likelihood_ratio}
\end{equation}
where here $\boldsymbol{\eta}$ represents the nuisance parameters, a single hat represents their global maximum-likelihood values, and a double hat represents their conditional values under $H_0$. Because the null lies on the physical boundary $\Sigma_M=0$, the significance and confidence thresholds should be calibrated with boundary-corrected asymptotics or Monte Carlo simulations~\cite{Chernoff1954,SelfLiang1987}.

If no significant excess is found, the profile likelihood gives upper limits of:
\begin{equation}
    \tau_M<\tau_{\mathrm{lim}},
    \qquad
    \epsilon_M<\epsilon_{\mathrm{lim}},
    \label{eq:response_upper_limits}
\end{equation}
at the chosen confidence level. For the temporal-response parameterization:
\begin{equation}
    \tau_M
    =
    \alpha_t\frac{\hbar}{E_M},
    \label{eq:temporal_scale_parameterization}
\end{equation}
where $\alpha_t$ is a dimensionless response coefficient, the temporal limit implies:
\begin{equation}
    E_M
    >
    \alpha_t\frac{\hbar}{\tau_{\mathrm{lim}}}.
    \label{eq:temporal_bound_on_EM}
\end{equation}

A nuclear-enhanced energy-response model may instead be written as:
\begin{equation}
    \frac{\epsilon_M}{E_{\mathrm{Th}}}
    =
    A_\epsilon
    \left(
        \frac{E_{\mathrm{ref}}}{E_M}
    \right)^2,
    \label{eq:energy_response_parameterization}
\end{equation}
where $E_{\mathrm{Th}}=\hbar\omega_M$ is the clock-transition energy, $E_{\mathrm{ref}}$ is the nuclear scale controlling the relevant matrix element, and $A_\epsilon$ is a dimensionless matching coefficient. The upper limit on $\epsilon_M$ then gives us:
\begin{equation}
    E_M
    >
    E_{\mathrm{ref}}
    \sqrt{
        \frac{
            A_\epsilon E_{\mathrm{Th}}
        }{
            \epsilon_{\mathrm{lim}}
        }
    },
    \label{eq:energy_bound_on_EM}
\end{equation}
where the coefficient $A_\epsilon$ must remain explicit unless it has been fixed by a microscopic nuclear calculation. Any reported constraint on $E_M$ is thus conditional on the specified temporal or nuclear-response matching model.

\section{Concluding Remarks}
\label{sec:conclusion}

We have presented a theoretical framework for using the new $^{229}$Th nuclear clocks as a probe of the nonlocal time--energy uncertainty principle. By identifying time with the temporal offset inferred from the collective nuclear phase. The conjugate energy observable is supplied by fluctuations of the ground--isomer population difference. So near the Ramsey operating point, these quantities form an effective canonical pair and obey the ordinary time--energy uncertainty relation.

For an ensemble of coherently interrogated nuclei, the clock transition can be described through collective-spin observables where an unsqueezed coherent spin state reaches the standard quantum limit, while a nonlinear collective interaction can prepare an entangled squeezed state with tunable squeezing strength. This squeezing reduces the clock-phase uncertainty and increases the conjugate energy-population uncertainty. So it redistributes the quantum uncertainty rather than removing it. 

Ramsey interrogation combined with controlled analysis rotations provides us with a direct method for reconstructing the complete transverse-spin covariance matrix. This reconstruction determines the squeezed and anti-squeezed variances, their correlation, and the orientation and area of the uncertainty ellipse in phase space. Mapping these observables into time--energy variables then yields a measurable covariance matrix that can be compared with the local quantum prediction across a tunable family of squeezed states to potentially validate our hypothesis.

Within the Moffat--Thompson formalism of nonlocal quantum field theory, nonlocal deformation of the free kinetic operator does not move the clock transition or broaden its line by itself. Here a physical correction comes into play only when the regulated interactions are matched onto the nuclear Hamiltonian. The resulting leading effect is a deterministic modification of the ground--isomer energy difference. 

A credible signal must persist under systematic variations and appear consistently in the reconstructed covariance matrix, since a phase floor by itself would not be decisive. This is because laser noise, decoherence, crystal fluctuations, imperfect preparation, finite readout resolution, and errors in the effective particle number can produce similar behavior, this will be covered in a dedicated follow-up.

A null result would still be informative as it would constrain the allowed response covariance and, after specifying the relevant microscopic matching coefficients, place a lower bound on the Moffat scale. Our present work so far establishes the theoretical and statistical structure of the test.

\section*{Acknowledgments}

We thank Rajibul Islam, Lauren Blanc, Martin Green, Laurent Freidel, and Savas Dimopoulos for insightful and stimulating discussions. Research at the Perimeter Institute for Theoretical Physics is supported by the Government of Canada through Industry Canada and by the Province of Ontario through the Ministry of Research and Innovation (MRI).

\end{document}